\documentclass[article, oneside]{memoir}
\usepackage{amsmath}
\usepackage{amssymb}
\usepackage{graphicx}
\usepackage{bm}
\usepackage[sort&compress, numbers]{natbib}

\graphicspath{{figs/}} %All figures in the figs subdirectory

\captionnamefont{\bfseries\small}
\captiontitlefont{\small}
\captiondelim{. }

\let\origfigure=\figure
\let\endorigfigure=\endfigure
\renewenvironment{figure}{\origfigure\centering}{\endorigfigure}

\newcommand{\bra}[1]{\ensuremath{\langle#1|}}
\newcommand{\ket}[1]{\ensuremath{|#1\rangle}}
\newcommand{\mtxel}[3]{\ensuremath{\langle#1|#2|#3\rangle}}
\newcommand{\innerprod}[2]{\ensuremath{\langle#1|#2\rangle}}
\renewcommand{\exp}[1]{\ensuremath{\mathrm{e}^{#1}}}

\newcommand{\vect}[1]{\ensuremath{\bm{#1}}}
\newcommand{\uvect}[1]{\ensuremath{\bm{\hat{#1}}}}

\newcommand{\kL}{\ensuremath{\uvect{k}_\mathrm{L}}}
\newcommand{\kM}{\ensuremath{\uvect{k}_\mathrm{M}}}

\renewcommand{\eqref}[1]{Eq.~(\ref{#1})}
\newcommand{\figref}[1]{Fig.~\ref{#1}}
\newcommand{\secref}[1]{Section~\ref{#1}}

\newcommand{\drot}[4]{\ensuremath{D_{#2#3}^{#1}(#4)}}
\newcommand{\Drot}[3]{\drot{#1}{#2}{#3}{\phi, \theta, \chi}}

\newcommand{\threej}[6]{\ensuremath{%
    \begin{pmatrix}
      #1&#2&#3\\
      #4&#5&#6
    \end{pmatrix}}}

\newcommand{\sixj}[6]{\ensuremath{%
    \begin{Bmatrix}
      #1&#2&#3\\
      #4&#5&#6
    \end{Bmatrix}}}

\newcommand{\ninej}[9]{\ensuremath{%
    \begin{Bmatrix}
      #1&#2&#3\\
      #4&#5&#6\\
      #7&#8&#9
    \end{Bmatrix}}}

\newcommand{\multipole}[4]{\ensuremath{%
    T(#1,#2)^\dagger_{#3#4}}}
\newcommand{\multmom}[4]{\ensuremath{%
    {\langle}T(#1,#2)^\dagger_{#3#4}{\rangle}}}

\newcommand{\alp}{\ensuremath{{\alpha_+}}}
\newcommand{\phase}[1]{\ensuremath{(-1)^{#1}}}

\newcommand{\Jpl}{\ensuremath{J_{\alpha_+}}}
\newcommand{\Kpl}{\ensuremath{K_{\alpha_+}}}
\newcommand{\Mpl}{\ensuremath{M_{\alpha_+}}}

\newcommand{\Ja}{\ensuremath{J_\alpha}}
\newcommand{\Ma}{\ensuremath{M_\alpha}}
\newcommand{\Ka}{\ensuremath{K_\alpha}}

\newcommand{\Jap}{\ensuremath{J'_{\alpha'}}}
\newcommand{\Map}{\ensuremath{M'_{\alpha'}}}
\newcommand{\Kap}{\ensuremath{K'_{\alpha'}}}

\begin{document}

\title{\textbf{TIME-RESOLVED PHOTOELECTRON SPECTROSCOPY OF
    NON-ADIABATIC
    DYNAMICS IN POLYATOMIC MOLECULES}\\
  \emph{\normalsize{Published in Advances in Chemical Physics, Volume
      139 (ed S. A. Rice), John Wiley \& Sons, Inc., Hoboken, NJ,
      USA. doi: 10.1002/9780470259498.ch6}}
}

\author{
  ALBERT STOLOW\\
  \emph{\normalsize{Steacie Institute for Molecular Sciences, 
      National Research Council Canada}}\\
  \emph{\normalsize{Ottawa, Ontario, K1A 0R6, Canada.}}\\
  \vspace{0.5em}\\
  JONATHAN G. UNDERWOOD\\
  \emph{\normalsize{Department of Physics and Astronomy, University College London}}\\
  \emph{\normalsize{Gower Street, London, WC1E 6BT}}
}

\date{}

\maketitle
\tableofcontents*

\chapter{Introduction}
The photodynamics of polyatomic molecules generally involves complex
intramolecular processes which rapidly redistribute both charge and
vibrational energy within the molecule. The coupling of vibrational and
electronic degrees of freedom leads to the processes known as radiationless
transitions, internal conversion, isomerization, proton and electron transfer
etc.~\cite{Bixon1968, Jortner1969, Henry1973, Freed1976, Stock1997, Worth2004,
  Klessinger1994, Koppel1984}. These non-adiabatic dynamics underlie the
photochemistry of almost all polyatomic molecules~\cite{Michl1990} and are
important in photobiological processes such as vision and
photosynthesis~\cite{Schoenlein1991}, and underlie many concepts in active
molecular electronics~\cite{Jortner1997}. The coupling of charge with energy
flow is often understood in terms of the breakdown of the Born-Oppenheimer
approximation (BOA), an adiabatic separation of electronic from nuclear
motions. The BOA allows the definition of the nuclear potential energy
surfaces that describe both molecular structures and nuclear trajectories,
thereby permitting a mechanistic picture of molecular dynamics. The breakdown
of the BOA is uniquely due to nuclear dynamics and occurs at the intersections
or near intersections of potential energy surfaces belonging to different
electronic states. Non-adiabatic coupling often leads to complex, broadened
absorption spectra due to the high density of nuclear states and strong
variations of transition dipole with nuclear coordinate. In this situation,
the very notion of distinct and observable vibrational and electronic states
is obscured. The general treatment of these problems remains one of the most
challenging problems in molecular physics, particularly when the state density
becomes high and multi-mode vibronic couplings are involved. Our interest is
in developing time-resolved methods for the experimental study of
non-adiabatic molecular dynamics. The development of femtosecond methods for
the study of gas-phase chemical dynamics is founded upon the seminal studies
of A.H. Zewail and co-workers, as recognized in 1999 by the Nobel Prize in
Chemistry~\cite{Zewail2000}. This methodology has been applied to chemical
reactions ranging in complexity from bond-breaking in diatomic molecules to
dynamics in larger organic and biological molecules.

Femtosecond time-resolved methods involve a pump-probe configuration in which
an ultrafast pump pulse initiates a reaction or, more generally, creates a
nonstationary state or wavepacket, the evolution of which is monitored as a
function of time by means of a suitable probe pulse. Time-resolved or
wavepacket methods offer a view complementary to the usual spectroscopic
approach and often yield a physically intuitive picture. Wave packets can
behave as zeroth-order or even classical-like states and are therefore very
helpful in discerning underlying dynamics. The information obtained from these
experiments is very much dependent on the nature of the final state chosen in
a given probe scheme. Transient absorption and nonlinear wave mixing are often
the methods of choice in condensed-phase experiments because of their
generality. In studies of molecules and clusters in the gas phase, the most
popular methods, laser-induced fluorescence and resonant multiphoton
ionization, usually require the probe laser to be resonant with an electronic
transition in the species being monitored. However, as a chemical reaction
initiated by the pump pulse evolves toward products, one expects that both the
electronic and vibrational structures of the species under observation will
change significantly and some of these probe methods may be restricted to
observation of the dynamics within a small region of the reaction coordinate.

We focus here upon gas-phase time-resolved photoelectron spectroscopy (TRPES)
of neutral polyatomic molecules. TRPES is particularly well suited to the
study of ultrafast non-adiabatic processes because photoelectron spectroscopy
is sensitive to both electronic configurations and vibrational
dynamics~\cite{Eland1984}.  Due to the universal nature of ionization
detection, TRPES has been demonstrated to be able to follow dynamics along the
entire reaction coordinate. In TRPES experiments, a time-delayed probe laser
generates free electrons via photoionization of the evolving excited state,
and the electron kinetic energy and/or angular distribution is measured as a
function of time. As a probe, TRPES has several practical and conceptual
advantages~\cite{Fischer1995}: (a) Ionization is always an allowed process,
with relaxed selection rules due to the range of symmetries of the outgoing
electron. Any molecular state can be ionized. There are no 'dark' states in
photoionization; (b) Highly detailed, multiplexed information can be obtained
by differentially analyzing the outgoing photoelectron as to its kinetic
energy and angular distribution; (c) Charged particle detection is extremely
sensitive; (d) Detection of the ion provides mass information on the carrier
of the spectrum; (e) Higher order (multiphoton) processes, which can be
difficult to discern in femtosecond experiments, are readily revealed; (f)
Photoelectron-photoion coincidence measurements can allow for studies of
cluster solvation effects as a function of cluster size and for time-resolved
studies of scalar and vector correlations in photodissociation
dynamics. Beginning in 1996, TRPES has been the subject of a number of
reviews~\cite{Stolow1996, Stolow1998, Hayden2000, Radloff2000, Takatsuka2000,
  Neumark2001, Suzuki2001, Seideman2002, Reid2003, Stolow2003, Stolow2003a,
  Suzuki2004, Wollenhaupt2005, Suzuki2006, Hertel2006} and these cover various
aspects of the field. An exhaustive review of the TRPES literature, including
dynamics in both neutrals and anions, was published
recently~\cite{Stolow2004}. Therefore, rather than a survey, our emphasis here
will be on the conceptual foundations of TRPES and the advantages of this
approach in solving problems of non-adiabatic molecular dynamics, amplified by
examples of applications of TRPES chosen mainly from our own work.

In the following sections we begin with a review of wavepacket dynamics. We
emphasize the aspects of creating and detecting wavepackets and the special
role of the final state which acts as a ``template'' onto which the dynamics
is projected. We then discuss aspects of the dynamical problem of interest
here, namely the non-adiabatic excited state dynamics of isolated polyatomic
molecules. We believe that the molecular ionization continuum is a
particularly interesting final state for studying time-resolved non-adiabatic
dynamics. Therefore, in some detail, we consider the general process of
photoionization and discuss features of single photon photoionization dynamics
of excited molecular state and its energy and angle-resolved detection. We
briefly review the experimental techniques that are required for laboratory
studies of TRPES. As TRPES is more involved than ion detection, we felt it
important to motivate the use of photoelectron spectroscopy as a probe by
comparing mass-resolved ion yield measurements with TRPES, using the example
of internal conversion dynamics in a linear hydrocarbon molecule. Finally, we
consider various applications of TRPES, with examples selected to illustrate
the general issues that have been addressed.

%%% Local Variables:
%%% mode: latex
%%% TeX-master: "trpes"
%%% End:

\chapter{Wavepacket dynamics}
\section{Frequency and Time domain perspectives}
Time-resolved experiments on isolated systems involve the creation and
detection of wavepackets which we define to be coherent superpositions of
exact molecular eigenstates $\ket{N}$.  By definition, the exact
(non Born-Oppenheimer) eigenstates are the solutions to the time-independent
Schr\"{o}dinger equation and are stationary. Time dependence, therefore, can
only come from superposition and originates in the differing quantum
mechanical energy phase factors $\exp{-iE_{N}t/\hbar}$ associated with each
eigenstate. Conceptually, there are three steps to a pump-probe wavepacket
experiment: (i) the preparation or pump step; (ii) the dynamical evolution;
and (iii) the probing of the non-stationary superposition state.

\begin{figure}
  \includegraphics[height=8cm]{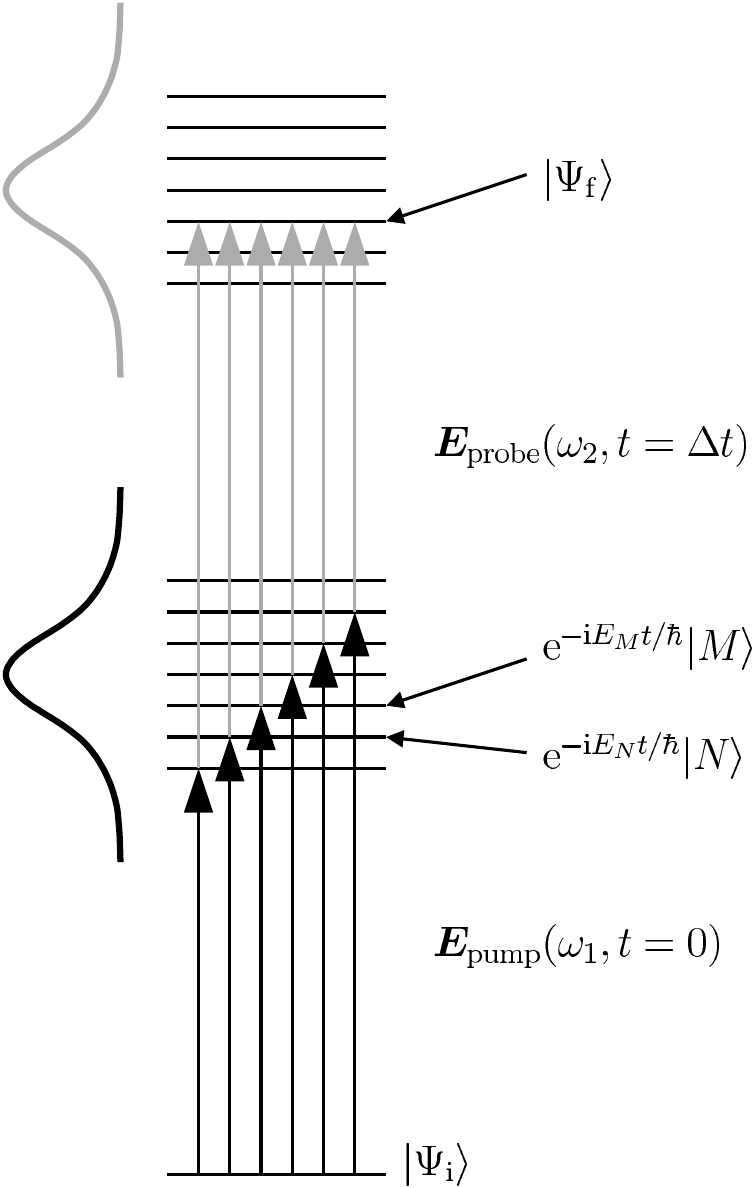}
  \caption{The creation, evolution and detection of wavepackets.  The pump
    laser pulse $\vect{E}_{\mathrm{pump}}$ (black) creates a coherent
    superposition of molecular eigenstates at $t=0$ from the ground state
    $\ket{\Psi_\mathrm{i}}$. The set of excited state eigenstates $\ket{N}$
    in the superposition (wavepacket) have different energy phase factors,
    leading to non-stationary behaviour (wavepacket evolution).  At time
    $t=\Delta t$ the wavepacket is projected by a probe pulse
    $\vect{{E}_\mathrm{probe}}$ (grey) onto a set of final states
    $\ket{\Psi_\mathrm{f}}$ which act as a ``template'' for the dynamics.  The
    time-dependent probability of being in a given final state
    $\ket{\Psi_\mathrm{f}}$ is modulated by the interferences between all
    degenerate coherent two-photon transition amplitudes leading to that final
    state.}
  \label{fig:cartoon}
\end{figure}

From a frequency domain point of view, a femtosecond pump-probe experiment,
shown schematically in \figref{fig:cartoon}, is a sum of coherent two-photon
transition amplitudes constrained by the pump and probe laser bandwidths. The
measured signal is proportional to the population in the final state
$\ket{\Psi_\mathrm{f}}$ at the end of the two pulse sequence. As these
two-photon transitions are coherent, we must therefore add the transition
amplitudes and then square in order to obtain the probability. As discussed
below, the signal contains interferences between all degenerate two-photon
transitions. When the time delay between the two laser fields is varied, the
phase relationships between the two-photon transition amplitudes changes,
modifying the interference in the final state. The amplitudes and initial
phases of the set of the initially prepared excited eigenstates are determined
by the amplitudes and phases of the pump laser field frequencies, and the
transition dipole amplitudes between the initial state and the excited state
of interest. Once the pump laser pulse is over, the wavepacket $\Psi(t)$
evolves freely according to relative energy phase factors in the superposition
as given by
\begin{equation}
  \label{eq:eigenpacket}
  \ket{\Psi(t)}=\sum_N A_N \exp{-iE_Nt/\hbar}\ket{N}.
\end{equation}
The complex coefficients $A_{N}$ contain both the amplitudes and initial
phases of the exact molecular eigenstates $\ket{N}$ which are prepared by the
pump laser, and the $E_{N}$ are the excited state eigenenergies.  The probe
laser field interacts with the wavepacket after the pump pulse is over,
projecting it onto a specific final state $|\Psi_{f}\rangle$ at some time
delay $\Delta t$.  This final state is the ``template'' onto which the
wavepacket dynamics are projected.  The time dependence of the differential
signal, S$_{f}(\Delta t)$, for projection onto a single final state can be
written as
\begin{equation}
  \label{eq:sig}
  S_f(t)=\left|
    \mtxel{\Psi_f}{\vect{E}_{\mathrm{probe}}(\omega)\cdot\vect{d}}{\Psi(t)}
  \right|^2
  =\left|
    \sum_N B_N\exp{-iE_N t/\hbar}
  \right|^2,
\end{equation}
where the complex coefficients $B_N$ contain both the wavepacket amplitudes
$A_N$ and the (complex) probe transition dipole matrix elements connecting
each eigenstate in the superposition $\ket{N}$ to the final state,
\begin{equation}
  B_N=A_N\mtxel{\Psi_f}{\vect{E}_{\mathrm{probe}}(\omega)\cdot\vect{d}}{N}.
\end{equation}
\eqref{eq:sig} may be re-written as
\begin{equation}
  S_f(t)=2\sum_{N}\sum_{M\le N}|B_N||B_M|
  \cos[(E_N-E_M)t/\hbar+\Phi_{NM}],
\end{equation}
where the phase factor $\Phi_{NM}$ contains the initial phase differences of
the molecular eigenstates, and the phase difference of the probe transition
dipole matrix elements connecting the states $\ket{N}$ and $\ket{M}$ to the
final state.  The most detailed information is in this final state resolved
differential signal S$_{f}(t)$. It arises from the coherent sum over all
two-photon transition amplitudes consistent with the pump and probe laser
bandwidths and contains interferences between all degenerate two-photon
transitions.  It can be seen that the signal as a function of $\Delta t$
contains modulations at frequencies $(E_{N} - E_{M})/\hbar$, corresponding to
the set of all level spacings in the superposition.  This is the relationship
between the wavepacket dynamics and observed pump-probe signal. It is the
interference between individual two-photon transitions arising from the
initial state, through different excited eigenstates and terminating in the
same single final state, which leads to these modulations. The Fourier
transform power spectrum of this time domain signal therefore contains
frequencies which give information about the set of level spacings in the
excited state. The transform, however, also yields the Fourier amplitudes at
these frequencies, each corresponding to a modulation depth seen in the time
domain data at that frequency. These Fourier amplitudes relate to the overlaps
of each excited state eigenfunction within the wavepacket with a specific,
chosen final state.  Different final states will generally have differing
transition dipole moment matrix elements with the eigenstates $\ket{N}$
comprising the wavepacket, and so in general each final state will produce a
signal S$_{f}$ which has different Fourier amplitudes in its power
spectrum. For example, if two interfering transitions have very similar
overlaps with the final state, they will interfere constructively or
destructively with nearly 100\% modulation and, hence, have a very large
Fourier amplitude at that frequency. Conversely, if one transition has much
smaller overlap with the final state (due to e.g. a ``forbidden'' transition
or negligible Franck-Condon overlap) than the other, then the interference
term will be small and the modulation amplitude at that frequency will be
negligible.  Clearly, the form of the pump probe signal will depend on how the
final state ``views'' the various eigenstates comprising the wavepacket.  An
important point is that by carefully choosing different final states, it is
possible for the experimentalist to emphasize and probe particular aspects of
the wavepacket dynamics. In general there will be a set of final states which
fall within the probe laser bandwidth. We must differentiate, therefore,
between integral and differential detection techniques. With integral
detection techniques (e.g. total fluorescence, ion yield etc.), the
experimentally measured total signal, $S(\Delta t)$, is proportional to the
total population in the set of all energetically allowed final states, $\sum_f
S_{f}(\Delta t)$, created at the end of the two-pulse sequence. Information is
clearly lost in carrying out this sum since the individual final states may
each have different overlaps with the wavepacket.  Therefore, differential
techniques such as dispersed fluorescence, translational energy spectroscopy
or photoelectron spectroscopy, which can disperse the observed signal with
respect to final state, will be important.  The choice of the final state is
of great importance as it determines the experimental technique and
significantly determines the information content of an experiment.

We now consider a pump-probe experiment from a time-domain perspective. The
coherent superposition of exact molecular eigenstates constructs, for a short
time, a zeroth order state.  Zeroth order states are often physically
intuitive solutions to a simpler Hamiltonian $H_0$, and can give a picture of
the basic dynamics of the problem.  The full Hamiltonian is then given by
$H=H_0+V$. Suppose we choose to expand the molecular eigenstates in a complete
zeroth-order basis of $H_0$ which we denote by $\ket{n}$
\begin{equation}
  \ket{N}=\sum_n a^N_n\ket{n},
\end{equation}
then the wavepacket described in \eqref{eq:eigenpacket} may be written in
terms of these basis states as
\begin{equation}
  \label{eq:wpkt_zero}
  \ket{\Psi(t)}=\sum_n C_n
  \exp{-\mathrm{i}(E_n+E_n^{\mathrm{int}})t/\hbar}
  \ket{n},
\end{equation}
where the coefficients in the expansion are given by $C_n=\sum_N a^N_nA_N$
(with $A_N$ the eigenstate coefficients in the wavepacket in
\eqref{eq:eigenpacket}). To zeroth-order, the eigenstate $\ket{N}$ is
approximated by $\ket{n}$. The time dependence of the wavepacket expressed in
the zeroth-order basis reflects the couplings between the basis states
$\ket{n}$ which are caused by terms in the full molecular Hamiltonian which
are not included in the model Hamiltonian, $H_0$. In writing
\eqref{eq:wpkt_zero}, the eigenenergies of the true molecular eigenstates have
been expressed in terms of the eigenenergies of the zeroth-order basis as
$E_N=E_n+E_n^{\mathrm{int}}$, where $E_n^{\mathrm{int}}$ is the interaction
energy of zeroth order state $\ket{n}$ with all other zeroth order states. The
wavepacket evolution, when considered in terms of the zeroth-order basis
contains frequency components corresponding to the couplings between states,
as well as frequency components corresponding to the energies of the
zeroth-order states.  To second order in perturbation theory, the interaction
energy (coupling strength) $E^{\mathrm{int}}_n$ between zeroth-order states is
given in terms of the matrix elements of $V$ by
\begin{equation}
  E_n^{\mathrm{int}}=\mtxel{n}{V}{n}+
  \sum_{m\not=n}\frac{\mtxel{m}{V}{n}^2}{E_m-E_n}.
\end{equation}
Just as the expansion in the zeroth-order states can describe the exact
molecular eigenstates, likewise an expansion in the exact states can be used
to prepare, for a short time, a zeroth-order state. If the perturbation $V$ is
small, and the model Hamiltonian $H_0$ is a good approximation to $H$, then
the initially prepared superposition of eigenstates will resemble a
zeroth-order state. The dephasing of the exact molecular eigenstates in the
wavepacket superposition subsequently leads to an evolution of the initial zeroth
order electronic character, transforming into a different zeroth order
electronic state as a function of time.

A well known example is found in the problem of intramolecular vibrational
energy redistribution (IVR). The exact vibrational states are eigenstates of
the full rovibrational Hamiltonian which includes all orders of couplings and
are, of course, stationary. An example of a zeroth order state would be a
normal mode, the solution to a parabolic potential. A short pulse could create
a superposition of exact vibrational eigenstates which, for a short time,
would behave as a normal mode (e.g. stretching).  However, due to the
dephasing of the exact vibrational eigenstates in the wavepacket, this zeroth
order stretching state would evolve into a superposition of other zeroth order
states (e.g. other normal modes such as bending). Examples of using TRPES to
study such vibrational dynamics will be given in \secref{sec:app_nuc_dyn}.

\section{Non-adiabatic molecular dynamics}
\label{sec:nonBO}
As discussed in the previous section, wavepackets allow for the development of
a picture of the time evolution of the zeroth order states, and with a
suitably chosen basis this provides a view of both charge and energy flow in
the molecule.  For the case of interest here, excited state non-adiabatic
dynamics, the appropriate zeroth order states are the Born-Oppenheimer (BO)
states~\cite{Bixon1968, Jortner1969, Henry1973, Freed1976, Stock1997,
  Worth2004, Klessinger1994, Koppel1984}.  These are obtained by invoking an
adiabatic approximation that the electrons, being much lighter than the
nuclei, can rapidly adjust to the slower time-dependent fields due to the
vibrational motion of the atoms. The molecular Hamiltonian can be separated
into kinetic energy operators of the nuclei $T_\mathrm{n}(\vect{R})$ and
electrons $T_\mathrm{e}(\vect{r})$, and the potential energy of the electrons
and nuclei, $V(\vect{R},\vect{r})$,
\begin{equation}
  \label{eq:fullH}
  H(\vect{r},\vect{R})=T_\mathrm{n}(\vect{R})+
  T_\mathrm{e}(\vect{r})+V(\vect{R},\vect{r}),
\end{equation}
where $\vect{R}$ denotes the nuclear coordinates, and $\vect{r}$ denotes
the electronic coordinates.  The Born-Oppenheimer basis is obtained by setting
$T_\mathrm{n}(\vect{R})=0$, such that $H$ describes the electronic motion in a
molecule with fixed nuclei, and solving the time-independent Schr\"odinger
equation treating the nuclear coordinates $\vect{R}$ as a
parameter~\cite{Worth2004}. In this approximation, the adiabatic BO electronic
states $\Phi_\alpha(\vect{r};\vect{R})$ and potential energy surfaces
$V_\alpha(\vect{R})$ are defined by
\begin{equation}
  \label{eq:HBO}
  [H_\mathrm{e}(\vect{r};\vect{R})-V_\alpha(\vect{R})]
  \Phi_\alpha(\vect{r};\vect{R})=0,
\end{equation}
where the ``clamped nuclei'' electronic Hamiltonian is defined by
$H_\mathrm{e}(\vect{r};\vect{R})=T_\mathrm{e}(\vect{r})+V(\vect{r},\vect{R})$. The
eigenstates of the full molecular Hamiltonian (\eqref{eq:fullH}) may be
expanded in the complete eigenbasis of BO electronic states defined by
\eqref{eq:HBO},
\begin{equation}
  \label{eq:BOwfnexp}
  \langle\vect{r};\vect{R}\ket{N}=
  \sum_\alpha \chi_\alpha(\vect{R})
  \Phi_\alpha(\vect{r};\vect{R}),
\end{equation}
where the expansion coefficients $\chi_\alpha(\vect{R})$ are functions of the
nuclear coordinates.. The zeroth-order BO electronic states
$\Phi_\alpha(\vect{r};\vect{R})$ have been obtained by neglecting the nuclear
kinetic energy operator $T_\mathrm{n}(\vect{R})$, and so will be coupled by
this term in the Hamiltonian. Substitution of the expansion
\eqref{eq:BOwfnexp} into the Schr\"odinger equation
$[H(\vect{r},\vect{R})-E_N]\ket{N}=0$ gives a system of coupled differential
equations for the nuclear wavefunctions~\cite{Koppel1984, Worth2004,
  Stock1997, Born1954}
\begin{equation}
  [T_\mathrm{n}(\vect{R})+V_\alpha(\vect{R})-E_N]\chi_\alpha(\vect{R})=
  \sum_\beta \Lambda_{\alpha\beta}(\vect{R})\chi_\beta(\vect{R})
\end{equation}
where $E_N$ is the eigenenergy of the exact moleculer eigenstate $\ket{N}$.
The non-adiabatic coupling parameters $\Lambda_{\alpha\beta}(\vect{R})$ are
defined as
\begin{equation}
  \Lambda_{\alpha\beta}(\vect{R})=T_\mathrm{n}(\vect{R})\delta_{\alpha\beta}
%  -\mtxel{\alpha}{T_\mathrm{n}(\vect{R})}{\beta}
  -\int\mathrm{d}\vect{r}\,\Phi_\alpha^\ast(\vect{r})
  T_\mathrm{n}(\vect{r})\Phi_\beta(\vect{r})
\end{equation}
The diagonal terms $\alpha=\beta$ are corrections to the frozen nuclei
potentials $V_{\alpha}(\vect{R})$ and together form the nuclear zeroth-order
states of interest here. The off-diagonal terms $\alpha\not=\beta$ are the
operators which lead to transitions (evolution) between zeroth order states.
The kinetic energy is a derivative operator of the nuclear coordinates and,
hence, it is the motion of the nuclei which leads to electronic transitions.
One could picture that it is the time-dependent electric field of the
oscillating (vibrating) charged nuclei which can lead to electronic
transitions.  When the Fourier components of this time-dependent field match
electronic level spacings, transitions can occur. As nuclei move slowly,
usually these frequencies are too small to induce any electronic
transitions. When the adiabatic electronic states become close in energy, the
coupling between them can be extremely large, the adiabatic approximation
breaks down, and the nuclear and electronic motions become strongly
coupled~\cite{Bixon1968, Jortner1969, Henry1973, Freed1976, Stock1997,
  Worth2004, Klessinger1994, Koppel1984}. A striking example of the result of
the non-adiabatic coupling of nuclear and electronic motions is a conical
intersection between electronic states, which provide pathways for interstate
crossing on the femtosecond timescale and have been termed ``photochemical
funnels''~\cite{Stock1997}.  Conical intersections occur when adiabatic
electronic states become degenerate in one or more nuclear coordinates, and
the non-adiabatic coupling becomes infinite.  This divergence of the coupling
and the pronounced anharmonicity of the adiabatic potential energy surfaces in
the region of a conical intersection causes very strong electronic couplings
as well as strong coupling between vibrational modes. Such non-adiabatic
couplings can have pronounced effects.  For example, analysis of the the
absorption band corresponding to the S$_2$ electronic state of pyrazine
demonstrated that the vibronic bands in this region of the spectrum have a
very short lifetime due to coupling of the S$_2$ electronic state with the
S$_1$ electronic state~\cite{Raab1999, Woywod1994}, and an early demonstration
of the effect of a conical intersection was made in the study of an unexpected
band in the photoelectron spectrum of butatriene~\cite{Cederbaum1981,
  Cederbaum1977}. Detailed examples are given in
Section~\ref{sec:applications}.

The nuclear function $\chi_\alpha(\vect{R})$ is usually expanded in terms of a
wavefunction describing the vibrational motion of the nuclei, and a rotational
wavefunction~\cite{Wilson1955, Bransden2003}. Analysis of the vibrational part
of the wavefunction usually assumes that the vibrational motion is harmonic,
such that a normal mode analysis can be applied~\cite{Wilson1955,
  Bunker1998}. The breakdown of this approximation leads to vibrational
coupling, commonly termed intramolecular vibrational energy redistribution,
IVR. The rotational basis is usually taken as the rigid rotor
basis~\cite{Wilson1955, Kroto1975, Zare1988, Bunker1998}. This separation
between vibrational and rotational motions neglects centrifugal and Coriolis
coupling of rotation and vibration~\cite{Wilson1955, Kroto1975, Zare1988,
  Bunker1998}.  In the following, we will write the wavepacket prepared by the
pump laser in terms of the zeroth-order BO basis as
\begin{equation}
  \ket{\Psi(t)}=\sum_{J_\alpha M_\alpha\tau_\alpha v_\alpha\alpha}
  C_{J_\alpha M_\alpha\tau_\alpha v_\alpha\alpha}(t)
  \ket{J_\alpha M_\alpha \tau_\alpha}
  \ket{v_\alpha}\ket{\alpha}.
\end{equation}
The three kets in this expansion describe the rotational, vibrational and
electronic states of the molecule respectively,
\begin{align}
 \innerprod{\phi, \theta, \chi}{J_\alpha M_\alpha \tau_\alpha} &=
 \psi_{J_\alpha M_\alpha \tau_\alpha}(\phi, \theta, \chi),\\
 \innerprod{\vect{R}}{v_\alpha}&=\psi_{v_\alpha}(\vect{R}),\\
 \innerprod{\vect{r};\vect{R}}{\alpha}&=\Phi_\alpha(\vect{r};\vect{R}),
\end{align}
where $(\phi, \theta, \chi)$ are the Euler angles~\cite{Zare1988} connecting
the lab fixed frame (LF) to the molecular frame (MF). The quantum numbers
$J_\alpha$ and $M_\alpha$ denote the total angular momentum and its projection
on the lab-frame $z$-axis, and $\tau_\alpha$ labels the $(2J_\alpha+1)$
eigenstates corresponding to each $(J_\alpha,M_\alpha)$~\cite{Zare1988,
  Kroto1975, Bunker1998}.  The vibrational state label $v_\alpha$ is a
shorthand label that denotes the vibrational quanta in each of the vibrational
modes of the molecule.  The time-dependent coefficients
$C_{J_{\alpha}M_{\alpha}\tau_{\alpha}v\alpha}(t)$ will in general include
exponential phase factors which reflect all of the couplings described above,
as well as the details of the pump step.

For a vibrational mode of the molecule to induce coupling between adiabatic
electronic states $\Phi_\alpha(\vect{r};\vect{R})$ and
$\Phi_\beta(\vect{r};\vect{R})$, the direct product of the irreducible
representations of $\Phi_\alpha(\vect{r};\vect{R})$,
$\Phi_\beta(\vect{r};\vect{R})$ and the vibrational mode must contain the
totally symmetric representation of the molecular point group,
\begin{equation}
  \Gamma_\alpha\otimes\Gamma_v\otimes\Gamma_\beta\supset A_1,
\end{equation}
where $\Gamma_v$ is the irreducible representation of the vibrational mode
causing the non-adiabatic coupling.  As discussed in the previous section, an
initially prepared superposition of the molecular eigenstates will tend to
resemble a zeroth-order BO state. This BO state will then evolve due to the
coupling provided by the nuclear kinetic energy operator which leads to this
evolution -- a process which is often called a radiationless transition. For
example, a short pulse may prepare the S$_{2}$ (zeroth-order) BO state which,
via non-adiabatic coupling, evolves into the S$_{1}$ (zeroth-order) BO state,
a process which is referred to as ``internal conversion''. For the remainder
of this article we will adopt the language of zeroth order states and their
evolution due to intramolecular couplings.

%%% Local Variables: 
%%% mode: latex
%%% TeX-master: "trpes"
%%% End: 

\chapter{Probing non-adiabatic dynamics with photoelectron spectroscopy}
\label{sec:photoionzn}
As discussed in the previous section, the excited state dynamics of polyatomic
molecules is dictated by the coupled flow of both charge and energy within the
molecule. As such, a probe technique which is sensitive to both nuclear
(vibrational) and electronic configuration is required in order to elucidate
the mechanisms of such processes. Photoelectron spectroscopy provides such a
technique, allowing for the disentangling of electronic and nuclear motions,
and in principle leaving no configuration of the molecule unobserved, since
ionization may occur for all molecular configurations. This is in contrast
to other techniques, such as absorption or fluorescence spectroscopy, which
sample only certain areas of the potential energy surfaces involved, as
dictated by oscillator strengths, selection rules and Franck-Condon factors.

The molecular ionization continuum provides a template for observing both
excited state vibrational dynamics, via Franck-Condon distributions, and
evolving excited state electronic configurations.  The latter are understood
to be projected out via electronic structures in the continuum, of which there
are two kinds -- that of the cation and that of the free electron. The
electronic states of the cation can provide a map of evolving electronic
structures in the neutral state prior to ionization -- in the independent
electron approximation emission of an independent outer electron occurs
without simultaneous electronic reorganization of the ``core'' (be it cation
or neutral) -- this is called the ``molecular orbital'' or Koopmans'
picture~\cite{Eland1984, Koopmans1933, Ellis2005}. These simple correlation
rules indicate the cation electronic state expected to be formed upon single
photon single active electron ionization of a given neutral state. The
probabilities of partial ionization into specific cation electronic states can
differ drastically with respect to the molecular orbital nature of the probed
electronic state. If a given probed electronic configuration correlates, upon
removal of a single active outer electron, to the ground electronic
configuration of the continuum, then the photoionization probability is
generally higher than if it does not. The electronic states of the free
electron, commonly described as scattering states, form the other electronic
structure in the continuum.  The free electron states populated upon
photoionization reflect angular momentum correlations and are therefore
sensitive to neutral electronic configurations and symmetries. This
sensitivity is expressed in the form of the photoelectron angular distribution
(PAD). Furthermore, since the active molecular frame ionization dipole moment
components are geometrically determined by the orientation of the molecular
frame within the laboratory frame, and since the free electron scattering
states are dependent upon the direction of the molecular frame ionization
dipole, the form of the laboratory frame PAD is sensitive to the molecular
orientation, and so will reflect the rotational dynamics of the neutral
molecules.

\begin{figure}
  \includegraphics[width=8cm]{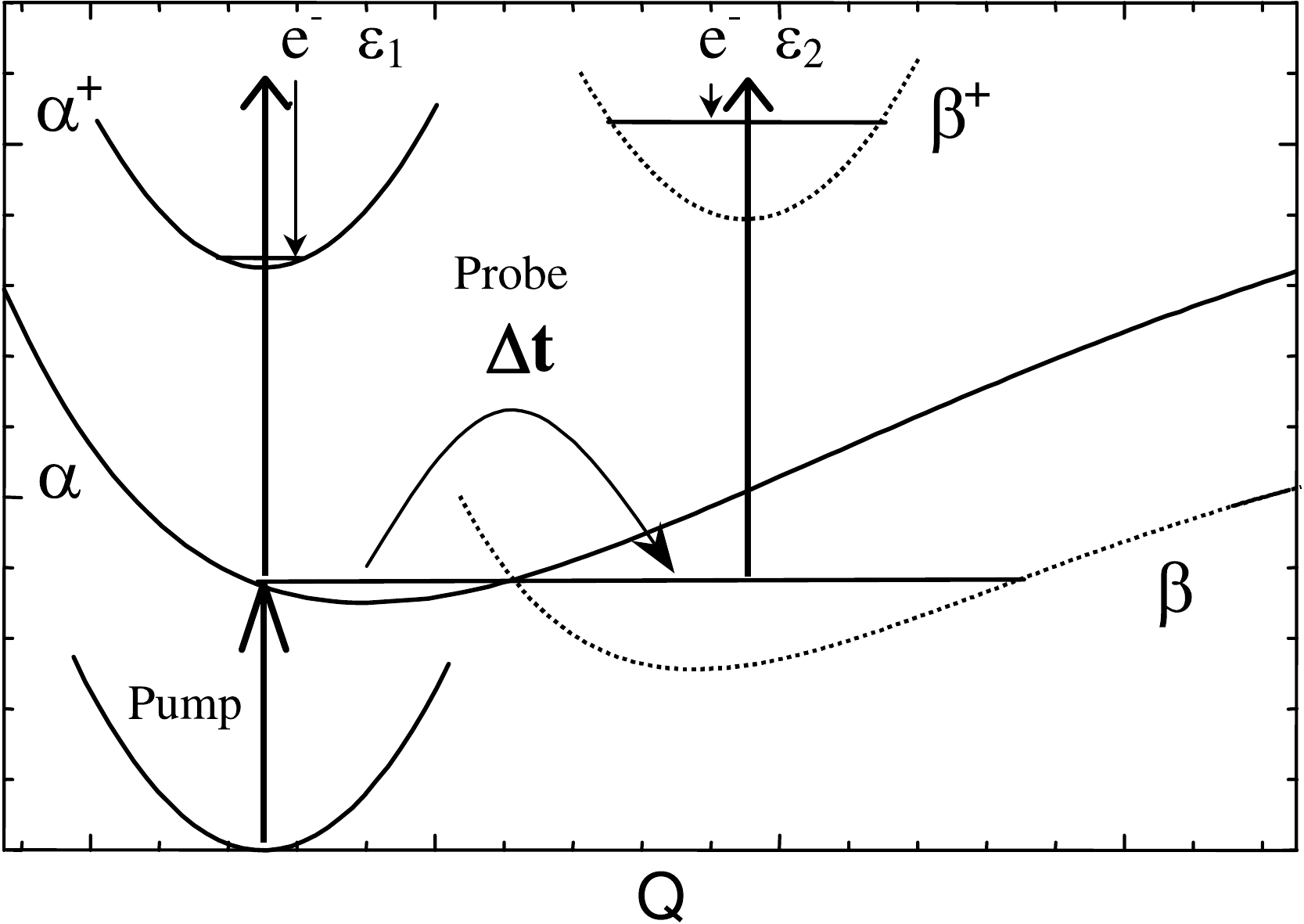}
  \caption{A TRPES scheme for disentangling electronic from vibrational
    dynamics in excited polyatomic molecules. A zeroth-order electronic state
    $\alpha$ is prepared by a femtosecond pump pulse. Via a non-adiabatic
    process it converts to a vibrationally hot lower-lying electronic state,
    $\beta$. The Koopmans type ionization correlations suggest that these two
    states will ionize into different electronic continua:
    $\alpha\rightarrow\alpha^++e^-(\varepsilon_1)$ and
    $\beta\rightarrow\beta^++e^-(\varepsilon_2)$.  When the wave packet has
    zeroth-order $\alpha$ character, any vibrational dynamics in the $\alpha$
    state will be reflected in the structure of the $\varepsilon_1$
    photoelectron band. After the non-adiabatic process, the wave packet has
    zeroth-order $\beta$ electronic character; any vibrational dynamics in the
    state will be reflected in the $\varepsilon_2$ band. This allows for the
    simultaneous monitoring of both electronic and vibrational excited-state
    dynamics.}
  \label{fig:Koopmans}
\end{figure}

We first consider a schematic example to illustrate how the cation electronic
structures can be used in (angle integrated) TRPES to disentangle electronic
from vibrational dynamics in ultrafast non-adiabatic processes, depicted in
\figref{fig:Koopmans}.  A zeroth- order bright state, $\alpha$, is coherently
prepared with a femtosecond pump pulse. According to the Koopmans'
picture~\cite{Eland1984, Koopmans1933, Ellis2005}, it should ionize into the
$\alpha^+$ continuum, the electronic state of the cation obtained upon removal
of the outermost valence electron (here chosen to be the ground electronic
state of the ion). This process produces a photoelectron band
$\varepsilon_1$. We now consider any non-adiabatic coupling process which
transforms the zeroth-order bright state $\alpha$ into a lower lying
zeroth-order dark state $\beta$, as induced by promoting vibrational modes
of appropriate symmetry. Again, according to the Koopmans picture, the state
should ionize into the $\beta^+$ ionization continuum (here assumed to be an
electronically excited state of the ion), producing a photoelectron band
$\varepsilon_2$. Therefore, for a sufficiently energetic probe photon (i.e.,
with both ionization channels open), we expect a switching of the electronic
photoionization channel from $\varepsilon_1$ to $\varepsilon_2$ during the
non-adiabatic process.  This simple picture suggests that one can directly
monitor the evolving excited-state electronic configurations (i.e., the
electronic population dynamics) during non-adiabatic processes while
simultaneously following the coupled nuclear dynamics via the vibrational
structure within each photoelectron band. The cation electronic structures can
act as a ``template'' for the disentangling of electronic from vibrational
dynamics in the excited state~\cite{Blanchet1998, Blanchet1999, Blanchet2000,
  Resch2001, Blanchet2001}.

More specifically, the BO electronic state $\Phi_\alpha(\vect{r};\vect{R})$
(which is an eigenfunction of the electronic Hamiltonian $H_e$) is a complex
multi-electron wave function. It can be expressed in terms of self consistent
field (SCF) wave functions $\ket{\Phi_n}$ which are comprised of a Slater
determinant of single electron molecular spin-orbitals~\cite{Ellis2005},
\begin{equation}
  \label{eq:parents}
  \ket{\alpha}=\sum_n A_n^\alpha\ket{\Phi_n}.
\end{equation}
Each $\ket{\Phi_n}$ corresponds to a single configuration, and the fractional
parentage coefficients $A_n^\alpha$ reflect the configuration interaction
(caused by electron correlation) for each BO electronic state. The
configuration interaction ``mixes in'' SCF wavefunctions of the same overall
symmetry, but different configurations. The correlations between the neutral
electronic state and the ion electronic state formed upon ionization are
readily understood in this independent electron picture~\cite{Seel1991,
  Seel1991a, Blanchet1999, Ellis2005}. In the Koopman's picture of
photoionization, a single active electron approximation is adopted, ionization
occurs out of a single molecular orbital, and the remaining core electron
configuration is assumed to remain unchanged upon ionization. As such, a
multi-electron matrix element reduces to a single electron matrix element for
each configuration that contributes to the electronic state, weighted by the
fractional parentage coefficients.

The two limiting cases for Koopmans-type correlations in TRPES experiments, as
initially proposed by Domcke~\cite{Seel1991, Seel1991a}, have been
demonstrated experimentally~\cite{Blanchet2001, Schmitt2001} and will be
further discussed in \secref{sec:trpes_continua}. The first case, Type (I), is
when the neutral excited states $\alpha$ and $\beta$ clearly correlate to
different cation electronic states, as in \figref{fig:Koopmans}. Even if there
are large geometry changes upon internal conversion and/or ionization,
producing vibrational progressions, the electronic correlations should favor a
disentangling of the vibrational dynamics from the electronic population
dynamics. An example of this situation is discussed in
Section~\ref{sec:trpes_continua}. The other limiting case, Type (II), is when
the neutral excited states $\alpha$ and $\beta$ correlate equally strongly to
the same cation electronic states, and so produce overlapping photoelectron
bands. An example of a Type (II) situation in which vastly different
Franck-Condon factors allow the states $\alpha$ and $\beta$ to be
distinguished in the PES is given in Section~\ref{sec:trpes_continua}, but
more generally Type (II) ionization correlations are expected to hinder the
disentangling of electronic from vibrational dynamics purely from the PES. It
is under these Type (II) situations when measuring the time
resolved PAD is expected to be of most utility -- as discussed below, the PAD
will reflect the evolution of the molecular electronic symmetry under
situations where electronic states are not readily resolved in the PES. The
continuum state accessed by the probe transition may be written as a direct
product of the cation and free electron states. As with any optical
transition, there are symmetry based ``selection rules'' for the
photoionization step. In the case of photoionization, there is the requirement
that the direct product of the irreducible representations of the state of the
ion, the free electron wavefunction, the molecular frame transition dipole
moment and the neutral state contains the totally symmetric irreducible
representation of the molecular point
group~\cite{Chandra1987,Burke1972}. Since the symmetry of the free electron
wavefunction determines the form of the PAD, the shape of the PAD will reflect
(i) the electronic symmetry of the neutral molecule and (ii) the symmetries of
the contributing molecular frame transition dipole moment components. Since
the relative contributions of the molecular frame transition dipole moments
are geometrically determined by the orientation of the molecule relative to
the ionizing laser field polarization, the form of the laboratory frame PAD
will reflect the distribution of molecular axis in the laboratory frame, and
so will reflect the rotational dynamics of the molecule~\cite{Reid2000,
  Reid1999, Underwood2000, Althorpe2000, Seideman2000a, Althorpe1999,
  Tsubouchi2001}.

We turn now to a more detailed description of the photoionization probe step
in order to clarify the ideas presented above. Time resolved photoelectron
spectroscopy probes the excited state dynamics using a time delayed probe
laser pulse which brings about ionization of the excited state wavepacket,
usually with a single photon
\begin{equation}
  \mathrm{AB}(\ket{\Psi(t)}) 
  +h\nu\rightarrow
  \mathrm{AB}^+(\ket{\Psi_+}) 
  +e^-(\epsilon\kL).
\end{equation}
Here and in what follows we use a subscript $+$ to denote the quantum numbers
of the ion core, $\epsilon$ to denote the kinetic energy of the electron, and
$\kL$ to denote the laboratory frame (LF) direction of the emitted
photoelectron, with magnitude $k=\sqrt{2m_e\epsilon}$. In the following
treatment we shall assume that the probe laser intensity remains low enough
that a perturbative description of the probe process is valid, and that the
pump and probe laser pulses are separated temporally. Full non-perturbative
treatments have been given in the literature for situations in which these
approximations are not appropriate~\cite{Seideman2002, Suzuki2002,
  Seideman2001, Althorpe1999, Seideman1997}.

The single particle wave function for the free photoelectron may be expressed
as an expansion in angular momentum partial waves characterised by an orbital
angular momentum quantum number $l$ and and associated quantum number
$\lambda$ for the projection of $l$ on the molecular frame (MF)
$z$-axis~\cite{Cooper1968, Buckingham1970, Smith1988, Reid2003, Seideman2002,
  Park1996},
\begin{equation}
  \innerprod{\vect{r}';\vect{R}}{\kM\epsilon}
  =\sum_{l\lambda}  
  \mathrm{i}^l
  \exp{-\mathrm{i}\sigma_l(\epsilon)}
  Y_{l\lambda}(\kM)
  \psi_{l\lambda}(\vect{r}';\epsilon,\vect{R}),
\end{equation}
where the asymptotic recoil momentum vector of the photoelectron in the MF is
denoted by $\kM$, and $Y_{l\lambda}(\kM)$ is a spherical
harmonic~\cite{Zare1988}. The radial wavefunction in this expansion,
$\psi_{l\lambda}(\vect{r}';\epsilon,\vect{R})$ depends upon the MF position
vector of the free photoelectron $\vect{r}'$, and parametrically upon the
nuclear coordinates $\vect{R}$, and also the photoelectron energy
$\epsilon$. The energy dependent scattering phase shift $\sigma_l(\epsilon)$
depends upon the potential of the ion core, and contains the Coulomb phase
shift. This radial wavefunction contains all details of the scattering of the
photoelectron from the non-spherical potential of the
molecule~\cite{Park1996}. In this discussion, we shall neglect the spin of the
free electron, assuming it to be uncoupled from the other (orbital and rotational)
angular momenta -- the results derived here are unaffected by other angular
momenta coupling schemes.

When considering molecular photoionization, it is useful to keep in mind the
conceptually simpler case of atomic ionization~\cite{Cooper1968,
  Smith1988}. For atomic ionization, the ionic potential experienced by the
photoelectron is a central field within the independent electron approximation
-- close to the ion core, the electron experiences a potential which is
partially shielded due to the presence of the other
electrons~\cite{Fano1986}. Far from the ion core, in the asymptotic region,
the Coulombic potential dominates. The spherically symmetric nature of this
situation means that the angular momentum partial waves of orbital angular
momentum $l$ form a complete set of independent ionization channels (i.e. $l$
remains a good quantum number throughout the scattering process). Single
photon ionization from a single electronic state of an atom produces a free
electron wave function comprising only two partial waves with angular momenta
$l_0\pm1$, where $l_0$ is the angular momentum quantum number of the electron
prior to ionization. In the molecular case, however, the potential experienced
by the photoelectron in the region of the ion core is non-central. As a
result, $l$ is no longer a good quantum number and scattering from the ion
core potential can cause changes in $l$. For linear and symmetric top
molecules, $\lambda$ remains a good quantum number, but for asymmetric top
molecules $\lambda$ also ceases to be conserved during scattering. The
multipolar potential felt by the electron in the ion core region falls off
rapidly such that in the asymptotic region, the Coulombic potential
dominates. As such, a partial wave description of the free electron remains
useful in the molecular case~\cite{Dixit1985, Buckingham1970}, but the partial
waves are no longer eigenstates of the scattering potential resulting in
multi-channel scattering amongst the partial wave states and a much richer
partial wave composition when compared to the atomic case~\cite{Park1996}. To
add to this richness of partial waves, the molecular electronic state is no
longer described by a single value of $l_0$. Nonetheless, a partial wave
description of the free electron wave function remains a useful description,
since, despite the complex scattering processes, the expansion is truncated at
relatively low values of $l$.

For polyatomic molecules Chandra showed that it is useful to re-express the
photoelectron wavefunction in terms of symmetry adapted spherical
harmonics~\cite{Chandra1987, Underwood2000, VonDerLage1947, Chandra1988,
  Chandra1991, Chandra1998, Burke1972},
\begin{equation}
  \label{eq:MFsymadapwfn}
  \innerprod{\vect{r}';\vect{R}}{\kM\epsilon}
  =\sum_{\Gamma\mu hl}
  \mathrm{i}^l
  \exp{-\mathrm{i}\sigma_l(\epsilon)}X^{\Gamma\mu\ast}_{hl}(\kM)
  \psi_{\Gamma\mu hl}(\vect{r}';\epsilon,\vect{R}).
\end{equation}
The symmetry adapted spherical harmonics (also referred to as generalized
harmonics), $X^{\Gamma\mu}_{hl}(\kM)$, satisfy the symmetries of the
molecular point group~\cite{Chandra1987} and are defined as
\begin{equation}
  \label{eq:sash}
  X^{\Gamma\mu}_{hl}(\uvect{k})=\sum_{\lambda}
  b^{\Gamma\mu}_{hl\lambda}
  Y_{l\lambda}(\uvect{k}),
\end{equation}
where $\Gamma$ defines an irreducible representation (IR) of the molecular
point group of the molecule plus electron system, $\mu$ is a degeneracy index
and $h$ distinguishes harmonics with the same values of $\Gamma\mu l$ indices.
The $b^{\Gamma\mu}_{hl\lambda}$ symmetry coefficients are found by
constructing generalized harmonics using the projection
theorem~\cite{Conte1984, Tsukerblat1994, McWeeny1963, Chrishol1976} employing
the spherical harmonics $Y_{lm}(\theta,\phi)$ as the generating function. In
using the molecular point group, rather than the symmetry group of the full
molecular Hamiltonian, we are assuming rigid behaviour. To go beyond this
assumption, it is necessary to consider the full molecular symmetry
group~\cite{Bunker1998}. Such a treatment has been given by Signorell and
Merkt~\cite{Signorell1997}.

Combining \eqref{eq:MFsymadapwfn} and \eqref{eq:sash}, the free electron
wavefunction \eqref{eq:MFsymadapwfn} may be re-expressed in the lab frame (LF)
using the properties of the spherical harmonics under rotation as
\begin{equation}
  \label{eq:LFwfn}
  \innerprod{\vect{r}';\vect{R}}{\kL\epsilon}
  =\sum_{l\lambda m}\sum_{\Gamma\mu h}
  \mathrm{i}^l\exp{-\mathrm{i}\sigma_l(\epsilon)}
  b^{\Gamma\mu}_{hl\lambda}
  \Drot{l\ast}{m}{\lambda}
  Y_{l m}^\ast(\uvect{k}_L)
  \psi_{\Gamma\mu hl}(\vect{r}';\epsilon,\vect{R}),
\end{equation}
where $\Drot{l}{m}{\lambda}$ is a Wigner rotation matrix
element~\cite{Zare1988}.

The partial differential photoionization cross section for producing
photoelectrons with a kinetic energy $\epsilon$ at time $t$ ejected in the
LF direction $\kL$ is then
\begin{equation}
  \label{eq:dcs1}
  \sigma(\epsilon, \kL;t)
  \propto
  \sum_{n_{\alpha_+}M_{\alpha_+}}
  \bigg|
  \sum_{n_\alpha M_\alpha}
  C_{n_{\alpha}M_{\alpha}}(t)
  \bra{\kL\epsilon n_{\alpha_+}M_{\alpha_+}}
  \vect{d}\cdot\uvect{e}
  \ket{n_{\alpha}M_{\alpha}}
  \mathcal{E}(n_{\alpha_+},n_{\alpha},\epsilon)
  \bigg|^2,
\end{equation}
where we have introduced the shorthand notation for quantum numbers
$n_{\alpha}=J_{\alpha}\tau_{\alpha}v_{\alpha}\alpha$.  We have implicitly
assumed that the coefficients $C_{n_\alpha M_\alpha}(t)$ do not vary over the
duration of the probe pulse.  We have taken the laser field of the probe pulse
to be of the form
\begin{equation}
  E(t)=\uvect{e}f(t)\cos(\omega_0 t+\phi(t)),
\end{equation}
where $f(t)$ is the pulse envelope, $\uvect{e}$ is the probe pulse polarization
vector, $\omega_0$ is the carrier frequency and $\phi(t)$ is a time-dependent
phase. In \eqref{eq:dcs1} $\mathcal{E}(n_{\alpha_+},n_{\alpha},\epsilon)$ is
the Fourier transform of the probe pulse at the frequency
$2\pi(E_{n_{\alpha_+}}-E_{n_{\alpha}}+\epsilon)/h$, as defined by
\begin{equation}
  E(\omega)=\int\exp{\mathrm{i}\omega t}E(t)\,\mathrm{d}t.
\end{equation}

In order to evaluate the matrix elements of the dipole moment operator in
\eqref{eq:dcs1}, it is convenient to separate out the geometrical aspects of
the problem from the dynamical parameters. To that end, it is convenient to
decompose the LF scalar product of the transition dipole moment $\vect{d}$ with
the polarization vector of the probe laser field $\uvect{e}$ in terms of the
spherical tensor components as~\cite{Zare1988}
\begin{equation}
  \label{eq:d_dot_e}
  \vect{d}\cdot\uvect{e} = \sum_{p=-1}^{1} (-1)^p d_p e_{-p}.
\end{equation}
The LF spherical tensor components of the electric field polarization
are defined as 
\begin{equation}
  e_0=e_z\qquad
  e_{\pm1}=\frac{\mp1}{\sqrt{2}}(e_x\pm\mathrm{i}e_y). 
\end{equation}
For linearly polarized light, it is convenient to define the lab frame
$z$-axis along the polarization vector, such that the only non-zero component
is $e_0$. For circularly polarized light, the propagation direction of the
light is usually chosen to define the LF $z$-axis such that the non-zero
components are $e_1$ for right circularly polarized light and $e_{-1}$ for
left circularly polarized light. Other polarizations states of the probe pulse
are described by more than a single non-zero component $e_p$, and for
generality, in what follows, we shall not make any assumptions about the
polarization state of the ionising pulse. The LF components of the dipole
moment $d_p$ are related to the MF components through a rotation,
\begin{equation}
  \label{eq:lfdip}
  d_p=\sum_{q=-1}^{1}\Drot{1\ast}{p}{q}d_q.
\end{equation}
The rotational wavefunctions appearing in \eqref{eq:dcs1} may be expressed in
terms of the symmetric top basis as~\cite{Zare1988}
\begin{equation}
  \label{eq:asymtop}
  \ket{JM\tau}=\sum_K a_K^{J\tau} \ket{JKM},
\end{equation}
where the symmetric top rotational basis functions are defined in terms of the
Wigner rotation matrices as 
\begin{equation}
  \label{eq:symtop}
  \innerprod{\bm{\Omega}}{JKM}=
  \left(\frac{2J+1}{8\pi^2}\right)^{1/2}
  \Drot{J\ast}{M}{K}.
\end{equation}
Using \eqref{eq:d_dot_e}-\eqref{eq:symtop}, the matrix elements of the dipole
moment operator in \eqref{eq:dcs1} may be written as
\begin{equation}
  \label{eq:mtxel}
  \begin{split}
    &\bra{\kL\epsilon n_{\alpha_+}M_{\alpha_+}}
    \vect{d}\cdot\uvect{e}
    \ket{n_{\alpha}M_{\alpha}}
    \\&=
    \frac{1}{8\pi^2}[J_\alpha,J_{\alpha_+}]^{1/2}
    \sum_{l\lambda m}
    (\mathrm{-i})^l\exp{\mathrm{i}\sigma_l(\epsilon)}
    Y_{l m}(\uvect{k}_L)
    \sum_{K_{\alpha}K_{\alpha_+}}
    a^{J_\alpha\tau_\alpha}_{K_{\alpha}}
    a^{J_{\alpha_+}\tau_{\alpha_+}}_{K_{\alpha_+}}
    \sum_{pq}(-1)^{p}e_{-p}
    \\&\times
    \int
    \Drot{l}{m}{\lambda}
    \Drot{J_{\alpha_+}}{M_{\alpha_+}}{K_{\alpha_+}}
    \Drot{1\ast}{p}{q}
    \Drot{J_\alpha\ast}{M_\alpha}{K_\alpha}
    \,\mathrm{d}\bm{\Omega}
    \\&\times
    \sum_{\Gamma\mu h}
    b^{\Gamma\mu}_{hl\lambda}
    D_{\Gamma\mu h l}^{\alpha v_{\alpha}\alpha_+v_\alp}(q),
  \end{split}
\end{equation}
where we have introduced the shorthand $[X,Y,\dots]=(2X+1)(2Y+1)\dots$.  The
dynamical functions in \eqref{eq:mtxel} are defined as
\begin{equation}
  \label{eq:dynamics1}
  D_{\Gamma\mu h l}^{\alpha v_{\alpha}\alpha_+v_\alp}(q)
  =\int
  \psi^\ast_{v_{\alpha_+}}(\vect{R})\psi_{v_{\alpha}}(\vect{R})
  d^{\alpha\alpha_+}_{\Gamma\mu h l}(q;\vect{R})
  \,\mathrm{d}\vect{R}.
\end{equation}
These dynamical parameters are integrals over the internuclear separations
$\vect{R}$, as well as the electronic coordinates $\vect{r}$ through the
electronic transition dipole matrix elements, $d^{\alpha\alpha_+}_{\Gamma\mu h
  l}(q;\vect{R})$. These electronic transition dipole matrix elements are
evaluated at fixed internuclear configurations~\cite{Dixit1985} and are
defined as
\begin{equation}
  \label{eq:dynamics2}
  d^{\alpha\alpha_+}_{\Gamma\mu h l}(q;\vect{R})=
  \int
  \Phi^\ast_{\alpha_+\Gamma\mu hl}(\vect{r};\epsilon,\vect{R})
  d_q
  \Phi_\alpha(\vect{r};\vect{R})
  \,\mathrm{d}\vect{r}.
\end{equation}
Here $\Phi_{\alpha_+\Gamma\mu hl}(\vect{r};\epsilon,\vect{R})$ is the
antisymmetrized electronic wavefunction which includes the free electron
radial wavefunction $\psi_{\Gamma\mu hl}(\vect{r}';\epsilon,\vect{R})$ and the
electronic wave function of the ion
$\Phi_{\alpha_+}(\vect{r}'';\vect{R})$~\cite{Chandra1987, Chandra1987a,
  Dixit1985, Park1996} (where $\vect{r}''$ are the position vectors of the ion
electrons). For the integral in \eqref{eq:dynamics2} to be non-zero, the
following condition must be met:
\begin{equation}
  \label{eq:direct_product}
  \Gamma\otimes\Gamma_{\alp}\otimes\Gamma_{q}\otimes\Gamma_{\alpha}
  \subset A_1.
\end{equation}
That is, the direct product of the IRs of the free electron, the ion, the
transition dipole moment and the neutral electronic state respectively must
contain the totally symmetric IR of the molecular point group, $A_1$. Clearly,
the symmetries of the contributing photoelectron partial waves will be
determined by the electronic symmetry of the BO electronic state undergoing
ionization, as well as the molecular frame direction of the ionization
transition dipole moment (which determines the possible $\Gamma_q$), and the
electronic symmetry of the cation. As such, the evolution of the photoelectron
angular distribution, which directly reflects the allowed symmetries of the
partial waves, will reflect the evolution of the molecular electronic
symmetry.

It is frequently the case that the electronic transition dipole matrix element
$d^{\alpha\alpha_+}_{\Gamma\mu h l}(q;\vect{R})$ is only weakly dependent upon
the nuclear coordinates $\vect{R}$ such that the Franck-Condon
approximation~\cite{Bransden2003} may be employed. Within this approximation,  
\begin{equation}
  D_{\Gamma\mu h l}^{\alpha v_{\alpha}\alpha_+v_\alp}(q)
  =
  \bar{d}^{\alpha\alpha_+}_{\Gamma\mu h l}(q)
  \int
  \psi^\ast_{v_{\alpha_+}}(\vect{R})\psi_{v_{\alpha}}(\vect{R})
  \,\mathrm{d}\vect{R},
\end{equation}
where $\bar{d}^{\alpha\alpha_+}_{\Gamma\mu h l}(q)$ is the value of
$d^{\alpha\alpha_+}_{\Gamma\mu h l}(q;\vect{R})$ averaged over
$\vect{R}$. Within this approximation, the overlap integral between the
molecular vibrational state and the cation vibrational state determines the
ionization efficiency to each cation vibrational state~\cite{AlJoboury1963,
  Frost1965, Frost1967, Turner1967, Branton1970, Blake1970, Turner1970,
  Eland1984, Ellis2005}. The Franck-Condon factors are determined by the
relative equilibrium geometries of the electronic states of the neutral
($\Phi_\alpha(\vect{r};\vect{R})$) and cation
($\Phi_{\alpha_+}(\vect{r}'';\epsilon,\vect{R})$)~\cite{Eland1984,
  Ellis2005}. If the neutral and cation electronic states have similar
equilibrium geometries, each neutral vibronic state will produce a single
photoelectron peak for each vibrational mode corresponding to $\Delta v=0$
transitions upon ionization. However, if there is a substantial difference in
the equilibrium geometries, a vibrational progression in the PES results from
ionization of each neutral vibronic state, corresponding to $\Delta
v=0,1,2\dots$ transitions upon ionization for each populated vibrational
mode. In either case, the photoelectron spectrum will reflect the vibronic
composition of the molecular wavepacket, and the time dependence of the
vibrational structure in the photoelectron spectrum directly reflects the
nuclear motion of the molecule. Of course, this Franck-Condon mapping of the
vibrational dynamics onto the PES will break down if the variation of the
electronic ionization dipole matrix elements varies significantly with
$\vect{R}$, for example in a region in which vibrational auto-ionization is
active~\cite{Berry1966, Blake1970, Eland1984, Ellis2005}.

In the Koopman's picture of photoionization~\cite{Eland1984, Koopmans1933,
  Ellis2005}, a single active electron approximation is adopted, ionization
occurs out of a single molecular orbital, and the remaining core electron
configuration is assumed to remain unchanged upon ionization. As such, the
multi-electron matrix element in \eqref{eq:dynamics2} reduces to a single
electron matrix element for each configuration that contributes to the
electronic state, weighted by the fractional parentage coefficients.  In the
limit of the electronic state $\ket{\alpha}$ being composed of a single
configuration, ionization will access the continuum corresponding to the ion
state $\alpha^+$ which has the same core electronic configuration. In the
single active electron approximation, for a single configuration, the
electronic transition dipole matrix element in \eqref{eq:dynamics2} may be
rewritten as~\cite{Chandra1987, Chandra1987a}
\begin{equation}
  \label{eq:dynamics3}
  d^{\alpha\alpha_+}_{\Gamma\mu h l}(q;\vect{R})=
  \int
  \psi^\ast_{\Gamma\mu hl}(\vect{r}';\epsilon,\vect{R})
  d_q
  \phi_i(\vect{r}';\vect{R})
  \,\mathrm{d}\vect{r}',
\end{equation}
where $\phi_i(\vect{r}';\vect{R})$ is the initial bound molecular orbital from
which photoionization takes place. In order for \eqref{eq:dynamics3} to be
non-zero, the following condition must be met~\cite{Chandra1987}:
\begin{equation}
  \Gamma\otimes\Gamma_{q}\otimes\Gamma_{i}
  \subset A_1.
\end{equation}

\begin{figure}
  \includegraphics[width=6cm]{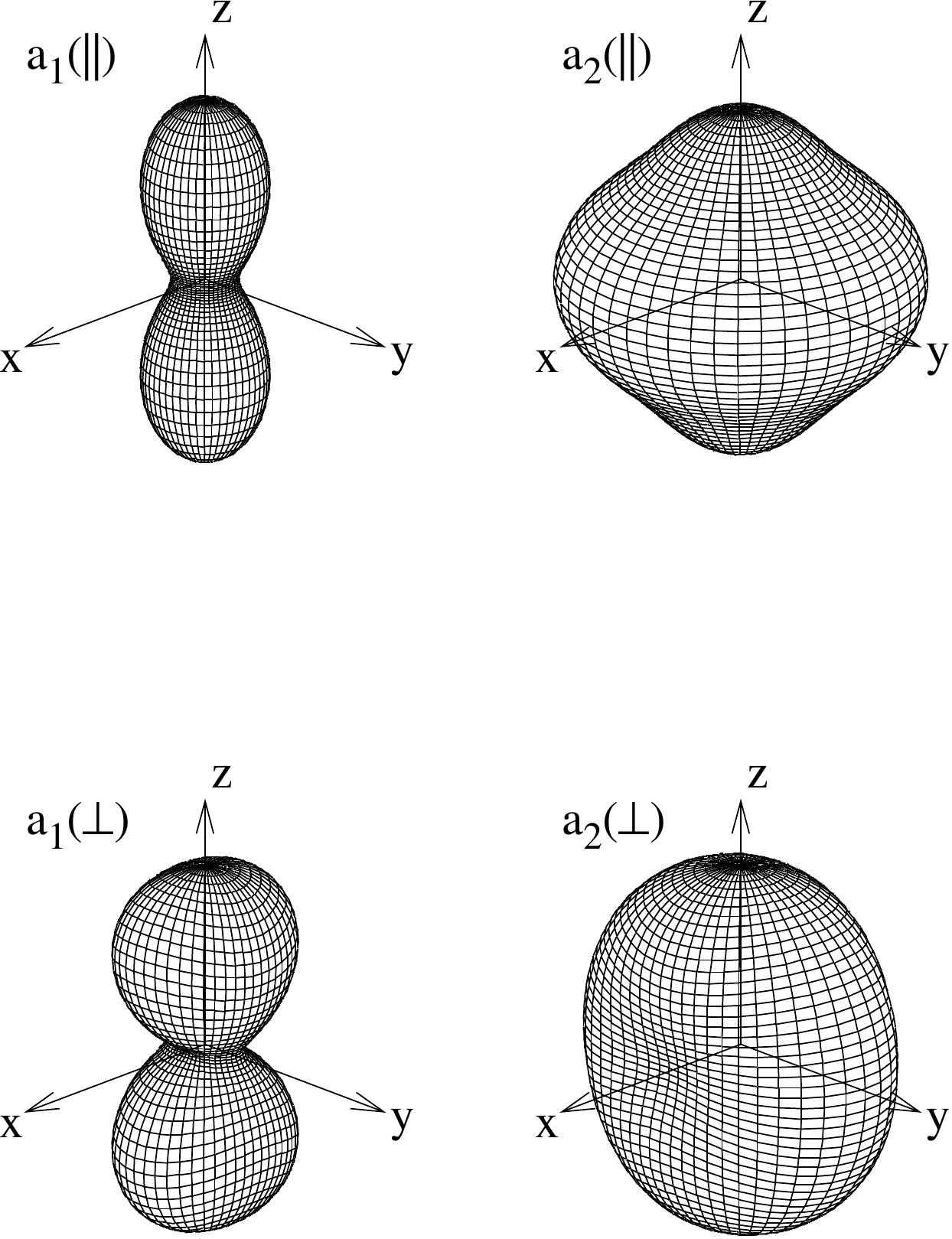}
  \caption{Calculated LF PADs for ionization of a model $C_{3v}$
    molecule. PADs are shown for ionization of $a_1$ and $a_2$ symmetry
    orbitals for the same set of dynamical parameters. The molecular axis
    distribution in these calculations was described as a $\cos^2\theta$
    distribution, where $\theta$ is the angle between the direction of linear
    polarization of the pump laser and the principle molecular axis. The
    linear probe polarization is along the $z$-axis. Panel (a) shows PADs
    for parallel pump and probe polarizations, while panel (b) shows
    PADs for perpendicular pump and probe polarizations. See
    ref.~\cite{Underwood2000} for dynamical parameters used in these
    calculations.} 
  \label{fig:cos2pad}
\end{figure}

Within the independent electron and single active electron approximations, the
symmetries of the contributing photoelectron partial waves will be determined
by the symmetry of the orbital(s) from which ionization occurs, and so the PAD
will directly reflect the evolution of the molecular orbital
configuration. Example calculations demonstrating this are shown in
\figref{fig:cos2pad} for a model $C_{3v}$ molecule, where a clear difference in
the PAD is observed according to whether ionization occurs from an $a_1$ or an
$a_2$ symmetry orbital~\cite{Underwood2000} (this is discussed in more detail
below). 

We return now to considering the detailed form of the photoelectron angular
distribution (PAD) in time-resolved pump-probe PES experiments.  It is
convenient to describe the excited state population dynamics in terms of the
density matrix, defined by~\cite{Blum1996, Zare1988}
\begin{equation}
  \label{eq:density_matrix}
  \rho(n_\alpha,n'_{\alpha'};t)_{M_{\alpha}M'_{\alpha'}}=
  C_{n_{\alpha}M_{\alpha}}(t)
  C^\ast_{n'_{\alpha'}M'_{\alpha'}}(t).
\end{equation}
The diagonal elements of the density matrix contain the populations of each of
the BO states, whereas off-diagonal elements contain the relative phases of
the BO states. The components of the density matrix with $\alpha=\alpha'$
describe the vibrational and rotational dynamics in the electronic state
$\alpha$, while the rotational dynamics within a vibronic state are described
by the density matrix elements with $\alpha=\alpha'$ and
$v_\alpha=v'_{\alpha'}$. The density matrix components with
$n_\alpha=n'_{\alpha'}$ describe the angular momentum polarization of the
state $J_\alpha$, often referred to as angular momentum orientation and
alignment~\cite{Zare1988, Blum1996, Greene1982, OrrEwing1994}. The density
matrix may be expanded in terms of multipole moments as
\begin{multline}
  \label{eq:multipole_expansion}
  \rho(n_\alpha,n'_{\alpha'};t)_{M_{\alpha}M'_{\alpha'}}=
  \sum_{KQ}(-1)^{J_\alpha-M_\alpha}[K]^{1/2}
  \threej{J_{\alpha}}{J'_{\alpha'}}{K}{M_{\alpha}}{-M'_{\alpha'}}{-Q}
  \\
  \multmom{n_{\alpha}}{n'_{\alpha'};t}{K}{Q},
\end{multline}
where $\threej{\cdot}{\cdot}{\cdot}{\cdot}{\cdot}{\cdot}$ is a Wigner $3j$
symbol. 

The multipole moments $\multmom{n_{\alpha}}{n'_{\alpha'};t}{K}{Q}$ are the
expectation values of the irreducible spherical tensor operators
$\multipole{n_{\alpha}}{n'_{\alpha'};t}{K}{Q}$ which transform under rotation
as the spherical harmonics~\cite{Zare1988, Blum1996} and are termed state
multipoles. From the properties of the Wigner $3j$ symbol, the possible range
of $K$ is given by $K=0\dots({J_{\alpha}}+{J'_{\alpha'}})$, and $Q=-K\dots
K$. The multipole moments with $K=Q=0$ contain the vibronic state populations
(terms with $\alpha=\alpha'$ and $v_\alpha=v'_{\alpha'}$) and coherences
(terms with $\alpha\not=\alpha'$ and/or $v_\alpha\not=v'_{\alpha'}$). The
multipole moments with $K=Q=0$ also contain the populations of the rotational
states, as well as the coherences between rotational states with the same
values of $J$ and $M$ in different electronic states. Multipole moments with
$K>0$ describe the angular momentum polarization and the coherence amongst
rotational states. In the perturbative limit, the maximum value of $K$ is
given by $2n$, where $n$ is the number of photons involved in the pump step --
e.g. a single photon pump step will prepared multipole moments with $K=0,2$.

The integral over the Euler angles in equation \eqref{eq:mtxel} is found
analytically using the Clebsch-Gordan series~\cite{Zare1988,
  Buckingham1970, Dixit1985}
\begin{multline}
  \label{eq:rotintegral}
    \int
    \Drot{l}{m}{\lambda}
    \Drot{J_{\alpha_+}}{M_{\alpha_+}}{K_{\alpha_+}}
    \Drot{1\ast}{p}{q}
    \Drot{J_\alpha\ast}{M_\alpha}{K_\alpha}
    \,\mathrm{d}\bm{\Omega}
    =\\
    (-1)^{M_{\alpha_+}-K_{\alpha_+}+p+q}
    \sum_{j_t}
    [j_t]
    \threej{J_{\alpha_+}}{J_\alpha}{j_t}{-M_{\alpha_+}}{M_\alpha}{m_t}
    \threej{J_{\alpha_+}}{J_\alpha}{j_t}{-K_{\alpha_+}}{K_\alpha}{k_t}
    \\\times
    \threej{l}{1}{j_t}{m}{-p}{m_t}
    \threej{l}{1}{j_t}{\lambda}{-q}{k_t},
\end{multline}
where $j_t$ corresponds to the angular momentum transferred to the ion during
the ionization process, with $m_t$ and $k_t$ denoting the projections of $j_t$
on the LF and MF $z$-axes respectively.

Expanding \eqref{eq:dcs1} and substituting in Eqs~(\ref{eq:mtxel}),
(\ref{eq:density_matrix}), (\ref{eq:multipole_expansion}) and
(\ref{eq:rotintegral}), and carrying out the summations over all LF projection
quantum numbers (see Appendix~\ref{app:betalm1}) gives an expression for the
LF PAD as an expansion in spherical harmonics,
\begin{equation}
  \label{eq:betalm}
  \sigma(\epsilon, \kL;t)=
  \frac{\sigma_{\mathrm{total}}(\epsilon;t)}{4\pi}
  \sum_{LM}\beta_{LM}(\epsilon;t)
  Y_{LM}(\kL),
\end{equation}
where $\sigma_{\mathrm{total}}(\epsilon;t)$ is the total cross section for
producing electrons with an energy $\epsilon$. The expansion coefficients
$\beta_{LM}(\epsilon;t)$ are given by
\begin{equation}
  \label{eq:betalm1}
  \begin{split}
    \beta_{LM}(\epsilon;t)&=-[L]^{1/2}
    \sum_{PR}
    (-1)^{P}[P]^{1/2}E_{PR}(\uvect{e})
    \sum_{KQ}
    (-1)^{K+Q}
    \\&\times
    \threej{P}{K}{L}{R}{-Q}{M}
    \sum_{n_{\alpha}n_{\alpha'}}
    (-1)^{J_\alpha}
    [J_\alpha, J'_{\alpha'}]^{1/2}
    \multmom{n_{\alpha}}{n'_{\alpha'};t}{K}{Q}
    \\&\times
    \sum_{ll'}
    (-1)^{l}
    [l,l']^{1/2}
    \threej{l}{l'}{L}{0}{0}{0}
    \sum_{j_tj_t'}
    (-1)^{j_t}[j_t,j_t']
    \ninej{1}{1}{P}{j_t}{j_t'}{K}{l}{l'}{L}
    \\&\times
    \sum_{qq'}
    \sum_{\lambda\lambda'}
    \sum_{k_tk_t'}
    (-1)^{q+q'}
    \threej{l}{1}{j_t}{\lambda}{-q}{k_t}
    \threej{l'}{1}{j_t'}{\lambda'}{-q'}{k_t'}
    \\&\times
    \sum_{J_\alp\tau_\alp}
    (-1)^{J_\alp}
    [J_\alp]
    \sixj{J_\alpha}{j_t}{J_\alp}{j_t'}{J'_{\alpha'}}{K}
    \\&\times
    \sum_{K_\alp}
    \left|
      a_{K_\alp}^{J_\alp\tau_\alp}
    \right|^2
    \sum_{K_\alpha K'_{\alpha'}}
    a_{K_\alpha}^{J_\alpha\tau_\alpha}
    a_{K'_{\alpha'}}^{J'_{\alpha'}\tau'_{\alpha'}}
    \\&\times
    \threej{J_\alp}{J_\alpha}{j_t}{-K_\alp}{K_\alpha}{k_t}
    \threej{J_\alp}{J'_{\alpha'}}{j_t'}{-K_\alp}{K'_{\alpha'}}{k_t'}
    \sum_{v_\alp\alp}
    \mathcal{E}(n_\alp, n_\alpha,\epsilon)
    \\&\times
    \mathcal{E}^\ast(n_\alp, n'_{\alpha'},\epsilon)
    \sum_{\Gamma\mu h}
    \sum_{\Gamma'\mu'h'}
    b^{\Gamma\mu}_{hl\lambda}
    b^{\Gamma'\mu'\ast}_{h'l'\lambda'}
    \mathrm{(-i)}^{l-l'}
    \exp{\mathrm{i}(\sigma_l(\epsilon)-\sigma_{l'}(\epsilon))}
    \\&\times
    D_{\Gamma\mu h l}^{{\alpha}v_{\alpha}{\alp}v_\alp}(q)
    D_{\Gamma'\mu' h' l'}^{{\alpha'}v'_{\alpha'}{\alp}v_{\alp}\ast}(q').
  \end{split}
\end{equation}
where $\sixj{\cdot}{\cdot}{\cdot}{\cdot}{\cdot}{\cdot}$ and
$\ninej{\cdot}{\cdot}{\cdot}{\cdot}{\cdot}{\cdot}{\cdot}{\cdot}{\cdot}$ are
Wigner $6j$ and $9j$ coefficients respectively~\cite{Zare1988}. 

The functions $E_{PR}(\uvect{e})$ describe the polarization of the probe laser
pulse, and are given by
\begin{equation}
  E_{PR}(\uvect{e})=[e\otimes e]^P_R=
  [P]^{1/2}\sum_p(-1)^{p}
  \threej{1}{1}{P}{p}{-(R+p)}{R}
  e_{-p}e^\ast_{-(R+p)}.
\end{equation}
From the properties of the Wigner $3j$ symbol, $P$ can take the values
$0,1,2$, and for linear polarization along the lab $z$-axis, $P=0,2$
only. The Wigner $3j$ symbol also restricts the values of $R$ to $-P\dots P$. 

If we make the assumption that the rotational states of the ion are not
resolved and the Fourier transform of the probe laser pulse remains constant
over the spectrum of transitions to ion rotational states, we can replace
$\mathcal{E}(n_\alp, n_\alpha,\epsilon)$ in \eqref{eq:betalm1} with an
averaged Fourier transform at a frequency $2\pi(\bar{E}_{\alp
  v_\alp}-\bar{E}_{\alpha v_\alpha}+\epsilon)/\hbar$, which we denote by
${\mathcal{E}}(\alpha,v_\alpha,\alp,v_\alp, \epsilon)$. This allows the
summations over the ion rotational states and also $j_t$, $k_t$ in
\eqref{eq:betalm1} to be carried out analytically (see
Appendix~\ref{app:betalm2}), yielding a simplified expression for the
coefficients $\beta_{LM}(\epsilon;t)$,
\begin{equation}
  \label{eq:betalm2}
  \begin{split}
    \beta_{LM}(\epsilon;t)&=[L]^{1/2}
    \sum_{\alpha v_{\alpha}}
    \sum_{\alpha' v'_{\alpha'}}
    \sum_{KQS}
    (-1)^{K+Q}
    A(\alpha,v_\alpha,\alpha',v'_{\alpha'};t)^K_{QS}
    \\&\times
    \sum_{P}
    (-1)^{P}[P]^{1/2}E_{PQ-M}(\uvect{e})
    \threej{P}{K}{L}{Q-M}{-Q}{M}
    \\&\times
    \sum_{qq'}
    (-1)^{q}
    \threej{1}{1}{P}{q}{-q'}{q'-q}
    \threej{P}{K}{L}{q'-q}{-S}{S+q-q'}
    \\&\times
    F_{LS}^{\alpha v_\alpha \alpha' v'_{\alpha'}}(q,q'),
  \end{split}
\end{equation}
where the dynamical parameters
$F_{LS}^{\alpha v_\alpha \alpha' v'_{\alpha'}}(q,q')$ describe the ionization
dynamics and are given by
\begin{equation}
  \label{eq:dynLS}
  \begin{split}
    F_{LS}^{\alpha v_\alpha \alpha' v'_{\alpha'}}(q,q')
    &=
    \sum_{ll'}
    [l,l']^{1/2}
    \threej{l}{l'}{L}{0}{0}{0}
    \sum_{\lambda\lambda'}
    (-1)^{\lambda'}
    \threej{l}{l'}{L}{-\lambda}{\lambda'}{S+q-q'}
    \\&\times
    \sum_{v_\alp\alp}
    {\mathcal{E}}(\alpha,v_\alpha,\alp,v_\alp, \epsilon)
    {\mathcal{E}}^\ast(\alpha',v'_{\alpha'},\alp,v_\alp, \epsilon)
    \\&\times
    \sum_{\Gamma\mu h}
    \sum_{\Gamma'\mu'h'}
    b^{\Gamma\mu}_{hl\lambda}
    b^{\Gamma'\mu'\ast}_{h'l'\lambda'}
    \mathrm{(-i)}^{l-l'}
    \exp{\mathrm{i}(\sigma_l(\epsilon)-\sigma_{l'}(\epsilon))}
    \\&\times
    D_{\Gamma\mu h l}^{{\alpha}v_{\alpha}{\alp}v_\alp}(q)
    D_{\Gamma'\mu' h' l'}^{{\alpha'}v'_{\alpha'}{\alp}v_{\alp}\ast}(q').
  \end{split}
\end{equation}
The parameters $A(\alpha,v_\alpha,\alpha',v'_{\alpha'};t)^K_{QS}$ in
\eqref{eq:betalm2} are defined as
\begin{equation}
  \label{eq:adm}
  \begin{split}
    A(\alpha,v_\alpha,\alpha',v'_{\alpha'};t)^K_{QS}&=
    \frac{[K]^{1/2}}{8\pi^2}
    \sum_{J_{\alpha}, \tau_\alpha}
    \sum_{J'_{\alpha'}, \tau'_{\alpha'}}
    (-1)^{2J_\alpha+J'_{\alpha'}}
    [J_\alpha, J'_{\alpha'}]^{1/2}
    \\&\times
    \sum_{K_{\alpha}K'_{\alpha'}}
    (-1)^{-K_{\alpha}}
    a_{K_\alpha}^{J_\alpha\tau_\alpha}
    a_{K'_{\alpha'}}^{J'_{\alpha'}\tau'_{\alpha'}}
    \threej{J_\alpha}{J'_{\alpha'}}{K}{-K_{\alpha}}{K'_{\alpha'}}{S}
    \\&\times
    \multmom{n_{\alpha}}{n'_{\alpha'};t}{K}{Q}.
  \end{split}
\end{equation}
The parameters in \eqref{eq:adm} have an immediate geometrical interpretation
-- they describe the LF distribution of molecular axes of the excited state
neutral molecules prior to ionization~\cite{Blum1996}. For this reason, we
refer to them as the axis distribution moments (ADMs). The molecular axis
distribution in a vibronic level may be expressed as an expansion of Wigner
rotation matrices with the coefficients
$A(\alpha,v_\alpha,\alpha,v_{\alpha};t)^K_{QS}$,
\begin{equation}
  P_{{\alpha}v_\alpha}(\phi, \theta, \chi;t) = 
  \sum_{KQS}
  A(\alpha,v_\alpha,\alpha,v_{\alpha};t)^K_{QS}
  \Drot{K^\ast}{Q}{S},
\end{equation}
and the molecular axis distribution of the whole excited state ensemble of
molecules is given by
\begin{equation}
  \label{eq:ensemble_axdis}
  P(\phi, \theta, \chi;t) = 
  \sum_{\alpha v_\alpha}
  \sum_{\alpha' v'_{\alpha'}}
  \sum_{KQS}
  A(\alpha,v_\alpha,\alpha',v'_{\alpha'};t)^K_{QS}
  \Drot{K^\ast}{Q}{S}.
\end{equation}
The ADMs connect the multipole moments
$\multmom{n_{\alpha}}{n'_{\alpha'};t}{K}{Q}$, which characterize the angular
momentum polarization and coherence, with the molecular axis
distribution. Non-zero ADMs with even $K$ characterize molecular axis
alignment, whereas non-zero ADMs with odd $K$ characterize molecular axis
orientation. A cylindrically symmetric molecular axis distribution along the
lab frame $z$-axis will have non-zero ADMs with $Q=S=0$ only. Linear and
symmetric top molecules, for which only the two angles $(\theta, \phi)$ are
required to characterise the molecular orientation~\cite{Zare1988, Blum1996},
require only ADMs with $S=0$ moments to fully characterize the molecular axis
distribution. Asymmetric top molecules may have non-zero ADMs with both
$Q\not=0$ and $S\not=0$ only when there is localization of all three Euler
angles. An isotropic distribution of molecular axes has the only non-zero ADMs
with $K=0$.

\eqref{eq:betalm2} explicitly expresses the sensitivity of the LF PAD to the
molecular axis distribution. In fact, an equivalent expression is obtained by
convolution of the molecular axis distribution with the vibronic transition
dipole matrix elements without explicit consideration of molecular
rotation~\cite{Underwood2000}. The general form of \eqref{eq:betalm2} is
obtained for other angular momentum coupling cases (e.g. in the presence of
strong spin-orbit coupling) -- the lack of resolution of the ion rotational
states essentially removes the details of the angular momentum coupling from
the problem~\cite{Buckingham1970} (although the expression for the ADMs
\eqref{eq:adm} may be different for other angular momentum coupling schemes).

From the properties of the Wigner $3j$ symbols, we see from \eqref{eq:betalm2}
that the maximum value of $L$ in the expansion in \eqref{eq:betalm} is the
smaller of $2l_{\mathrm{max}}$, where $l_{\mathrm{max}}$ is the largest value
of $l$ in the partial wave expansion \eqref{eq:LFwfn}, and
$(K_{\mathrm{max}}+2)$, where $K_{\mathrm{max}}$ is the maximum value of $K$
in the axis distribution in \eqref{eq:ensemble_axdis}. Each
$\beta_{LM}(\epsilon;t)$ is sensitive to ADMs with values of $K$ from $(L-2)$
(or zero if $(L-2)$ is negative) to $(L+2)$ (since the maximum value of $P$ is
2 and $L$, $K$ and $P$ must satisfy the triangle condition for non-zero Wigner
$3j$ coefficients).  In other words, the more anisotropic the molecular axis
distribution is, the higher the anisotropy of the LF PAD. The distribution of
molecular axes geometrically determines the relative contributions of the
molecular frame ionization transition dipole components that contribute to the
LF PAD. The sensitivity of the LF PAD to the molecular orientation is
determined by the relative magnitudes of the dynamical parameters
$F_{LS}^{\alpha v_\alpha \alpha' v'_{\alpha'}}(q,q')$ which reflect the
anisotropy of the ionization transition dipole and PAD in the MF.  While the
total cross section for ionization (i.e. $\beta_{00}(\epsilon;t)$) is sensitive
to ADMs with $K=0\dots2$, as is the total cross section of all one-photon
absorption processes, we see that measuring the PAD reveals information
regarding ADMs with higher $K$ values than single-photon absorption would
normally. We note also that the PAD from aligned/oriented molecules yields far
more information regarding the photoionization dynamics, and as such provides
a route to performing ``complete'' photoionization experiments~\cite{Reid1992}.

TRPES experiments frequently employ linearly polarized pump and probe pulses,
with the excited state ro-vibronic wavepacket prepared via a single photon
resonant transition and for this reason we shall briefly discuss this
situation. The linearly polarised pump pulse excites molecules with their
transition dipole moment aligned towards the direction of the laser
polarization, due to the scalar product interaction
$\vect{d}\cdot\uvect{e}_{\mathrm{pump}}$ of the transition dipole moment with
the polarization vector of the pump pulse. Since the transition dipole
typically has a well defined direction in the MF, this will create an ensemble
of axis-aligned excited state molecules. Since the pump pulse is linearly
polarized, the excited stated molecular axis distribution possesses
cylindrical symmetry, and so is described by ADMs with $Q=S=0$ in a LF whose
$z$-axis is defined by the pump polarization. The single photon nature of this
pump step limits the values of $K$ to 0 and 2. The fact that only even $K$
moments are prepared, and the molecular axis distribution is aligned and not
oriented, reflects the up-down symmetry of the pump
interaction~\cite{Blum1996}. Since the maximum value of $K$ is 2, the maximum
value of $L\le4$ in \eqref{eq:betalm}. This means that the PAD contains
information concerning the interference of partial waves with $l$ differing be
at most 4 -- the LF PAD will contain terms with at most $|l-l'|=0,2,4$, and
does not contain any information regarding the interference of odd and even
partial waves. The rotational wavepacket created by the pump pulse will
subsequently evolve under the field-free Hamiltonian of the molecule,
initially causing a reduction of the molecular axis alignment, and
subsequently causing a re-alignment of the molecules when the rotational
wavepacket re-phases~\cite{Baskin1986, Felker1987, Baskin1987, Felker1992,
  Felker1994, Riehn2002}. If the probe pulse is timed to arrive when the
molecules are strongly aligned there will be a strong dependence of the LF PAD
upon the direction of the probe pulse polarization relative to the pump pulse
polarization. If the pump and probe polarizations are parallel, then the LF
PAD will maintain cylindrical symmetry, and ionization transition dipole
moments along the molecular symmetry axis will be favoured. Rotating the probe
polarization away from that of the pump will remove the cylindrical symmetry
of the PAD and increase the contributions of the ionization transition dipole
moments perpendicular to the symmetry axis. An example of this effect can be
seen in the model calculations shown in \figref{fig:cos2pad}, where the shape
of the PADs clearly depend upon the molecular axis distribution in the frame
defined by the ionizing laser polarization. An experimental example of this
effect will also be discussed in \secref{sec:trpes_continua} below.  Clearly,
the LF PAD will be extremely sensitive to the rotational motion of the
molecule, and is able to yield detailed information pertaining to molecular
rotation and molecular axis alignment.~\cite{Reid2000, Underwood2000,
  Seideman2000a}. Furthermore, since the coupling of vibrational and
rotational motion will cause changes in the evolution of the molecular axis
alignment, the LF PAD can provide important information regarding such
couplings~\cite{Reid1999, Althorpe2000}. 

We note also that, in an experiment in which the species ionized by the probe
laser is a product of photodissociation initiated by the pump, the PAD will be
sensitive to the LF photofragment axis distribution, and as such will provide
a probe of the photofragment angular momentum coherence and
polarization. Furthermore, with a suitably designed experiment, PADs will
allow measurement of product vector correlations, such as that between the
photofragment velocity and angular momentum polarization. Such vector
correlations in molecular photodissociation have long been studied in a non
time-resolved fashion~\cite{OrrEwing1994, Dixon1986, Houston1987, Hall1989},
and have provided detailed information concerning the photodissociation
dynamics. Whereas these studies have focused upon the rotational state
resolved angular momentum polarization, time-resolved measurements may yield
information regarding the rotational coherence of the photofragments as well
as the angular momentum polarization.

\begin{figure}
  \includegraphics[width=6cm]{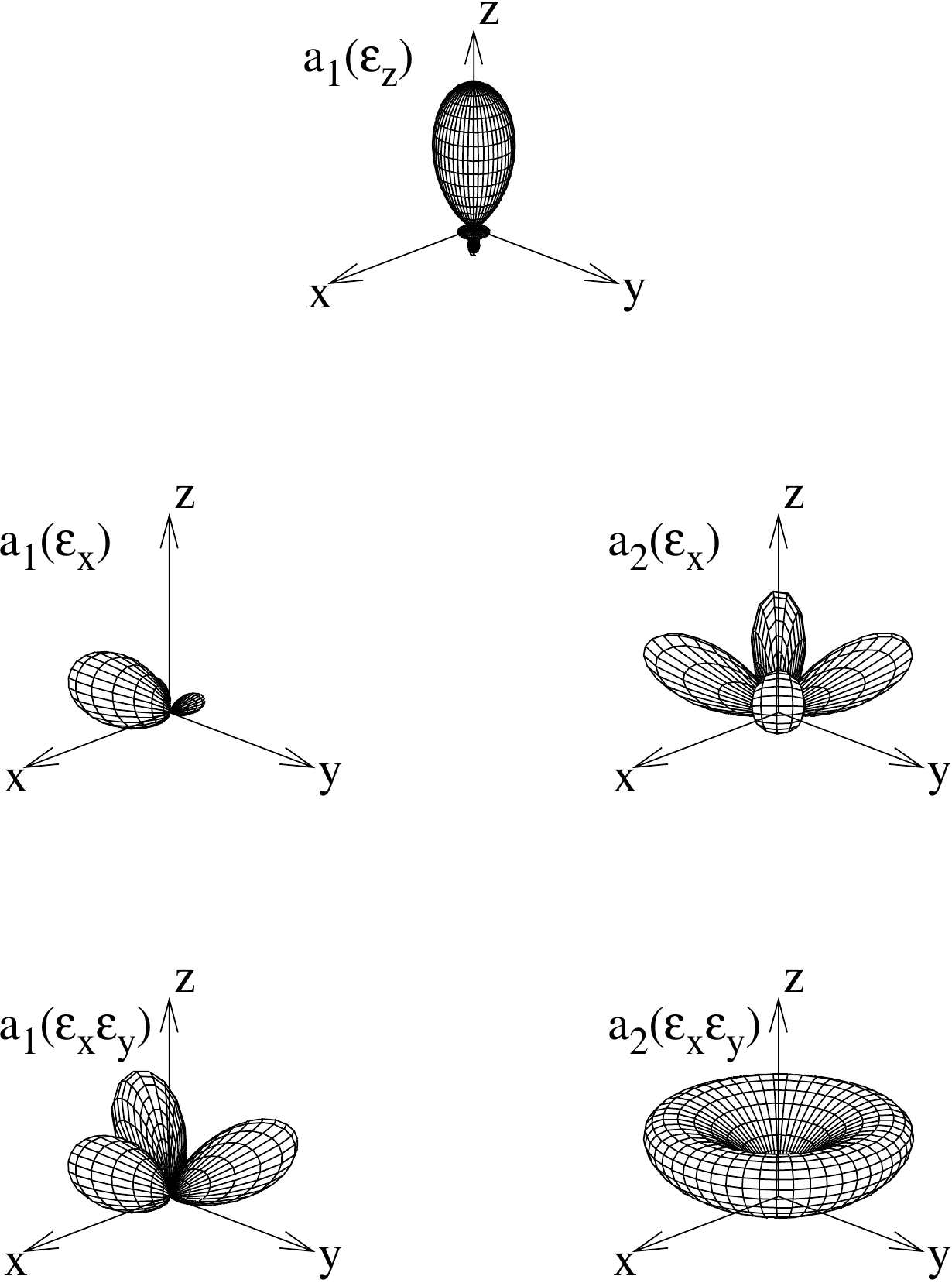}
  \caption{MF PADs for single photon ionization of $a_1$ and $a_2$ symmetry
    orbitals of a model $C_{3v}$ molecule for light linearly polarized along
    different axes of the molecule (indicated in parentheses). Note that no
    photoionization can occur from the $a_2$ orbital for light polarized along
    the $z$-axis (molecular symmetry axis). The same dynamical parameters as
    for the calculations of the LF PADs shown in \figref{fig:cos2pad} were
    used. For further details see ref.~\cite{Underwood2000}.}
  \label{fig:mfpad}
\end{figure}

In the preceding discussion we have discussed the form of the PAD as measured
in the LF, i.e. relative to the polarization direction of the ionizing
laser. However, by employing experimental techniques to measure the
photoelectron in coincidence with a fragment ion following dissociative
ionization, it is possible to measure the PAD referenced to the MF rather than
the LF, removing all averaging over molecular orientation. These experimental
techniques are described in \secref{sec:expt_trcis}, and examples of such
measurements are subsequently given in \secref{sec:app_photodis}. The form of
the MF PAD can be expressed in a form similar to the LF PAD,
\begin{equation}
  \label{eq:dcsMF}
  \sigma(\epsilon, \kM;t)
  \propto
  \sum_{\alp v_\alp}
  \bigg|
  \sum_{{\alpha}v_{\alpha}}
  C_{{\alpha}v_{\alpha}}(t)
  \bra{\kM\epsilon \alp v_\alp}
  \vect{d}\cdot\uvect{e}
  \ket{\alpha v_{\alpha}}
  \mathcal{E}(\alpha,v_\alpha,\alp,v_\alp, \epsilon)
  \bigg|^2,
\end{equation}
where $C_{{\alpha}v_{\alpha}}(t)$ are the time dependent complex coefficients
for each excited state vibronic level. The scalar product of the transition
dipole moment $\vect{d}$ with the polarization vector of the probe laser field
$\uvect{e}$ in terms of the spherical tensor components in the MF
as~\cite{Zare1988}
\begin{equation}
  \label{eq:MFd_dot_e}
  \vect{d}\cdot\uvect{e} = \sum_{q=-1}^{1} (-1)^q d_q e_{-q}.
\end{equation}
The electric field polarization is conveniently described in the LF. The MF
spherical tensor components of the electric field polarization tensor are
related to the components in the LF through a rotation
\begin{equation}
  \label{eq:erotMF}
  e_{-q}=\sum_{p=-1}^{1}\Drot{1}{-p}{-q}e_p.
\end{equation}
Substitution of \eqref{eq:MFsymadapwfn}, \eqref{eq:erotMF} and
\eqref{eq:MFd_dot_e} into \eqref{eq:dcsMF} yields an expression similar to
\eqref{eq:betalm} for the MF PAD (see
Appendix~\ref{app:mfpad})~\cite{Dill1976, Chandra1987, Underwood2000},
\begin{equation}
  \label{eq:MFbetalm}
  \sigma(\epsilon, \kM;t)=
  \frac{\sigma_{\mathrm{total}}(\epsilon;t)}{4\pi}
  \sum_{LM}\beta_{LM}^{\mathrm{M}}(\epsilon;t)
  Y_{LM}(\kM),
\end{equation}
where the $\beta_{LM}^{\mathrm{M}}(\epsilon;t)$ coefficients are given
by~\cite{Dill1976, Chandra1987}
\begin{equation}
  \label{eq:MFbetalm1}
  \begin{split}
    \beta_{LM}^{\mathrm{M}}(\epsilon;t)&=
    [L]^{1/2}
    \sum_{PR}
    \sum_{qq'}
    [P]^{1/2}
    (-1)^{q'}
    \threej{1}{1}{P}{q}{-q'}{q'-q}
    \Drot{P}{R}{q'-q}
    \\&\times
    E_{PR}(\uvect{e})
    \sum_{ll'}
    [l,l']^{1/2}
    \threej{l}{l'}{L}{0}{0}{0}
    \sum_{\lambda\lambda'}
    (-1)^{\lambda}
    \threej{l}{l'}{L}{\lambda}{-\lambda'}{M}
    \\&\times
    \mathrm{(-i)}^{l-l'}
    \exp{\mathrm{i}(\sigma_l(\epsilon)-\sigma_{l'}(\epsilon))}
    \sum_{{\alpha}v_{\alpha}}
    \sum_{{\alpha'}v'_{\alpha'}}
    C_{{\alpha}v_{\alpha}}(t)
    C_{{\alpha'}v'_{\alpha'}}^\ast(t)
    \sum_{\Gamma\mu h}
    \sum_{\Gamma'\mu'h'}
    \\&\times
    b^{\Gamma\mu}_{hl\lambda}
    b^{\Gamma'\mu'\ast}_{h'l'\lambda'}
    \sum_{\alp v_\alp}
    D_{\Gamma\mu h l}^{{\alpha}v_{\alpha}{\alp}v_\alp}(q)
    D_{\Gamma'\mu' h' l'}^{{\alpha'}v'_{\alpha'}{\alp}v_{\alp}\ast}(q')
    \\&\times
    \mathcal{E}(\alpha,v_\alpha,\alp,v_\alp, \epsilon)
    \mathcal{E}^\ast(\alpha',v'_{\alpha'},\alp,v_\alp, \epsilon).
  \end{split}
\end{equation}
Naturally, since this expression is for a property measured in the reference
frame connected to the molecule, molecular rotations do not appear in this
expression.  The range of $L$ in the summation \eqref{eq:MFbetalm} is
$0\dots2l_{\mathrm{max}}$, and includes both odd and even values. In general
the MF PAD is far more anisotropic than the LF PAD, for which $L=0,2,4$ in a
two-photon linearly polarized pump-probe experiment in the perturbative
limit. Clearly, the MF PAD contains far more detailed information than the LF
PAD concerning the ionization dynamics of the molecule as well as the
structure and symmetry of the electronic state from which ionization occurs,
since the partial waves that may interfere are no longer geometrically limited
as they are for the LF PAD. The contributing MF ionization transition dipole
components are determined by the laser polarization and the Euler angles
between the LF and MF. Example calculations for a model $C_{3v}$ molecule are
shown in \figref{fig:mfpad} employing the same dynamical parameters as for the
calculations of the LF PADs shown in \figref{fig:cos2pad} which demonstrate
the much higher anisotropy of the MF PAD. These calculations demonstrate the
dependence of the MF PAD upon the probe pulse MF polarization direction (or
equivalently the MF ionization transition dipole direction). The LF PAD
corresponds to a coherent summation over such MF PADs, weighted by the
molecular axis distribution in the LF.

Finally, in closing we note that significant advances have recently been made
towards defining the direction of the molecular orientation in the LF using
strong non-resonant laser fields~\cite{Lee2006, Friedrich1995, Seideman1995,
  Peronne2003, Lee2004, Bisgaard2004, Berakdar2003, Dooley2003, Larsen2000,
  Machholm2001, Stapelfeldt2003, Sugny2004, Underwood2003, Underwood2005,
  Jauslin2005, Tanji2005, Sakai2003}. This will provide the opportunity to
make PAD measurements which approximate well the MF PAD without coincident
detection of dissociative ionization. The alignment and orientation achievable
in with such techniques produces extremely high $K$ valued ADMs due to the
highly non-linear (multi-photon) nature of the matter-laser interaction which
produces the alignment and orientation. The prospects offered by these
techniques are exciting and will herald a new generation of PAD measurements.

%%% Local Variables: 
%%% mode: latex
%%% TeX-master: "trpes"
%%% End: 

\chapter{Experimental techniques}
\section{Photoelectron spectrometers}
In a photoelectron spectroscopy measurement the observables are the electron
kinetic energy distribution, the photoelectron angular distribution (PAD), the
electron spin and the set of scalar and vector correlations between these
electron distributions and those of the ion. Spectrometers for femtosecond
TRPES have modest energy resolution requirements as compared to modern
standards for photoelectron spectrometers.  For Gaussian optical pulses, the
time-bandwidth product is $\Delta\nu\Delta t=0.441$ and, therefore, the
bandwidth (FWHM) of a Gaussian 100~fs pulse is $\sim150$ cm$^{-1}$. A
pump-probe measurement typically involves the convolution of two such pulses,
leading to an effective bandwidth of $\sim25$~meV. This limits the energy
resolution required in measuring the energy of the photoelectrons. We
emphasize that in femtosecond pump-probe experiments, the laser intensity must
be kept below multiphoton ionization thresholds. This simply requires a
reduction of the laser intensity until one-photon processes dominate. At this
level the ionization probabilities are small and, usually, single particle
counting techniques are requred. Therefore, TRPES experiments are very
data-intensive and require the collection of many photoelectron spectra.  As a
result, most neutral TRPES experiments performed to date make use of high
efficiency electron energy analyzers in which a large fraction of the
photoelectrons are collected.

A commonly used analyzer in TRPES experiments is been the ``magnetic bottle''
time-of-flight (TOF) spectrometer~\cite{Kruit1983, Lange1995,
  Lochbrunner2000}. This technique uses a strong inhomogeneous magnetic field
(1~T) to rapidly parallelize electron trajectories into a flight tube,
followed by a constant magnetic field (10~G) to guide the electrons to the
detector. The collection efficiency of magnetic bottle spectrometers can
exceed 50\%, while maintaining an energy resolution essentially equivalent to
the fs laser bandwidth. Highest resolution is obtained for electrons created
within a small volume ($<100$~$\mu$m) at the very center of the interaction
region. Magnetic bottle analyzers have been used in many neutral TRPES
experiments. They are relatively simple, have high collection efficiency and
rapid data readout. Magnetic bottles have the general disadvantage that they
can only be used to determine electron kinetic energy distributions; the
complex electron trajectories in magnetic bottle analyzers make it impractical
to extract angular information.

Time-resolved two dimensional (2D) photoelectron imaging techniques, in which
position-sensitive detection is used to measure the photoelectron kinetic
energy and angular distributions simultaneously, is becoming increasingly
popular due to its sensitivity and ease of
implementation~\cite{Suzuki2006}. When used, as is common, with CCD camera
systems for image collection, particle hits are usually averaged on the CCD
chip because rapid CCD readout at kHz rates remains very challenging. The most
straightforward 2D imaging technique is photoelectron velocity-map imaging
(VMI)~\cite{Eppink1997, Parker1997}, a variant of the photofragment imaging
method~\cite{Heck1995}. Typically, a strong electric field images nascent
charged particles onto a microchannel plate (MCP) detector. The ensuing
electron avalanche falls onto a phosphor screen, which is imaged by the CCD
camera. Analysis of the resultant image allows for the extraction of both
energy and angle-resolved information. In this case a two dimensional
projection of the three dimensional distribution of recoil velocity vectors is
measured; various image re-construction techniques~\cite{Whitaker2000,
  Dribinski2002} are then used to recover the full three dimensional
distribution. Photoelectron VMI thus yields close to the theoretical limit for
collection efficiency, along with simultaneous determination of the
photoelectron energy and angular distributions.  The 2D particle imaging
approach may be used when the image is a projection of a cylindrically
symmetric distribution whose symmetry axis lies parallel to the
two-dimensional detector surface. This requirement precludes the use of pump
and probe laser polarization geometries other than parallel polarizations. It
may therefore be preferable to adopt fully three dimensional (3D) imaging
techniques based upon ``time-slicing''~\cite{Gebhardt2001, Townsend2003} or
full time-and-position sensitive detection~\cite{Continetti2003}, where the
full 3D distribution is obtained directly without mathematical reconstruction
(see below).  In femtosecond pump-probe experiments where the intensities must
be kept below a certain limit (requiring single particle counting methods),
``time-slicing'' may not be practical, leaving only full time-and-position
sensitive detection as the option.

Modern MCP detectors permit the direct 3D measurement of lab frame (LF) recoil
momentum vectors by measuring both spatial position $(x,y)$ on -- and time of
arrival $(z)$ at -- the detector face~\cite{Continetti2003}. Importantly, this
development does not require inverse transformation to reconstruct 3D
distributions and so is not restricted to experimental geometries with
cylindrical symmetry, allowing any desired polarization geometry to be
implemented. In 3D particle imaging, a weak electric field is used to extract
nascent charged particles from the interaction region. Readout of the $(x,y)$
position (i.e., the polar angle) yields information about the velocity
distributions parallel to the detector face, equivalent to the information
obtained from 2D detectors.  However, the additional timing information allows
measurement of the third $(z)$ component (i.e., the azimuthal angle) of the
LF velocity, via the ``turn-around'' time of the particle in the weak
extraction field.  Thus, these detectors allow for full 3D velocity vector
measurements, with no restrictions on the symmetry of the distribution or any
requirement for image re-construction techniques.  Very successful methods for
full time-and-position sensitive detection are based upon interpolation
(rather than pixellation) using either charge-division (such as ``wedge-and-
strip'')~\cite{Davies1999} or crossed delay-line anode timing MCP
detectors~\cite{Hanold1996}. In the former case, the avalanche charge cloud is
divided amongst three conductors -- a wedge, a strip and a zigzag. The $(x,y)$
positions are obtained from the ratios of the wedge and strip charges to the
zigzag (total) charge.  Timing information can be obtained from a capacitive
pick-off at the back of the last MCP plate.  In the latter case, the anode is
formed by a pair of crossed, impedance-matched delay-lines (i.e. $x$ and $y$
delay lines). The avalanche cloud that falls on a delay line propagates in
both directions towards two outputs. Measurement of the timing difference of
the output pulses on a given delay line yields the $x$ (or $y$) positions on
the anode. Measurement of the sum of the two output pulses (relative to, say,
a pickoff signal from the ionization laser or the MCP plate itself) yields the
particle arrival time at the detector face.  Thus, direct anode timing yields
a full 3D velocity vector measurement.  An advantage of delay line anodes over
charge division anodes is that the latter can tolerate only a single hit per
laser shot, precluding the possibility of multiple coincidences, and
additionally make the experiment sensitive to background scattered UV light.

\section{Coincidence techniques}
\label{sec:expt_trcis}
Photoionization always produces two species available for analysis - the ion
and the electron. By measuring both photoelectrons and photoions in
coincidence, the kinetic electron may be assigned to its correlated parent ion
partner (which may be identified by mass spectrometry). The extension of the
photoelectron-photoion-coincidence (PEPICO) technique to the femtosecond
time-resolved domain was shown to be very important for studies of dynamics in
clusters~\cite{Stert1997, Stert1999}. In these experiments, a simple yet
efficient permanent magnet design ``magnetic bottle'' electron spectrometer
was used for photoelectron TOF measurements.  A collinear TOF mass
spectrometer was used to determine the mass of the parent ion. Using
coincidence electronics, the electron TOF (yielding electron kinetic energy)
is correlated with an ion TOF (yielding the ion mass). In this manner, TRPES
experiments may be performed on neutral clusters, yielding time-resolved
spectra for each parent cluster ion (assuming cluster fragmentation plays no
significant role). Signal levels must be kept low (much less than one
ionization event per laser shot) in order to minimize false coincidences. The
reader is referred to a recent review for a detailed discussion on TRPEPICO
methods~\cite{Hertel2006}.

Coincident detection of photoions and photoelectrons has long been recognized
as a route to recoil or molecular frame photoelectron angular distributions in
non-time-resolved studies~\cite{Eland1979, Low1985, Powis1992}. For the case
of nanosecond laser photodetachment, correlated photofragment and
photoelectron velocities can provide a complete probe of the dissociation
process~\cite{Hanold1996, Garner1997}. The photofragment recoil measurement
defines the energetics of the dissociation process and the alignment of the
recoil axis in the LF, the photoelectron energy provides spectroscopic
identification of the products and the photoelectron angular distribution can
be transformed to the recoil frame in order to make measurements approaching
the MF PAD. Measuring the recoil frame PAD can also provide vector
correlations such as that between the photofragment angular momentum
polarization and the recoil vector. Time and angle-resolved PEPICO
measurements showing the evolution of photoion and photoelectron kinetic
energy and angular correlations will undoubtedly shed new light on the
photodissociation dynamics of polyatomic molecules. The integration of
photoion-photoelectron timing-imaging (energy and angular correlation)
measurements with femtosecond time-resolved spectroscopy was first
demonstrated, using wedge-and-strip anode detectors, in 1999~\cite{Davies1999,
  Davies2000}. This Time-Resolved Coincidence-Imaging Spectroscopy (TRCIS)
method allows the time evolution of complex dissociation processes to be
studied with unprecedented detail~\cite{Gessner2006} and was first
demonstrated for the case of the photodissociation dynamics of
NO$_{2}$~\cite{Davies1999} (discussed in more detail in
\secref{sec:app_photodis}).

TRCIS allows for kinematically complete energy- and angle-resolved detection
of both electrons and ions in coincidence and as a function of time,
representing the most differential TRPES measurements made to date. This
time-resolved 6D information can be projected, filtered and/or averaged in
many different ways, allowing for the determination of various time-resolved
scalar and vector correlations in molecular photodissociation. For example, an
interesting scalar correlation is the photoelectron kinetic energy plotted as
a function of the coincident photofragment kinetic energy. This 2D correlation
allows for the fragment kinetic energy distributions of specific channels to
be extracted. For experimentalists, an important practical consequence of this
is the ability to separate dissociative ionization (i.e. ionization followed
by dissociation) of the parent molecule from photoionization of neutral
fragments (i.e. dissociation followed by ionization). In both cases the same
ionic fragment may be produced and the separation of these very different
processes may be challenging. TRCIS, via the 2D energy-energy correlation map,
does this naturally. The coincident detection of the photoelectron separates
these channels: in one case (dissociative ionization) the photoelectron comes
from the parent molecule, whereas in the other case (neutral
photodissociation) the photoelectron comes from the fragment. In most cases,
these photoelectron spectra will be very different, allowing complete
separation of the two processes.

A very interesting vector correlation is the recoil direction of the
photoelectron as a function of the recoil direction of the coincident
photofragment. Although for each dissociation event the fragment may recoil in
a different laboratory direction, TRCIS determines this direction and,
simultaneously, the direction of the coincident electron. Therefore,
event-by-event detection via TRCIS allows the PAD to be measured in the
fragment recoil frame rather than the usual LF. In other words, it is
time-resolved dynamics from the molecule's point of view. This is important
because the usual LF PADs are generally averaged over all molecular
orientations, leading to a loss of information.  Specifically, for a
one-photon pump, one-photon probe TRPES experiment on a randomly aligned
sample, conservation of angular momentum in the LF limits the PAD anisotropy,
as discussed in \secref{sec:photoionzn}. In the recoil frame, these
limitations are relaxed, and an unprecedentedly detailed view of the excited
state electronic dynamics obtains.  Other types of correlations, such as the
time evolution of photofragment angular momentum polarization, may also be
constructed from the 6D data of TRCIS.

\section{Femtosecond Laser Technology}
Progress in femtosecond TRPES benefits from developments in femtosecond laser
technology, since techniques for photoelectron spectroscopy have been highly
developed for some time. There are several general requirements for such a
femtosecond laser system.  Most of the processes of interest are initiated by
absorption of a photon in the wavelength range $\sim$200--350~nm, produced via
non-linear optical processes such as harmonic generation, frequency mixing and
parametric generation. Thus the output pulse energy of the laser system must
be high enough for efficient use of nonlinear optical techniques and ideally
should be tunable over a wide wavelength range. Another important
consideration in a femtosecond laser system for time-resolved photoelectron
spectroscopy is the repetition rate.  To avoid domination of the signal by
multiphoton processes, the laser pulse intensity must be limited, thus also
limiting the available signal per laser pulse.  As a result, for many
experiments a high pulse repetition rate can be more beneficial than high
energy per pulse. Finally, the signal level in photoelectron spectroscopy is
often low in any case and, for time-resolved experiments, spectra must be
obtained at many time delays. This requires that any practical laser system
must run very reliably for many hours at a time.

Modern Ti:Sapphire based femtosecond laser oscillators have been the most
important technical advance for performing almost all types of femtosecond
time-resolved measurements~\cite{Krausz1992}. Ti:Sapphire oscillators are
tunable over a 725--1000~nm wavelength range, have an average output power of
several hundred mW or greater and can produce pulses as short as 8~fs, but
more commonly 50--130~fs, at repetition rates of 80--100~MHz. Broadly tunable
femtosecond pulses can be derived directly from amplification and frequency
conversion of the fundamental laser frequency.

The development of chirped-pulse amplification and Ti:Sapphire regenerative
amplifier technology now provides mJ pulse energies at repetition rates of
greater than 1~kHz with $<100$~fs pulse widths~\cite{Squier1991}. Chirped
pulse amplification typically uses a grating stretcher to dispersively stretch
fs pulses from a Ti:Sapphire oscillator to several hundred picoseconds. This
longer pulse can now be efficiently amplified in a Ti:Sapphire amplifier to
energies of several mJ while avoiding nonlinear propagation effects in the
solid-state gain medium.  The amplified pulse is typically recompressed in a
grating compressor.

The most successful approach for generating tunable output is optical
parametric amplification (OPA) of spontaneous parametric fluorescence or a
white light continuum, using the Ti:Sapphire fundamental or second harmonic as
a pump source. Typically, an 800~nm pumped femtosecond OPA can provide a
continuous tuning range of 1200--2600~nm~\cite{Nisoli1994}. Non-collinear OPAs
(NOPAs)~\cite{Wilhelm1997} pumped at 400~nm provide $\mu$J-level $\sim$10--20
fs pulses which are continuously tunable within a range of 480-750 nm,
allowing for measurements with extremely high temporal resolution. A computer
controlled stepper motor is normally used to control the time delay between
the pump and probe laser systems.  The development of femtosecond laser
sources with photon energies in the vacuum ultraviolet (VUV, 100--200~nm),
extreme ultraviolet (XUV, $<100$~nm) and beyond (soft x-ray) opens new
possibilities for TRPES, including the preparation of high lying molecular
states, the projection of excited states onto a broad set of cation electronic
states and, in the soft x-ray regime, time-resolved inner shell photoelectron
spectroscopy. High harmonic generation in rare gases is a well-established and
important method for generating femtosecond VUV, XUV~\cite{L'Huillier1995} and
soft x-ray radiation~\cite{Spielmann1997, Rundquist1998,
  Bartels2000}. Harmonics as high as the $\sim$300th order have been reported,
corresponding to photon energies in excess of 500~eV. Both pulsed rare gas
jets and hollow-core optical waveguides~\cite{Rundquist1998, Durfee1999} have
been used for high harmonic generation.  Lower harmonics of the Ti:sapphire
laser have been used in TRPES experiments~\cite{Sorensen2000,
  Nugent-Glandorf2002, Nugent-Glandorf2001, Nugent-Glandorf2002a}. As these
techniques become more commonplace, the range of applicability of TRPES will
be increased significantly.

%%% Local Variables:
%%% mode: latex
%%% TeX-master: "trpes"
%%% End:

\chapter{Comparison of Time-Resolved Ion with TRPES Measurements}
\section{Mass-resolved Ion Yield Measurements}
A powerful version of femtosecond pump-probe spectroscopy combines
photoionization with mass-resolved ion detection (i.e., mass spectrometry).
The mass of the parent ion directly identifies the species under interrogation
and measurement of fragment ions can provide information on dissociation
pathways in the excited molecule~\cite{Zewail2000}. However, fragmentation may
also be a consequence of the photoionization dynamics (i.e., dynamics in the
ionic continuum upon photoionization). As photoionization dynamics are
revealed by photoelectron spectroscopy, it is worth comparing time-resolved
mass spectrometry with TRPES in more detail. As a vehicle for this comparison,
we first discuss the illustrative example of excited state dynamics in linear
polyenes. Non-adiabatic dynamics in linear polyenes generally leads to the
fundamental process of cis-trans photoisomerization.
All-trans-(2,4,6,8)-decatetraene (DT, C$_{10}$H$_{14}$) provides a classic
example of internal conversion in a linear polyene~\cite{Blanchet1999,
  Blanchet2001}. In DT, the lowest excited state is the one-photon forbidden
S$_{1}$ $2^{1}A_{g}$ state whereas the second excited state is the one-photon
allowed S$_{2}$ $1^{1}B_{u}$ state (a classic $\pi\rightarrow\pi^{*}$
transition). When the energy gap between S$_{2}$ and S$_{1}$ is large, the
density of S$_{1}$ vibronic levels can be very large compared to the
reciprocal electronic energy spacing and the ``dark'' state forms an
apparently smooth quasicontinuum (the statistical limit for the radiationless
transition problem). The S$_{2}$--S$_{1}$ energy gap in DT is 5764~cm$^{-1}$
(0.71~eV) placing this large molecule (with 66 vibrational modes) in this
statistical limit.

\begin{figure}
  \includegraphics[width=6cm]{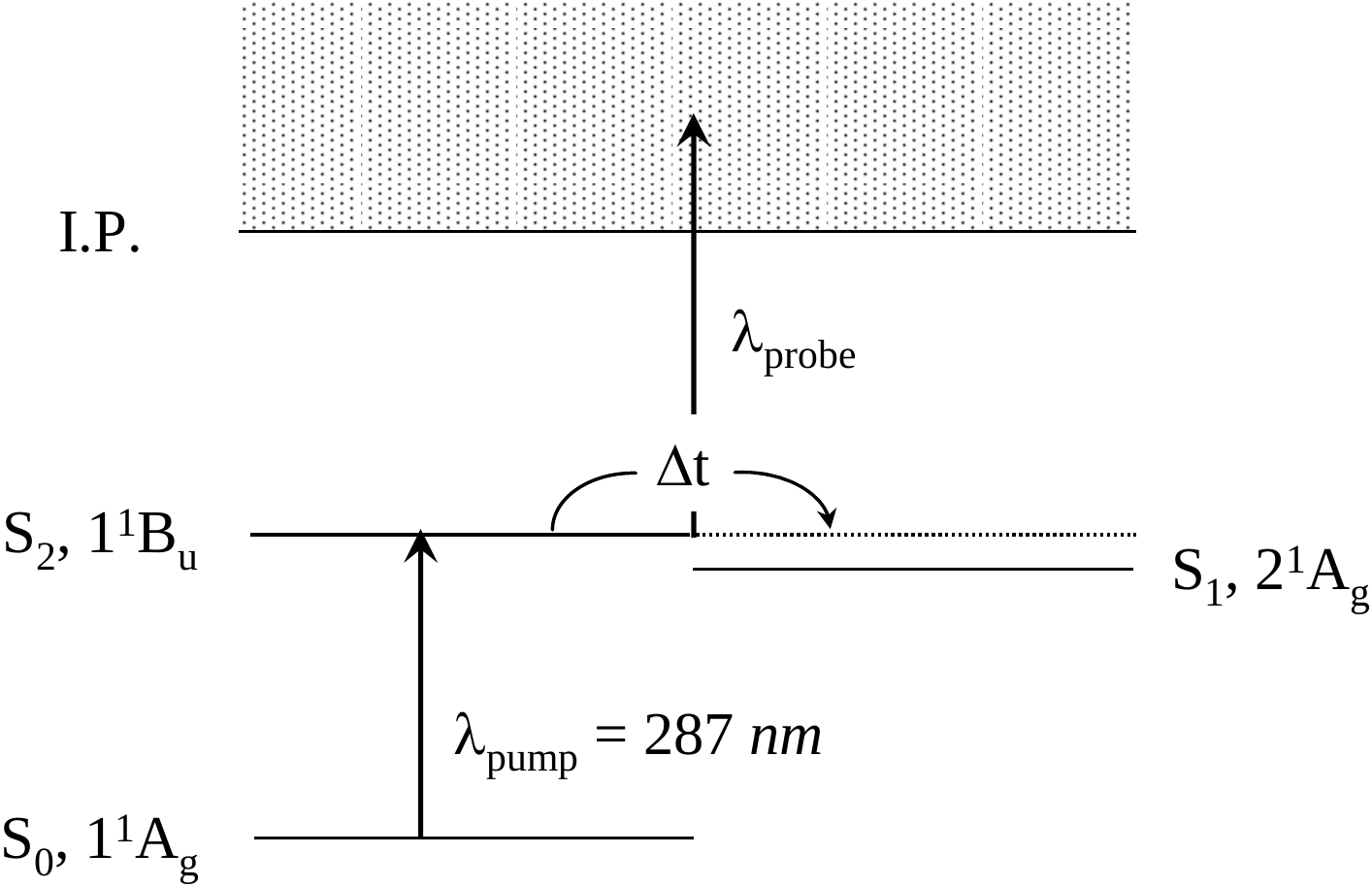}
  \caption{A femtosecond pump-probe photoionization scheme for studying
    excited state dynamics in all-trans decatetraene (DT). The molecule is
    excited to its S$_{2}$ electronic origin with a pump pulse at 287~nm
    (4.32~eV).  Due to non-adiabatic coupling, DT undergoes rapid internal
    conversion to the lower lying S$_{1}$ state (3.6~eV). The excited state
    evolution is monitored via single photon ionization. As the ionization
    potential is 7.29~eV, all probe wavelengths below 417~nm permit single
    photon ionization of the excited state.}
  \label{fig:DT_energetics}
\end{figure}

\begin{figure}
  \includegraphics[width=4in]{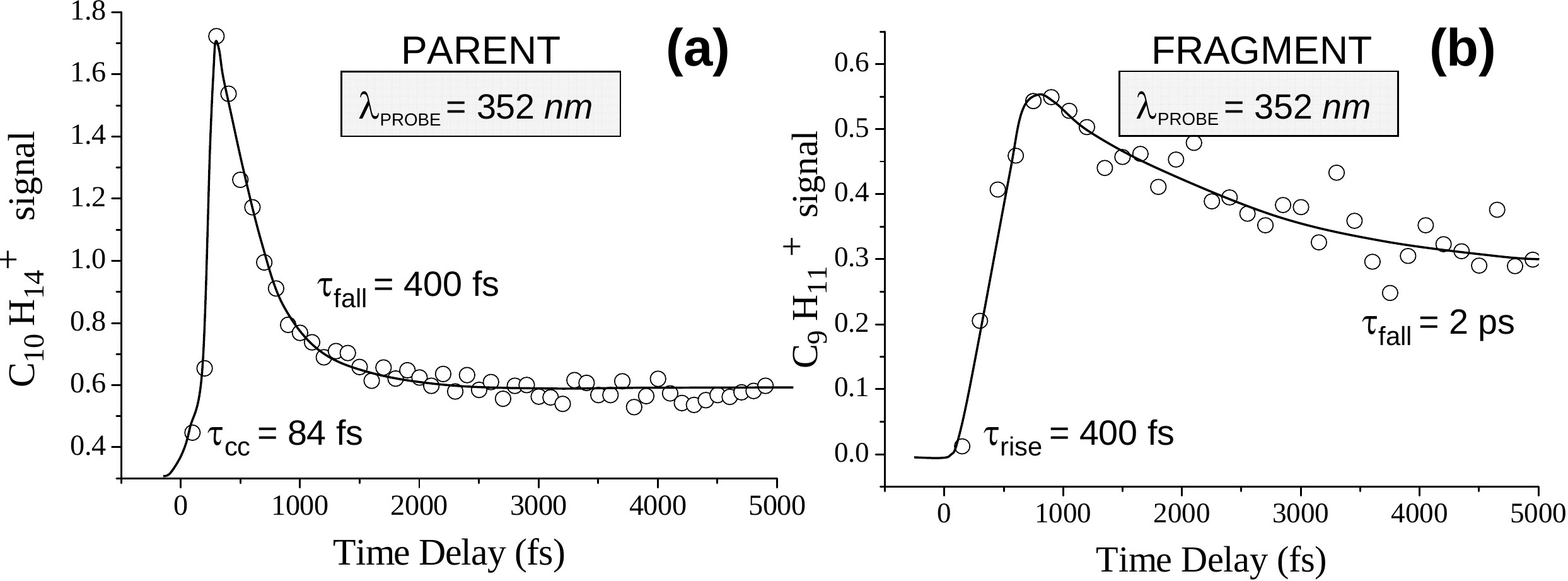}
  \caption{Time-resolved Mass Spectrometry. All-trans decatetraene (DT) was
    excited to its S$_{2}$ electronic origin with a fs pulse at
    $\lambda_{\mathrm{pump}}=287$~nm.  The excited state evolution was probed
    via single photon ionization using a femtosecond pulse at
    $\lambda_{\mathrm{probe}}=352$~nm. The time-resolution in these
    experiments was 80~fs (0.08~ps).  (a) Time-evolution of the parent ion
    C$_{10}$H$_{14}^{+}$ signal. The parent ion signal rises with the pump
    laser and then decays with a single exponential time constant of 0.4~ps,
    suggesting that this is the lifetime of the S$_{2}$ state. The fate of the
    molecule subsequent to this decay is unknown from these data. (b)
    Time-evolution of the fragment ion C$_{9}$H$_{11}^{+}$ signal,
    corresponding to methyl loss. The rise time of this fragment signal is
    0.4~ps, matching the decay time of the S$_{2}$ state. This might suggest
    that the methyl loss channel follows directly from internal conversion to
    S$_{1}$. The C$_{9}$H$_{11}^{+}$ signal subsequently decays with a time
    constant of about 2~ps, suggesting some further step in the excited state
    dynamics.}
  \label{fig:DT_ion_352}
\end{figure}

We consider in the following a time-resolved photoionization experiment using
mass-resolved ion detection, as illustrated in \figref{fig:DT_energetics}. DT
was excited to its S$_{2}$ origin and the ensuing dynamics followed by probing
via single photon ionization. In DT the S$_{2}$ electronic origin is at
4.32~eV (287~nm) and the ionization potential is 7.29~eV. Hence, all probe
laser wavelengths below 417~nm permit single photon ionization of the excited
state. Using $\lambda_{\mathrm{pump}}=287$~nm and
$\lambda_{\mathrm{probe}}=352$~nm one could therefore perform a time-resolved
experiment using mass-resolved ion detection, as shown in
\figref{fig:DT_ion_352}. The time-resolution (pump-probe cross correlation) in
these experiments was 80~fs. It can be seen in \figref{fig:DT_ion_352}(a) that
the parent ion C$_{10}$H$_{14}^{+}$ signal rises with the laser
cross-correlation and then decays with a 0.4~ps time constant. This suggests
that the S$_{2}$ state lifetime is 0.4~ps. The fate of the molecule following
this ultrafast internal conversion, however, cannot be discerned from these
data.  As mass-resolved ion signals are the observable in this experiment, one
could therefore look for the appearance of any potential reaction products
(e.g. fragments) following the internal conversion.

In \figref{fig:DT_ion_352}(b) we present the time evolution of a fragment ion
C$_{9}$H$_{11}^{+}$ which corresponds to the loss of a methyl group from the
parent molecule. The rise time of this signal is 0.4~ps, matching the decay of
the parent molecule. It might be concluded from these data that the 0.4~ps
decay of the S$_{2}$ state leads directly to methyl loss on the S$_{1}$
manifold.  The subsequent $\sim2$~ps decay of this signal would then be the
signature of some competing process in the S$_{1}$ state, perhaps internal
conversion to the S$_{0}$ ground state. In the following we will go on to show
that this conclusion is, in fact, incorrect.

\begin{figure}
  \includegraphics[width=3in]{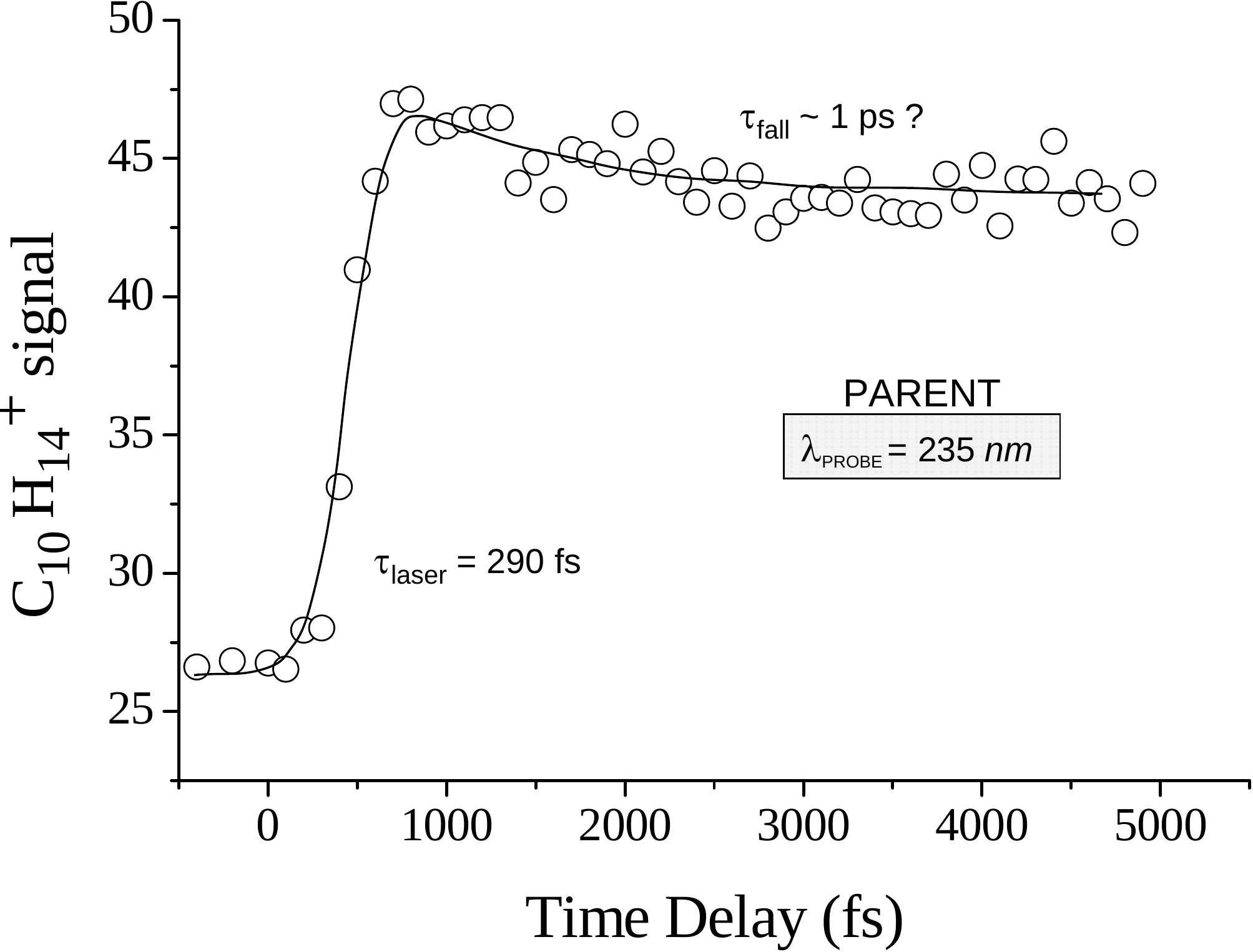}
  \caption{Time-resolved Mass Spectrometry. All-trans decatetraene (DT) was
    excited to its S$_{2}$ electronic origin with a fs pulse at
    $\lambda_{\mathrm{pump}}=287$~nm.  The excited state evolution was probed
    via single photon ionization using a fs pulse at
    $\lambda_{\mathrm{probe}}=235$~nm. The time-resolution in these
    experiments was 290~fs (0.3~ps).  The parent ion C$_{10}$H$_{14}^{+}$
    signal rises with the pump laser but then seems to stay almost constant
    with time. The modest decay observed can be fit with a single exponential
    time constant of $\sim1$~ps. Note that this result is in apparent
    disagreement with the same experiment performed at
    $\lambda_{\mathrm{probe}}=352$~nm which yielding a lifetime of 0.4~ps for
    the S$_{2}$ state.  The disagreement between these two results can be only
    reconciled by analyzing the time-resolved photoelectron spectrum.}
  \label{fig:DT_ion_235}
\end{figure}

One might think that the specific wavelength of the photoionization laser is
of little import as long as it sufficiently exceeds the ionization
potential. In \figref{fig:DT_ion_235} the results of the same experiment but
repeated this time using a probe laser wavelength of 235~nm are presented. As
the pump laser remained invariant, the same excited state wavepacket was
prepared in these two experiments. Contrasting with the 352~nm probe
experiment, we see that the parent ion signal does not decay in 0.4~ps but
rather remains almost constant, perhaps decaying slightly with a $\sim1$~ps
time constant. Furthermore, no daughter ion fragments were observed.  These
very different results seem hard to reconcile. Which probe laser wavelength
gives the right answer and why? As discussed below, the time evolution of
mass-resolved ion signals can be misleading.

The solution to this apparent paradox lies in the photoionization dynamics.
Clearly the form of the parent ion signal depends strongly upon the specific
photoionization dynamics and, in order to avoid misleading conclusions, must
be analyzed for each specific case. The (Koopmans') photoionization
correlations of excited state electronic configurations with those of the
cation play a critical role.

\section{TRPES: The role of electronic continua}
\label{sec:trpes_continua}
\begin{figure}
  \includegraphics[height=4in]{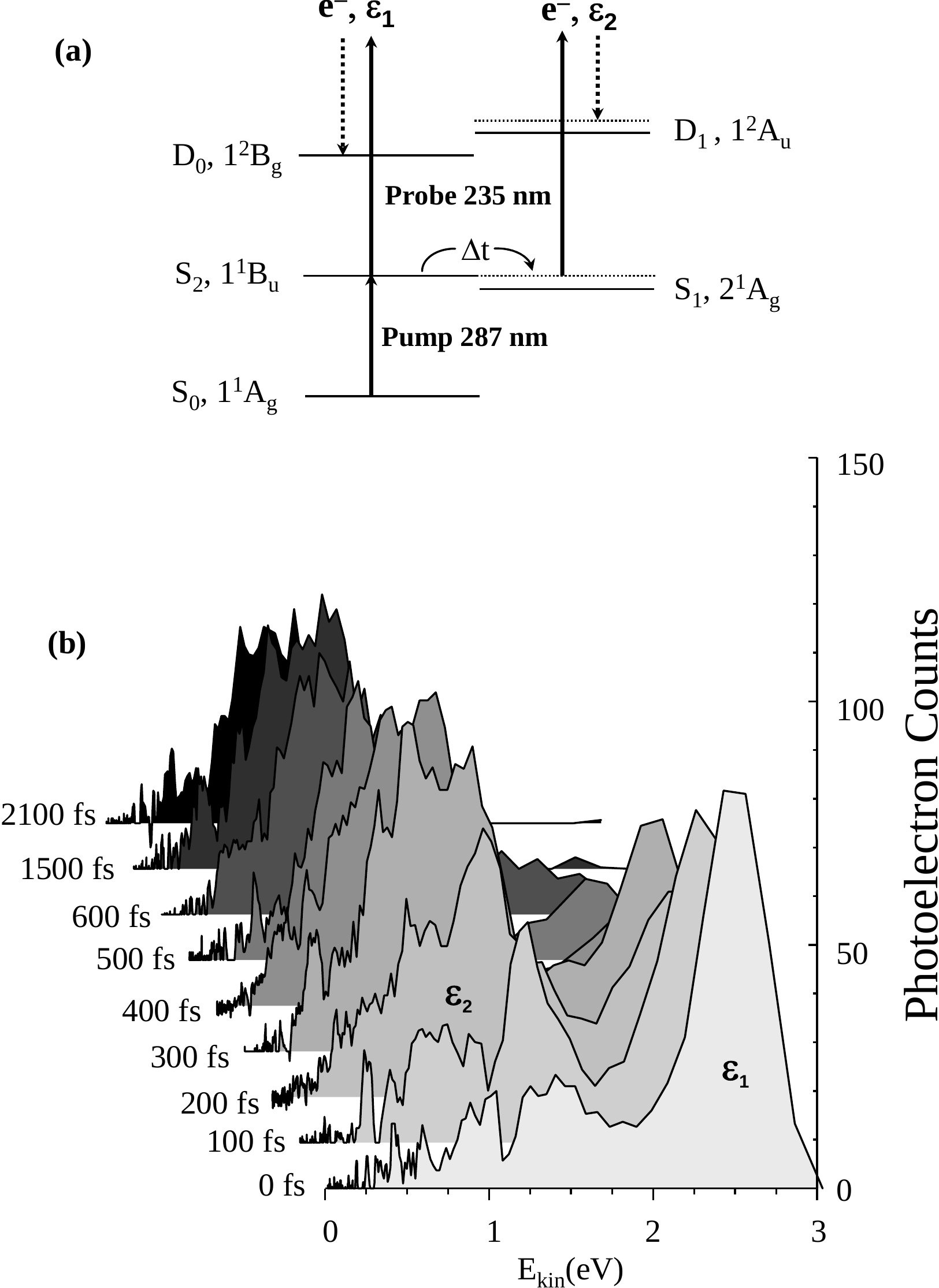}
  \caption{Time-resolved photoelectron spectra revealing vibrational and
    electronic dynamics during internal conversion in all trans decatetraene
    (DT).  (a) Level scheme in DT for one-photon probe ionization. The pump
    laser prepares the optically bright state S$_{2}$. Due to ultrafast
    internal conversion, this state converts to the lower lying state S$_{1}$
    with 0.7~eV of vibrational energy.  The expected ionization propensity
    rules are shown: S$_{2} \rightarrow D_{0} + e^{-}(\varepsilon_{1})$ and
    S$_{1} \rightarrow D_{1} + e^{-}(\varepsilon_{2})$. (b) Femtosecond
    time-resolved photoelectron kinetic energy spectra of DT pumped at
    $\lambda_{\mathrm{pump}}=287$~nm and probed at
    $\lambda_{\mathrm{probe}}=235$~nm.  There is a rapid ($\sim400$~fs) shift
    in the distribution: from an energetic peak $\varepsilon_{1}$ at 2.5~eV
    due to photoionization of S$_{2}$ into the D$_{0}$ cation ground
    electronic state; to a broad, structured band $\varepsilon_{2}$ at lower
    energies due to photoionization of vibrationally hot S$_{1}$ into the
    D$_{1}$ cation first excited electronic state. These results show a
    disentangling of electronic population dynamics from vibrational
    dynamics. The structure in the low energy band reflects the vibrational
    dynamics in S$_{1}$.}
  \label{fig:DT_TRPES_235}
\end{figure}

The above pump-probe experiments on DT were repeated using TRPES rather than
mass-resolved ion detection. More detailed discussions of these experimental
studies of Koopmans'-type correlations with TRPES can be found in the
literature~\cite{Blanchet1999, Blanchet2001, Schmitt2001}. The ultrafast
internal conversion of DT provides an example of Type (I) Koopmans'
correlations, and below we'll also discuss two experimental TRPES studies of
Type (II) correlations, namely the internal conversion in the polyaromatic
hydrocarbon phenanthrene, and in 1,4-diazabicyclo[2.2.2]octane (DABCO).  As
discussed in \secref{sec:photoionzn}, Type (I) ionization correlations are
defined as being the case when the neutral excited states $\alpha$ and $\beta$
in \figref{fig:Koopmans} correlate to different ion electronic continua, and
are referred to as \emph{complementary} ionization correlations. By contrast,
Type (II) correlations are defined as being the case when the neutral excited
states $\alpha$ and $\beta$ correlate to the same ion electronic continua, a
situation labeled \emph{corresponding} ionization correlations. As detailed
elsewhere~\cite{Blanchet1999, Blanchet2001}, the S$_{2}$ $1^{1}B_{u}$ state of
DT is a singly excited configuration and has Koopmans' correlations with the
D$_{0}$ $^{2}B_{g}$ electronic ground state of the cation. The dipole
forbidden S$_{1}$ $2^{1}A_{g}$ arises from configuration interaction between
singly and doubly excited $A_{g}$ configurations and has preferential
Koopmans' correlations with the D$_{1}$ $^{2}A_{u}$ first excited state of the
cation. These Koopmans' correlations are illustrated in
\figref{fig:DT_TRPES_235}(a). In \figref{fig:DT_TRPES_235}(b), we present
femtosecond TRPES results on DT for 287~nm pump excitation followed by 235~nm
probe laser ionization. The experimental photoelectron kinetic energy spectra
reveal a rapid shift of electrons from an energetic peak ($\varepsilon_{1}$ =
2.5~eV) to a broad, structured low energy component ($\varepsilon_{2}$). The
2.5~eV band is due to ionization of S$_{2}$ into the $D_{0}$ ion state.  The
broad, low energy band arises from photoionization of S$_{1}$ that correlates
with the $D_{1}$ ion state.  Its appearance is due to population of the
S$_{1}$ state by internal conversion.  Integration of each photoelectron band
directly reveals the S$_{2}$ to S$_{1}$ internal conversion time scale of
386$\pm$65 fs. It is important to note that these results contain more
information than the overall (integrated) internal conversion time. The
vibrational structure in each photoelectron band yields information about the
vibrational dynamics, which promote and tune the electronic population
transfer. In addition, it gives a direct view of the evolution of the ensuing
intramolecular vibrational energy redistribution (IVR) in the ``hot molecule''
which occurs on the ``dark'' S$_{1}$ potential surface~\cite{Blanchet2001}.

It is instructive to compare these TRPES results with the mass-resolved ion
yield experiment at the same pump and probe laser wavelengths, discussed above
(see \figref{fig:DT_ion_235}). The parent ion signal would be the same as
integrating the photoelectron spectra in \figref{fig:DT_TRPES_235}(b) over all
electron kinetic energies. In doing so, we would sum together a decaying
photoelectron band $\varepsilon_{1}$ with a growing photoelectron band
$\varepsilon_{2}$, leading to a signal which is more or less constant in time
and provides little information about the decay dynamics. This is the reason
why the parent ion signal in \figref{fig:DT_ion_235} does not show the 0.4~ps
decay that corresponds to the lifetime of the S$_{2}$ state. It provides a
clear example of how the parent ion signal as a function of time can be
misleading.

Why then does the time-resolved parent ion signal at 352~nm probe give the
correct 0.4~ps S$_{2}$ lifetime (\figref{fig:DT_ion_352}(a))?  It turns out that
352~nm is just below the energy threshold for reaching the D$_{1}$ state of
the cation. Therefore, upon internal conversion the formed S$_{1}$ state
cannot be easily ionized via a single photon since, as discussed above, it
does not have Koopmans' correlations with the D$_{0}$ ground state of the
cation. Therefore, single photon ionization probes only the decaying S$_{2}$
state which does have Koopmans'-allowed ionization into the D$_{0}$ ion ground
state. Hence, the correct 0.4~ps decay is observed in the parent ion
signal. The fragment ion signal has a 0.4~ps growth curve, indicating that it
arises from photoionization of the S$_{1}$ state formed by internal
conversion. Importantly, the observation of the fragment ion has nothing to do
with a possible neutral channel dissociation in the S$_{1}$ excited state --
there is none. Why then is a fragment ion observed? As both the D$_{0}$ and
D$_{1}$ states of the ion are stable with respect to dissociation, it must be
the case that a second probe photon is absorbed and higher lying
(predissociative) states of the cation are accessed. A probe laser power study
supported this point, yielding a quadratic dependence for the fragment ion and
a linear dependence for the parent ion signal. But why, under invariant laser
intensity, is a second probe photon absorbed only by the S$_{1}$ state and not
the S$_{2}$ state?  Again a consideration of the photoionization dynamics is
required.

\begin{figure}
  \includegraphics[height=4.0in]{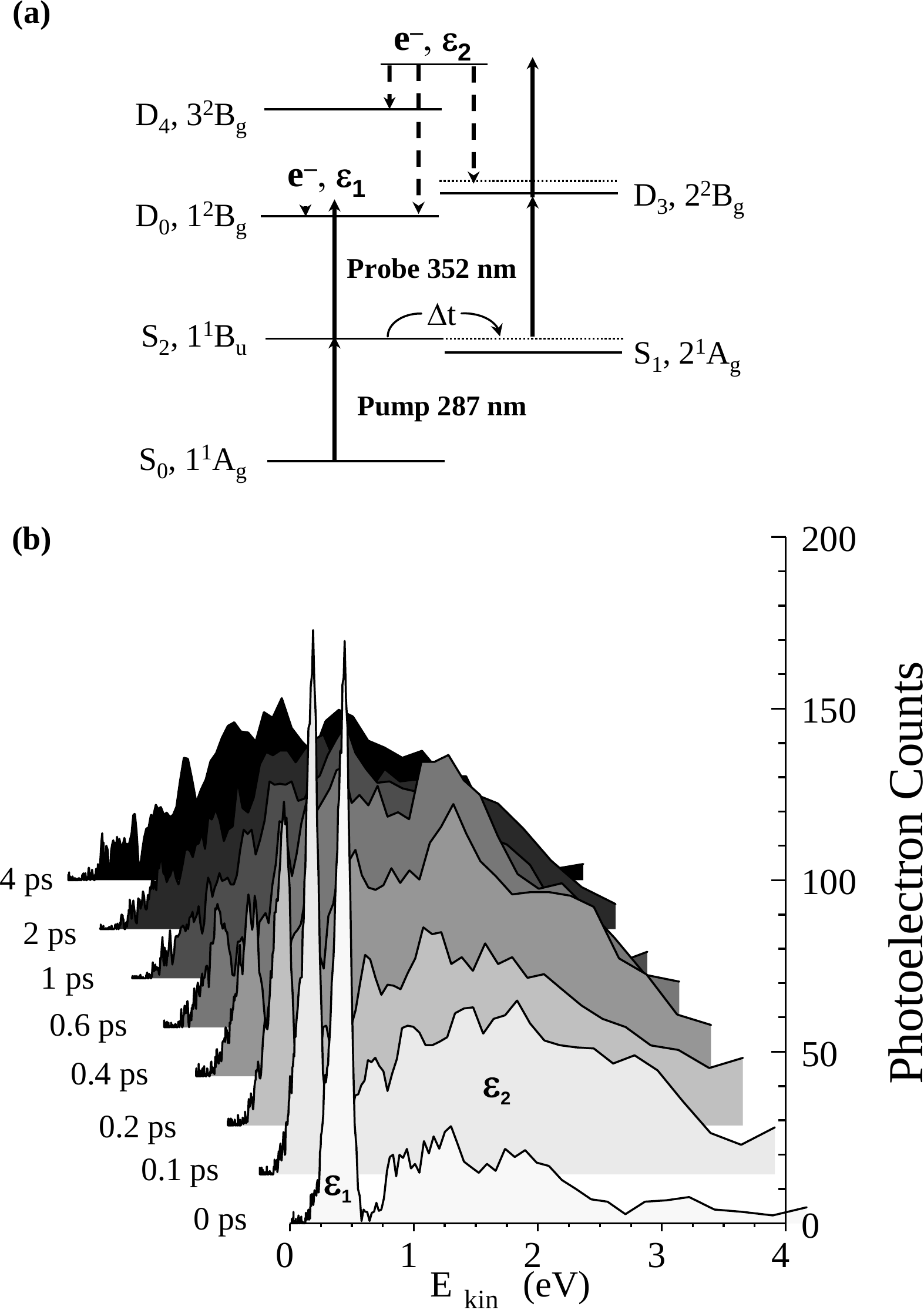}
  \caption{Time-resolved vibrational and electronic dynamics during internal
    conversion for all trans decatetraene (DT) pumped at
    $\lambda_{\mathrm{pump}}=287$~nm and probed at
    $\lambda_{\mathrm{probe}}=352$~nm.  (a) Level scheme in DT for one- and
    two-photon probe ionization. The pump laser is identical to that in
    \figref{fig:DT_TRPES_235} and prepares the identical S$_{2}$ state
    wavepacket.  The expected ionization propensity rules are: S$_{2}
    \rightarrow D_{0} + e^{-}(\varepsilon_{1})$ for 1-photon ($u
    \leftrightarrow g$) ionization and S$_{1} \rightarrow D_{0}, D_{3}, D_{4}
    + e^{-}(\varepsilon_{2})$ for 2-photon ($g \leftrightarrow g$) ionization.
    (b) Femtosecond time-resolved photoelectron kinetic energy spectra of DT
    pumped at 287~nm and probed at 352~nm, using both 1 and 2-photon probes.
    At 352~nm, the D$_{1}$ ion state is not energetically accessible from the
    S$_{1}$ state via a single photon transition.  Confirming the results of
    \figref{fig:DT_TRPES_235}, there is a rapid shift ($\sim400$~fs) in the
    distribution: from $\varepsilon_{1}$ a peak at 0.4~eV due to 1-photon
    ionization of S$_{2}$ into the D$_{0}$ cation ground electronic state; to
    $\varepsilon_{2}$ a broad, structured band at higher energies (1 - 3.5~eV)
    due to 2-photon ionization of the vibrationally hot S$_{1}$ into the
    D$_{0}$ cation ground and excited electronic states. The photoionization
    channel switches from a 1-photon to a 2-photon process during the internal
    conversion indicating again that the electronic structure of the
    ionization continuum is selective of the evolving electronic symmetry in
    the neutral state.}
  \label{fig:DT_TRPES_352}
\end{figure}

In \figref{fig:DT_TRPES_352}, we show the time-resolved photoelectron spectrum
for 352~nm probe laser ionization. Initially, the spectrum is characterized by
a low energy band, $\varepsilon_{1}$ at 0.56~eV which decays with time. As
indicated in \figref{fig:DT_TRPES_352}, the $\varepsilon_{1}$ band is due to
one-photon ionization of S$_{2}$ into D$_{0}$ and corresponds exactly with the
235~nm $\varepsilon_{1}$ band of \figref{fig:DT_TRPES_235}, simply shifted to
lower energy by the reduction in probe photon energy. This peak is also
narrower due to the improved kinetic resolution at low energy.  A broad
energetic band, $\varepsilon_{2}$, ranging from 0.6~eV to 4~eV grows with time
as the $\varepsilon_{1}$ band decays and therefore arises from photoionization
of the formed S$_{1}$ state.  The $\varepsilon_{2}$ band must, via energy
conservation, arise from two-photon probe ionization. As can be seen from
\figref{fig:DT_TRPES_352}(a), due to the symmetry of the two-photon dipole
operator, the ion continua accessed via two-photon ionization may also include
D$_{0}$, D$_{3}$ and D$_{4}$. This explains the broad range and high kinetic
energy of the photoelectrons in the $\varepsilon_{2}$ band. Integration of the
$\varepsilon_{1}$ and $\varepsilon_{2}$ bands provides yet another independent
confirmation of the internal conversion time scale, 377$\pm$47 fs, fully in
agreement with the 235~nm probe results. It is interesting to consider why, at
invariant probe laser intensity, the photoionization process switches from
single photon ionization of S$_{2}$ to two-photon ionization of S$_{1}$. We
note that in the both cases, the first probe photon is sufficient to ionize
the excited state, and therefore the S$_1$ ionization is due to absorption of
a second photon in the ionization continuum. This can be rationalized by a
consideration of the relative rates of two competing processes: second photon
absorption vs. autoionization. For the case of S$_{2}$, the photoionization
correlation is with D$_{0}$ and therefore the ionization is direct. In other
words, the ``autoionization'' is extremely rapid and second photon absorption
cannot compete. For the case of S$_{1}$, the photoionization correlation is
with D$_{1}$. The D$_{1}$ state, however, is energetically inaccessible and
therefore the transition is most likely into Rydberg series converging on the
D$_{1}$ threshold. For these to emit an electron into the open D$_{0}$
continuum channel, there must be an electronic rearrangement, for which there
is a finite autoionization rate. In this case, the absorption of a second
photon competes effectively with autoionization. These two-photon experiments
not only confirm the one photon results, but also demonstrate the symmetry
selectivity of the photoionization process itself.

\begin{figure}
  \includegraphics[height=5in]{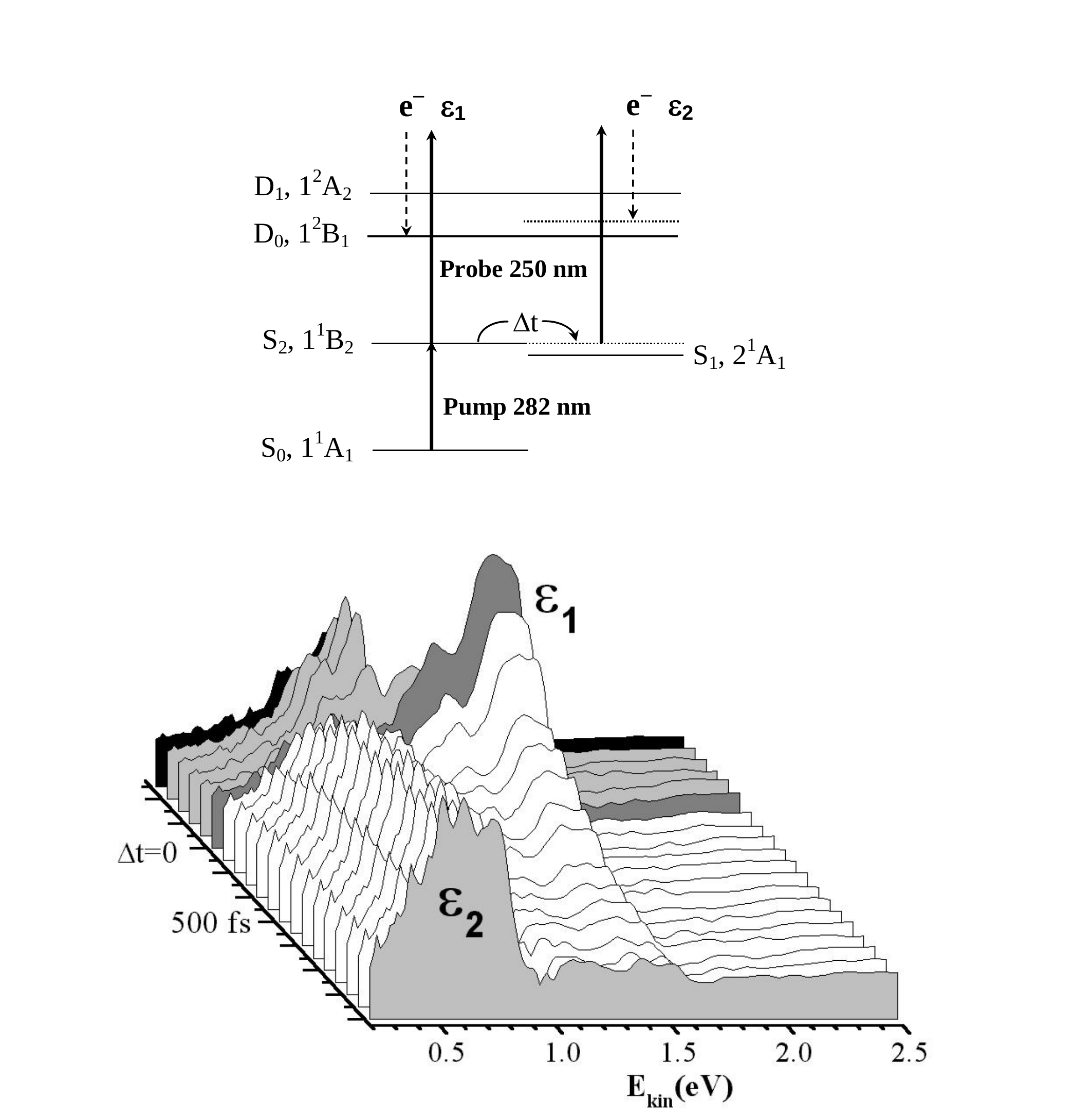}
  \caption{Energy level scheme for TRPES of phenanthrene (PH), an example of a
    Type (II) ionization correlation. (a) The pump laser prepares the
    optically bright state S$_{2}$. Due to ultrafast internal conversion, this
    state converts to the lower lying state S$_{1}$ with $\sim0.74$~eV of
    vibrational energy. The expected corresponding Type (II) Koopmans'
    correlations are shown: S$_{2} \rightarrow D_{0} + e^{-}(\varepsilon_{1})$
    and S$_{1} \rightarrow D_{0} + e^{-}(\varepsilon_{2})$.  (b) TRPES spectra
    of phenanthrene for a pump wavelength of $\lambda_{\mathrm{pump}}=282$~nm
    and a probe wavelength of $\lambda_{\mathrm{probe}}=250$~nm. The
    disappearance of band $\varepsilon_{1}$ at $\sim1.5$~eV and growth of band
    at $\varepsilon_{2}$ at $\sim0.5$~eV represents a direct measure of the
    S$_{2}-S_{1}$ internal conversion time (520 fs). Despite the unfavourable
    Type(II) ionization correlations, the rigidity of this molecule allows for
    direct observation of the internal conversion via vibrational propensities
    alone.}
  \label{fig:PH_TRPES}
\end{figure}

The other limiting Koopmans' case, Type (II), is where the one-electron
correlations upon ionization correspond to the same cationic states. An
example of Type (II) correlations is seen in the S$_{2}$--S$_{1}$ internal
conversion in the polyaromatic hydrocarbon phenanthrene (PH), discussed in
more detail elsewhere~\cite{Schmitt2001}. In the case of PH, both the S$_{2}$
and the S$_{1}$ states correlate similarly with the electronic ground state as
well as the first excited state of the cation. In this experiment, PH was
excited from the S$_{0}$ $^{1}A_{1}$ ground state to the origin of the S$_{2}$
$^{1}B_{2}$ state with a 282~nm (4.37~eV) femtosecond pump pulse and then
ionized after a time delay $\Delta t$ using a 250~nm (4.96~eV) probe photon.
The S$_{2}$ $^{1}B_{2}$ state rapidly internally converted to the lower lying
S$_{1}$ $^{1}A_{1}$ state at 3.63~eV, transforming electronic into vibrational
energy. In PH, both the S$_{2}$ $^{1}B_{2}$ and S$_{1}$ $^{1}A_{1}$ states can
correlate with the D$_{0}$ $^{2}B_{1}$ ion ground state.  The time-resolved
photoelectron spectra for PH, shown in \figref{fig:PH_TRPES}, revealed a
rapidly decaying but energetically narrow peak at $\varepsilon_{1}\sim1.5$~eV
due to photoionization of the vibrationless S$_{2}$ $^{1}B_{2}$ state into the
ionic ground state D$_{0}$ $^{2}B_{1}$, resulting in a decay time constant of
520$\pm$8 fs.  A broad photoelectron band, centered at about $\sim0.7$~eV, in
these photoelectron spectra was due to ionization of vibrationally hot
molecules in the S$_1$ state, formed by the S$_{2}$--S$_{1}$ internal
conversion. At times t $>1500$~fs or so (i.e. after internal conversion), the
photoelectron spectrum is comprised exclusively of signals due to S$_{1}$
ionization.  The S$_{1}$ state itself is long lived on the time scale of the
experiment.  Despite the fact that Type (II) molecules present an unfavorable
case for disentangling electronic from vibrational dynamics, in PH a dramatic
shift in the photoelectron spectrum was seen as a function of time. This is
due to the fact that PH is a rigid molecule and the S$_{2}$, S$_{1}$ and
D$_{0}$ states all have similar geometries.  The photoionization probabilities
are therefore dominated by small $\Delta v$ transitions. Hence, the 0.74~eV
vibrational energy in the populated S$_{1}$ state should be roughly conserved
upon ionization into the D$_{0}$ ionic state. Small geometry changes favor
conservation of vibrational energy upon ionization and thereby permit the
observation of the excited state electronic population dynamics via a
photoelectron kinetic energy analysis alone.  In general, however, significant
geometry changes will lead to overlapping photoelectron bands, hindering the
disentangling of vibrational from electronic dynamics.

\begin{figure}
  \includegraphics[width=4in]{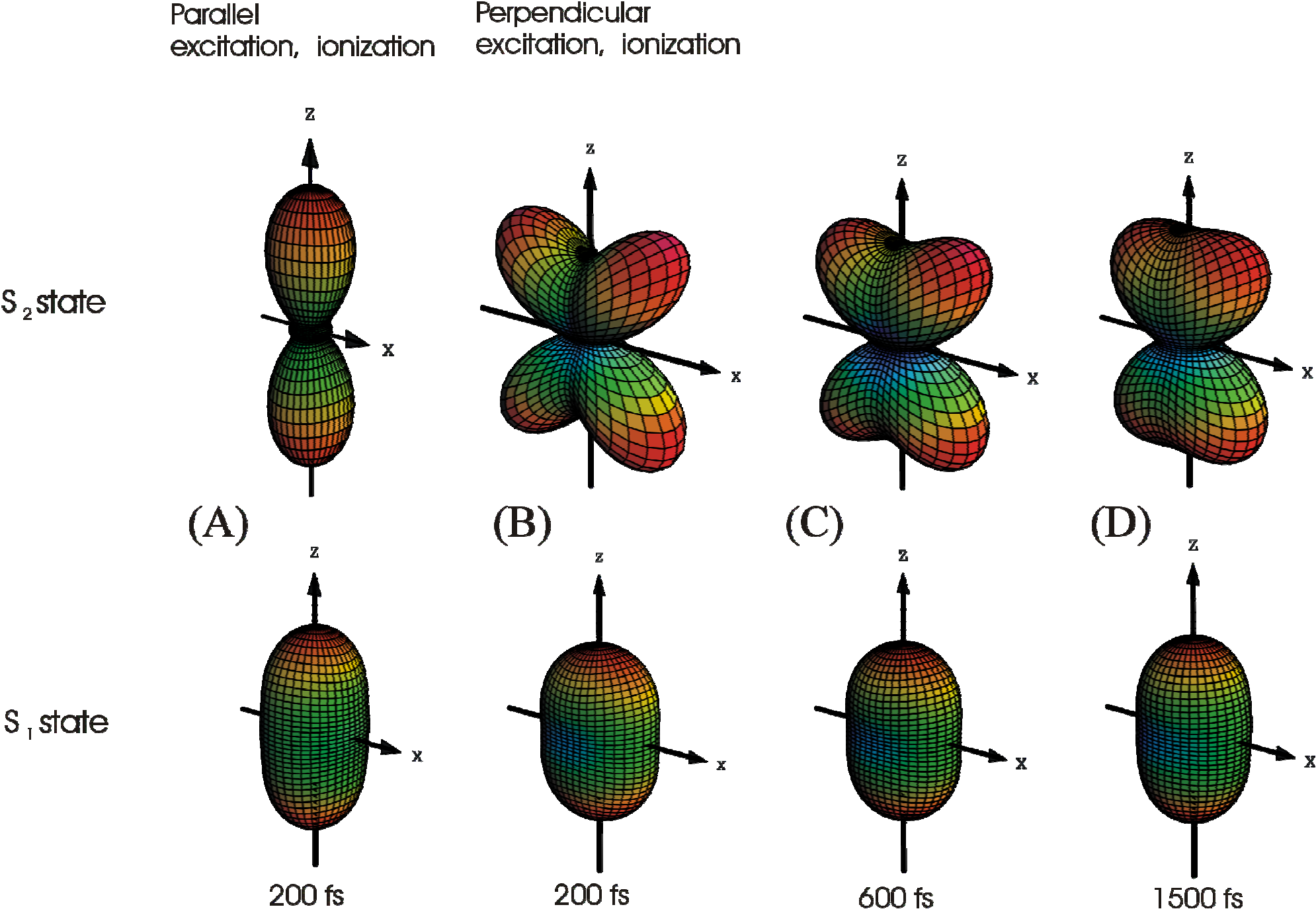}
  \caption{Time-resolved photoelectron angular distributions (PADs) from
    ionization of DABCO for linearly polarized pump and probe pulses.. Here
    the optically bright S$_2$ $^1E’$ state internally converts to the dark
    S$_1$ $^1A'_1$ state on picosecond timescales. (A) PADs at 200 fs time
    delay for pump and probe polarization vector both parallel to the
    spectrometer axis. The difference in electronic symmetry between S$_2$ and
    S$_1$ leads to significant changes in the form of the PAD. (B) PADs at 200
    fs time delay for pump polarization parallel and probe polarization
    perpendicular to the spectrometer axis, showing the effects of lab frame
    molecular alignment. (C) and (D) The PADs evolve as a function of time due
    to molecular axis rotational wavepacket dynamics.  Printed with permission
    from C .C. Hayden, unpublished.}
  \label{fig:dabco}
\end{figure}

As mentioned in \secref{sec:photoionzn}, where a molecule has Type (II)
ionization correlations, it might be expected that the coupled electronic
states of the neutral molecule would not be resolved in the photoelectron
spectrum -- the rigidity of PH and the large energy gap between the S$_2$ and
S$_1$ origins allowed for the resolution of the excited state dynamics in the
PES, but this is by no means a general situation. In such circumstances, the
measurement of time-resolved PADs (TRPADS) offers an complementary approach to
unravelling the dynamics. As discussed in \secref{sec:photoionzn}, the
requirement that the direct product of the irreducible representations of the
neutral electronic state, the transition dipole component, the ion electronic
state and the free electron contains the totally symmetric representation (see
\eqref{eq:direct_product}) means that, if the coupled electronic states are of
different symmetry, the PAD will differ for the two electronic states. The
evolution of the PAD can therefore be expected to provide a mechanism for
unravelling the electronic dynamics. A first demonstration of a TRPADs
measurement of non-adiabatically coupled electronic states was provided by
Hayden and co-workers who studied the molecule 1,4-diazabicyclo[2.2.2]octane
(DABCO). In this experiment, 251~nm pump pulse excited the origin of the
optically bright S$_2$ $^1E’$ electronic state, a 3p Rydberg state centred on
the nitrogen atoms~\cite{Parker1979}, via a single photon excitation. The S$_2$
state is coupled to the lower lying optically dark S$_1$ $^1A'_1$ state, a 3s
Rydberg state centered on the nitrogen atoms, and internal conversion takes
place on a $\sim1$~ps timescale. In the experiment, it was possible to resolve
the S$_1$ and S$_2$ states directly in the PES. In \figref{fig:dabco} we show
PADs measured for each electronic state at different time delays, measured
with a photoelectron imaging apparatus.

Although the S$_2$ and S$_1$ states of DABCO are easily resolved in the PES,
due again to the rigidity of the molecule, these measurements clearly
demonstrate the sensitivity of the PAD to the electronic symmetry of the
excited state. Furthermore, the data displayed in \figref{fig:dabco}
demonstrates the sensitivity of the PAD to the molecular axis alignment. The
single-photon pump step in this experiment produces a coherent superposition
of rotational states of the molecules and an anisotropic distribution of
molecular axes exhibiting alignment. Comparing panels (A) and (B) in
\figref{fig:dabco} we see that the PAD changes dramatically when the probe
polarization is changed from being parallel to being perpendicular to the pump
polarization. When the pump and probe polarizations are parallel, the LF PAD
possesses cylindrical symmetry. When the pump and probe polarizations are
perpendicular, cylindrical symmetry is lost, although the LF PAD exhibits
reflection symmetry in the plane containing the laser polarizations. As the
pump-probe time delay increases, the rotational wavepacket in the excited
state dephases and the molecular axis alignment decreases accordingly. As the
anisotropy of the distribution of molecular axes decreases, the PAD is also
seen to become less anisotropic, and also less sensitive to the probe laser
polarization direction, as discussed in \secref{sec:photoionzn}.

In closing this section we note that, although the Koopmans' picture is a
simplification of the ionization dynamics, it provides a very useful zeroth
order picture from which to consider the TRPES results.  Any potential failure
of this independent electron picture can always be experimentally tested
directly through variation of the photoionization laser frequency: resonance
structures should lead to variations in the form of the spectra with electron
kinetic energy, although the effect of resonances is more likely to be
prominent in PAD measurements, and indeed an observation of a shape resonance
in para-difluorobenzene has been reported~\cite{Bellm2003, Bellm2005}.

%%% Local Variables: 
%%% mode: latex
%%% TeX-master: "trpes"
%%% End: 

\chapter{Applications}
\label{sec:applications}
As discussed in the Introduction, a natural application of TRPES is to
problems of excited state non-adiabatic dynamics.  Non-adiabatic dynamics
involve a breakdown of the adiabatic (Born-Oppenheimer) approximation, which
assumes that electrons instantaneously follow the nuclear dynamics.  This
approximation is exact provided that the nuclear kinetic energy is negligible
and its breakdown is therefore uniquely due to the nuclear kinetic energy
operator. Spin-orbit coupling, leading to intersystem crossing, is not a
non-adiabatic process in this sense: the Born-Oppenheimer states could be
chosen to be the fully relativistic eigenstates and hence would be perfectly
valid adiabatic states.  Nevertheless, the description of intersystem crossing
as a non-adiabatic process is seen in the literature and we therefore include
spin-orbit coupling problems in this section. Furthermore, in this section we
have chosen to include examples from work which highlight some of the most
recent advances in the use of TRPES for studying molecular dynamics. These
include the application of TRCIS to the study of photodissociation dynamics,
as well as the use of PADs to measure molecular axis distributions, as
suggested by \eqref{eq:betalm2}. Examples are also given of the utility of
TRPES for the study of excited state vibrational dynamics.

\section{Internal Conversion: Electronic relaxation in substituted
  Benzenes}
Internal conversion, also referred to as spin-conserving electronic relaxation
or a radiationless transition, is one of the most important non-adiabatic
processes in polyatomic molecules, and is often the trigger for any ensuing
photochemistry~\cite{Bixon1968, Jortner1969, Henry1973, Freed1976, Stock1997,
  Worth2004, Klessinger1994, Koppel1984}. As will be discussed in more detail
in \secref{sec:app_switches}, in order to establish rules relating molecular
structure to function --- a concept central to the development of molecular
scale electronics --- it is first necessary to develop an understanding of the
relationship between molecular structure and excited state dynamics.  The
underlying ``rules'' governing these processes have yet to be fully
established. A phenomenological approach to such rules involves the study of
substituent effects in electronic relaxation (internal conversion)
dynamics. For this reason a series of monosubstituted benzenes was studied as
model compounds~\cite{Lee2002}. The focus of this work was on the first and
second $\pi\pi^\ast$ states of these aromatic systems and on substituents
which were expected to affect the electronic structure and relaxation rates of
the $\pi\pi^\ast$ states. The photophysics of benzenes is well understood, and
therefore the major purpose of this study was to establish the quantitative
accuracy of the internal conversion rates determined via TRPES.

\begin{figure}
  \includegraphics[width=8cm]{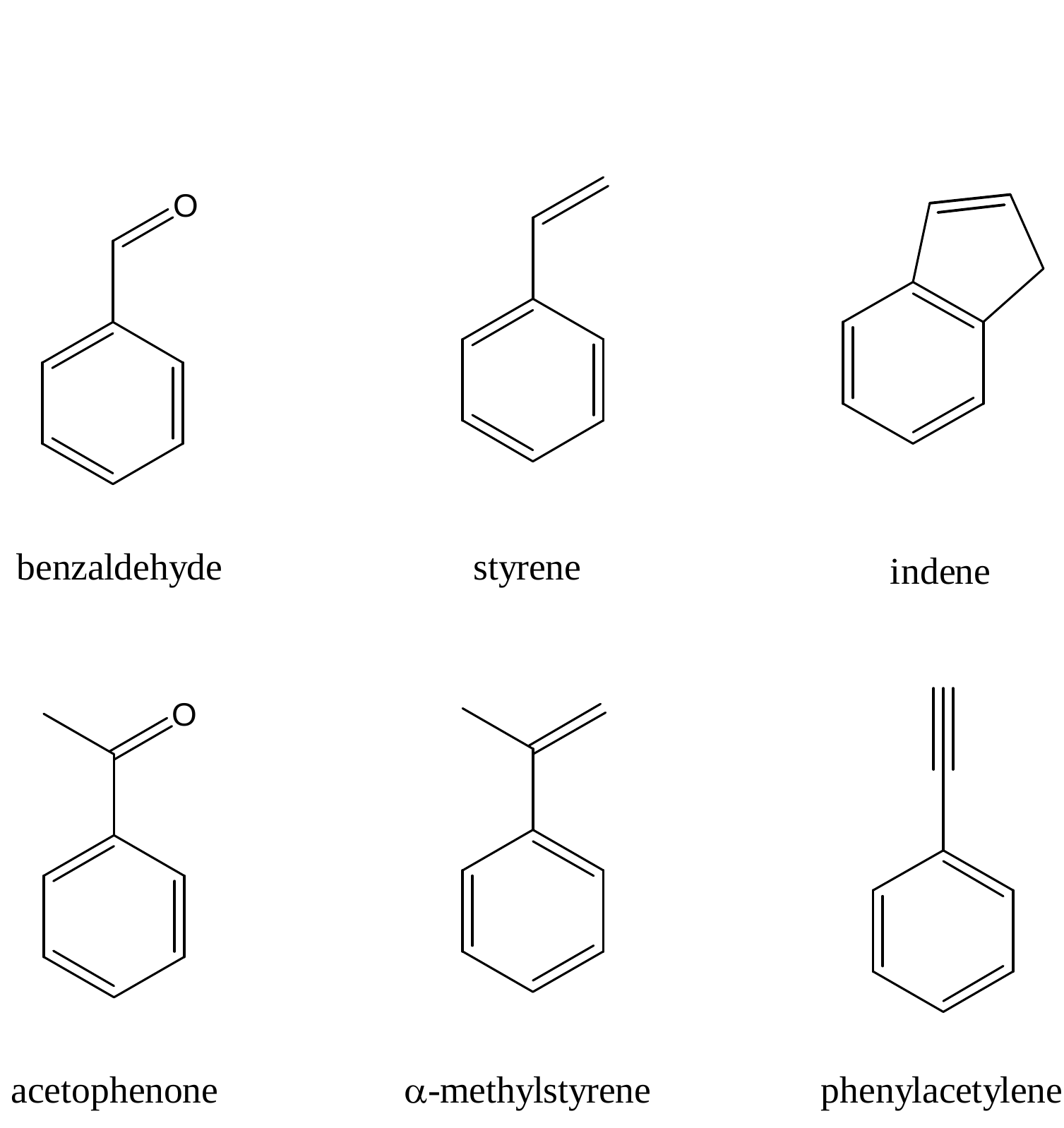}
  \caption{Molecular structures of some monosubstituted benzenes studied via
    TRPES in order to determine the quantitative accuracy of the extracted
    internal conversion rates.  Three different electronic substituents were
    used, C=O, C=C and C$\equiv$C, leading to different state
    interactions. The effects of vibrational dynamics were investigated via
    the use of methyl group (floppier), as in $\alpha$-MeSTY and ACP, or a
    ring structure (more rigid), as in IND, side group additions. BZA and ACP
    have favourable Type (1) ionization correlations whereas STY, IND,
    $\alpha$-MeSTY and ACT have unfavourable Type (2) ionization
    correlations.}
  \label{fig:subs_benzenes_ALL}
\end{figure}

As shown in \figref{fig:subs_benzenes_ALL}, six benzene derivatives were
studied: benzaldehyde (BZA), styrene (STY), indene (IND), acetophenone (ACP),
$\alpha$-methylstyrene ($\alpha$-MeSTY) and phenylacetylene ($\phi$ACT).  This
choice of substituents addressed several points: (i) the effect of the
substituent on the electronic states and couplings; (ii) the effect of the
substituent on the rigidity or floppiness of the MF; (iii) a comparison of
Type (I) with Type (II) Koopmans' systems; (iv) to investigate the potential
effects of autoionization resonances (i.e. non-Koopmans' behavior) on the
observed dynamics. Three electronically distinct substituents were chosen:
C=O, C=C and C$\equiv$C. For the C=C, potential off-axis conjugation effects
with the ring in STY was contrasted with the lack of these in the C$\equiv$C
of $\phi$ACT.  For the heteroatomic substituent C=O, the influence of the
additional n$\pi^\ast$ state on the $\pi\pi^\ast$ dynamics was investigated by
comparing BZA with STY and ACP with $\alpha$-MeSTY.  The effects of
vibrational dynamics and densities-of-states on the electronic relaxation
rates were studied via both methyl (``floppier'') and alkyl ring (``more
rigid'') substitution: STY was compared with both the floppier $\alpha$-MeSTY
and the much more rigid IND; BZA was compared with the floppier ACP. Both BZA
and ACP have Type (I) Koopmans' ionization correlations, the rest have Type
(II) correlations, allowing for another comparison of these two cases. In
order to investigate the potential effects of autoionization resonances, the
ionization probe photon energy was varied ($\sim0.4$~eV).  In these systems,
the form of the photoelectron spectra and the fits to the lifetime data were
invariant with respect to probe laser frequency~\cite{Lee2002}.

\begin{figure}
  \includegraphics[height=5in]{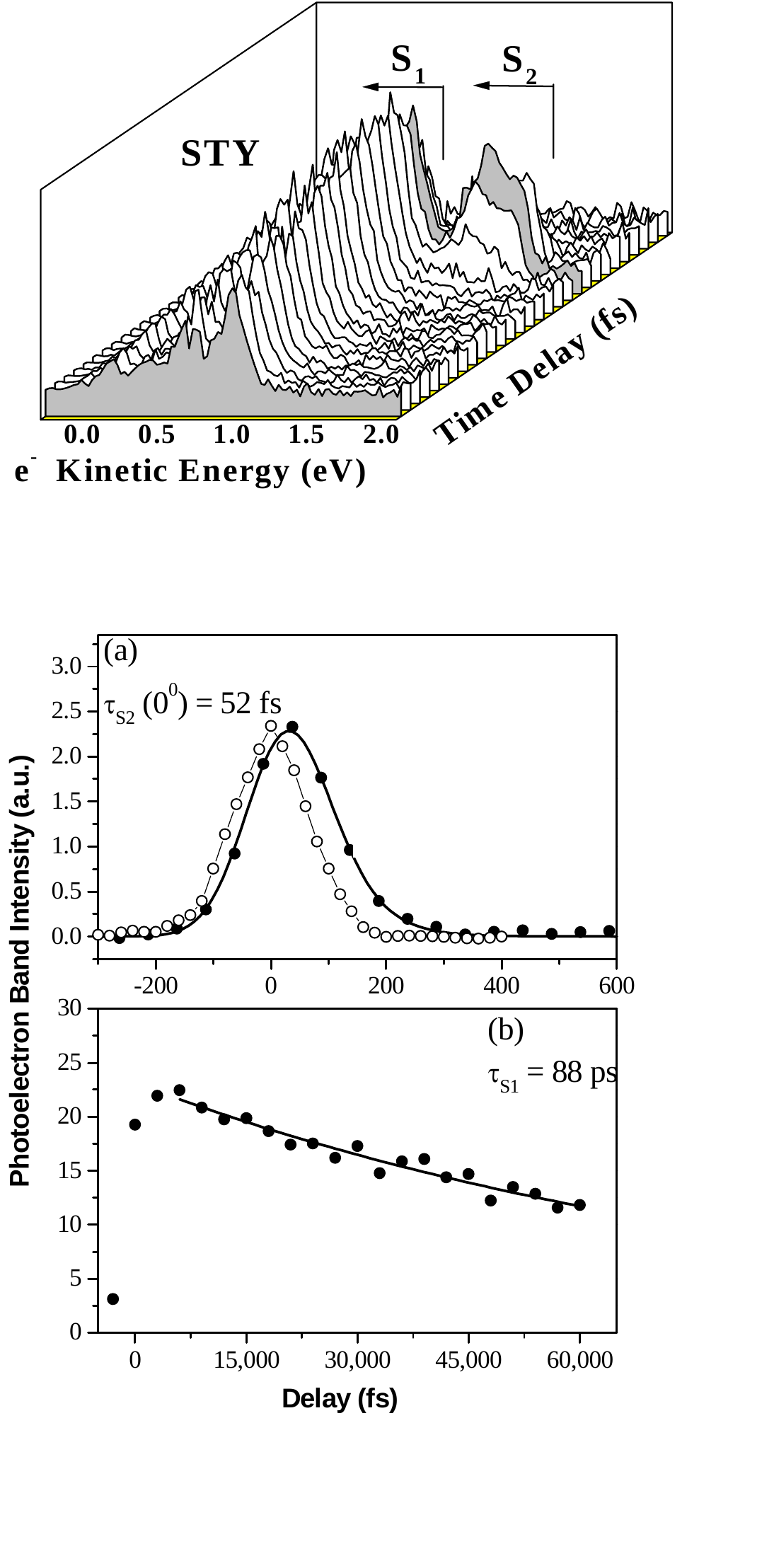}
  \caption{(top) TRPES spectra of substituted benzenes, shown here for styrene
    (STY) with $\lambda_{\mathrm{pump}}$ = 254.3~nm and photoionization
    $\lambda_{\mathrm{probe}}$ = 218.5~nm. The energetics and Koopmans'
    correlations allow for assignment of the photoelectron bands to ionization
    of S$_{2}(\pi\pi^\ast)$ and S$_{1}(\pi\pi^\ast)$, as indicated. The
    S$_{2}$ state decays on ultrafast time scales. The S$_{1}$ state decays on
    a much longer (ps) time scale. (Bottom) (a) Time-dependent
    S$_{2}(\pi\pi^\ast)$ 0$^{0}$ photoelectron band integral yields for STY,
    yielding a decay time constant of 52~fs.  Open circles represent the laser
    cross-correlation at these wavelengths. (b) Time-dependent
    S$_{1}(\pi\pi^\ast)$ photoelectron band integral yields for STY obtained
    from a fit to the long time delay part of the data (not shown).}
  \label{fig:subs_benzenes_STY}
\end{figure}

A sample TRPES spectrum, STY at $\lambda_{\mathrm{pump}}=254.3$~nm and
$\lambda_{\mathrm{probe}}=218.5$~nm, is shown in
\figref{fig:subs_benzenes_STY}.  The S$_{1}(\pi\pi^\ast)$ component grows in
rapidly, corresponding to the ultrafast internal conversion of the
S$_{2}(\pi\pi^\ast)$ state. The S$_{1}(\pi\pi^\ast)$ component subsequently
decays on a much longer picosecond time scale (not shown).  It can be seen
that despite STY being an unfavorable Type (II) case, the two bands are well
enough resolved to allow for unambiguous separation of the two channels and
determination of the sequential electronic relaxation time scales. Energy
integration over each band allows for extraction of the electronic relaxation
dynamics.  The time-dependent S$_{2}(\pi\pi^\ast)$ photoelectron band integral
yields for STY are also shown in \figref{fig:subs_benzenes_STY}. The open
circles represent the pump-probe cross-correlation (i.e. the experimental time
resolution) at these wavelengths. Integration over the S$_{2}(\pi\pi^\ast)$
photoelectron band is shown as the solid circles.  The solid line is the best
fit to the S$_{2}(\pi\pi^\ast)$ channel, yielding a lifetime of
52$\pm$5~fs. In \figref{fig:subs_benzenes_STY}, the time-dependent
S$_{1}(\pi\pi^\ast)$ photoelectron band integral yields for STY are shown,
obtained from a fit to the long delay part of the data, yielding a lifetime of
88$\pm$8~ps for the state S$_{1}(\pi\pi^\ast)$.

\begin{figure}
  \includegraphics[width=8cm]{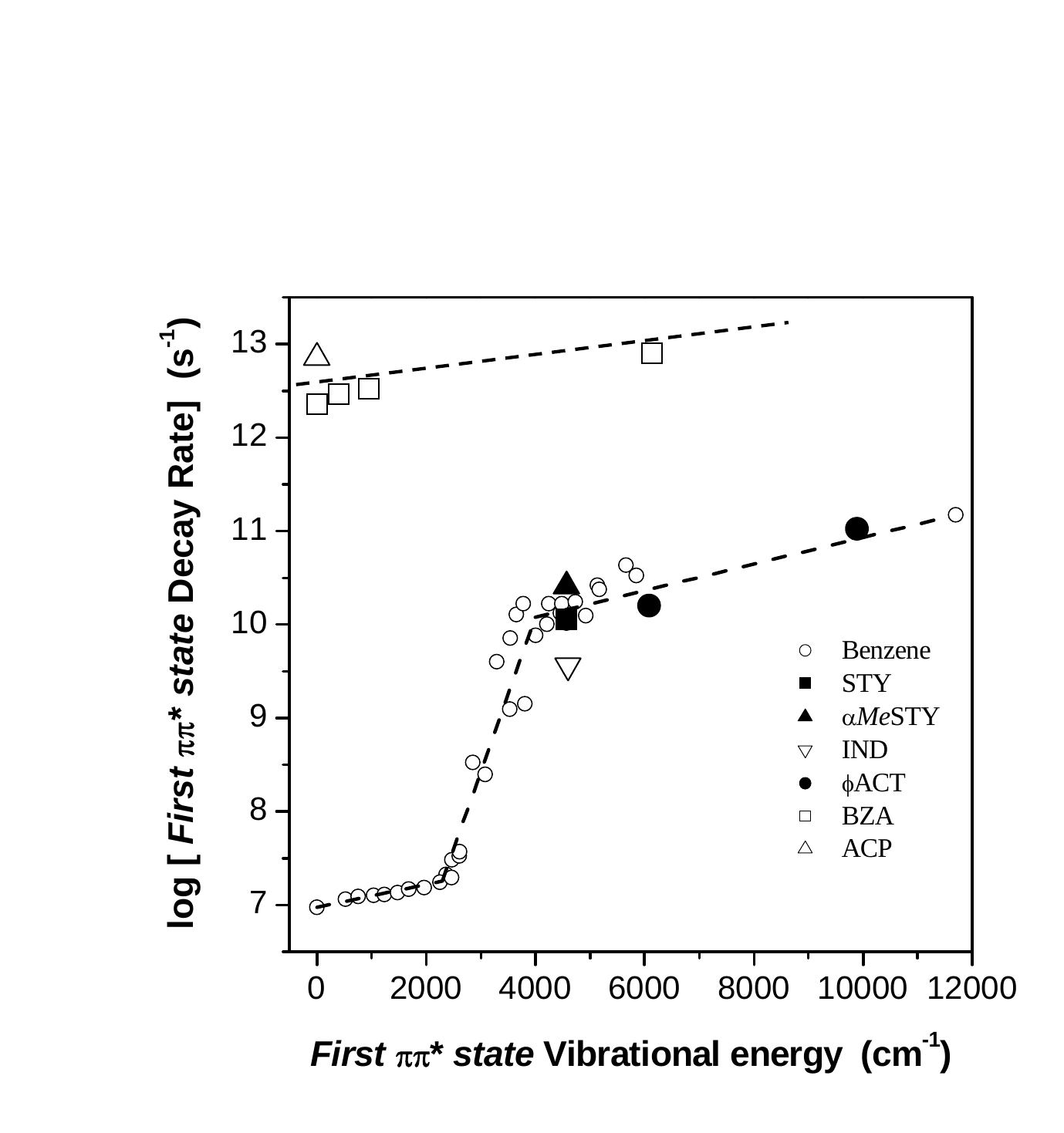}
  \caption{Excess vibrational energy dependence of the internal conversion
    rates of the first $\pi\pi^\ast$ state of benzene and its derivatives.}
  \label{fig:subs_benzenes_ALLrates}
\end{figure}

The excess vibrational energy dependence of the internal conversion rates of
the first $\pi\pi^\ast$ state of benzene and its derivatives is shown in
\figref{fig:subs_benzenes_ALLrates}. These data suggest that the first
$\pi\pi^\ast$ states of STY, $\alpha$-MeSTY, IND and ACT internally convert
essentially via benzene ring dynamics. By contrast, the first $\pi\pi^\ast$
states of BZA and ACP internally convert orders of magnitude faster,
indicating a completely different mechanism due to the presence of low lying
n$\pi^\ast$ states in BZA and ACP, absent in the other systems, which lead to
ultrafast intersystem crossing and the formation of triplet states. Overall,
these results demonstrate that the TRPES method is well-suited to the
quantitative study of electronic relaxation processes, producing direct and
accurate measurements of electronic relaxation rates which are in quantitative
agreement with the currently accepted understanding of aromatic
photophysics~\cite{Lee2002}.

\section{Excited state nuclear dynamics}
\label{sec:app_nuc_dyn}
As discussed in \secref{sec:photoionzn}, TRPES is sensitive to vibrational and
rotational dynamics, as well as electronic dynamics. In this section, we will
give examples of the use of TRPES to the study of intramolecular vibrational
energy redistribution (IVR), and the use of time-resolved PAD measurements as
a probe of rotational dynamics. 

A problem central to chemical reaction dynamics is that of
IVR~\cite{Felker1995, Keske2000}, the flow of energy between zeroth-order
vibrational modes. Indeed, IVR generally accompanies (and mediates)
non-adiabatic dynamics such as internal conversion and isomerization. The
description of separated rigid rotational and normal-mode vibrational motions
employed in \secref{sec:nonBO} and \secref{sec:photoionzn} provides an
adequate description only in regions of low state density. As the state
density increases, the vibrational dynamics become ``dissipative'' as
normal-mode vibrations become mixed and energy flows between these
zeroth-order states. Much work has been undertaken studying IVR using
e.g. fluroscence techniques~\cite{Moss1993, Felker1995, Keske2000}. However,
such techniques generally monitor flow of energy out of the initially excited
vibrational states, but do not directly observe the optically dark ``bath''
vibrational modes into which vibrational energy flows. TRPES provides a window
to these dark states, allowing for direct monitoring of IVR in molecules due
to the Franck-Condon correlations described in \secref{sec:photoionzn}.

\begin{figure}
  \includegraphics[width=8cm]{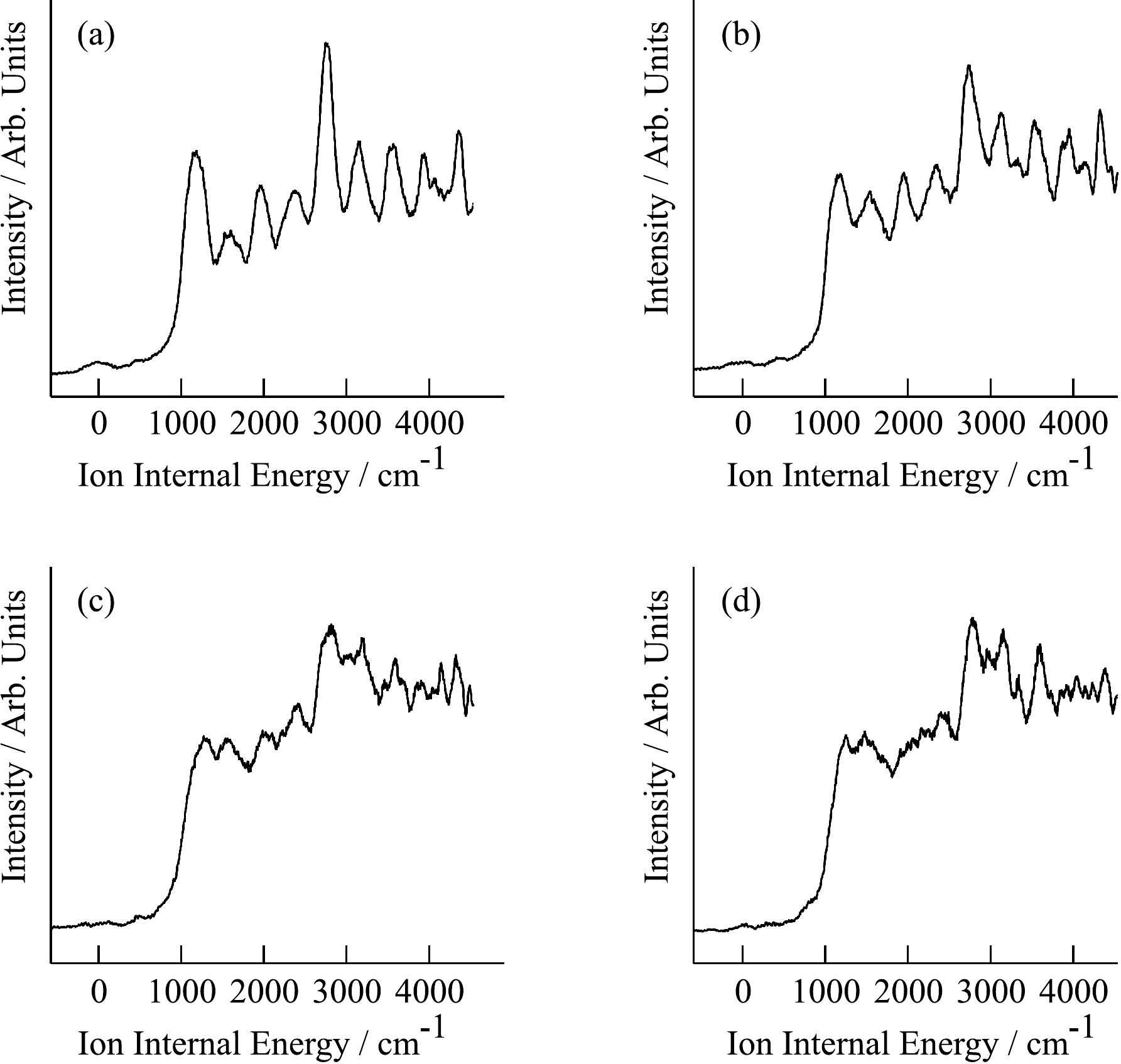}
  \caption{TRPES of the S$_1$ state of para-fluortoluene prepared in the $7^1$
    state, i.e. with one quanta of excitation in the C--F stretching mode. At
    early times, the PES contains a well resolved Franck-Condon progression
    corresponding to ionization of the localized C--F stretching mode. At
    later times, the onset of IVR obscures the PES as many more vibrational
    modes become populated. Reproduced with permission from
    ref.~\cite{King2005}.}
  \label{fig:ivr1}
\end{figure}

\begin{figure}
  \includegraphics[width=7cm]{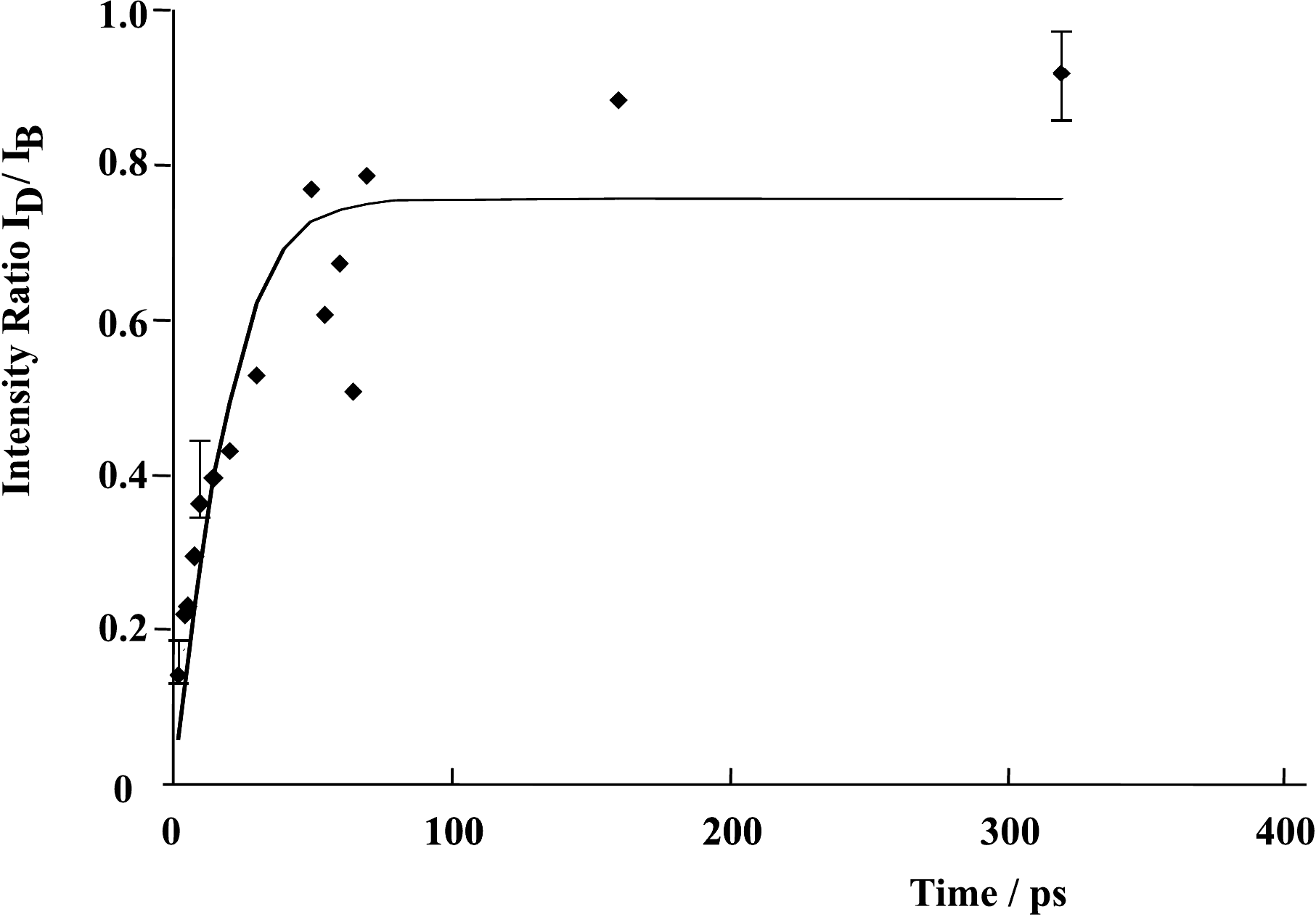}
  \caption{Ratio of the population of the initially prepared vibrational state
  to the population of the ``dark'' vibrational states populated through IVR
  for the data shown in \figref{fig:ivr1}. Reproduced with permission from
    ref.~\cite{King2005}.}
  \label{fig:ivr2}
\end{figure}

Reid and co-workers reported a picosecond TRPES study of IVR in the
electronically excited S$_1$ state of para-fluorotoluene~\cite{King2005}. By
selective excitation of specific vibrational modes in S$_1$ and measuring the
evolution of the PES as a function of time delay, information regarding
vibrational population dynamics was obtained. Analysis of this TRPES data also
employed high resolution PES data and required a detailed understanding of the
Franck-Condon factors for ionization. Reid and co-workers were able to measure
the rates of IVR for the the initially prepared $7^1$, $8^1$ and $11^1$
vibrational states (using Mulliken notation) of the first electronically
excited state S$_1$ of para-fluorotoluene~\cite{King2005}. By selective
population of vibrational states in S$_1$ it is possible to use TRPES to test
for vibrational mode-specificity in IVR, as well as varying the excess
vibrational energy. In \figref{fig:ivr1} example TRPES data is shown for
excitation of the $7^1$ mode of para-fluorotoluene, corresponding to one
quanta of excitation in the C--F stretching mode, and representing
1230~cm$^{-1}$ vibrational energy in S$_1$. The evolution of the PES shown in
\figref{fig:ivr1}, and the disappearance of resolved structure at long time
delays is a direct measure of the IVR of energy out of the C--F stretching
mode and into other modes of the molecule. By analysing spectra such as those
in \figref{fig:ivr1} in terms of the population in the initially prepared
state compared to the populations in all other ``dark'' modes, the rate of IVR
may be extracted. An example of such an analysis is shown in
\figref{fig:ivr2}.

\begin{figure}
  \includegraphics[width=6cm]{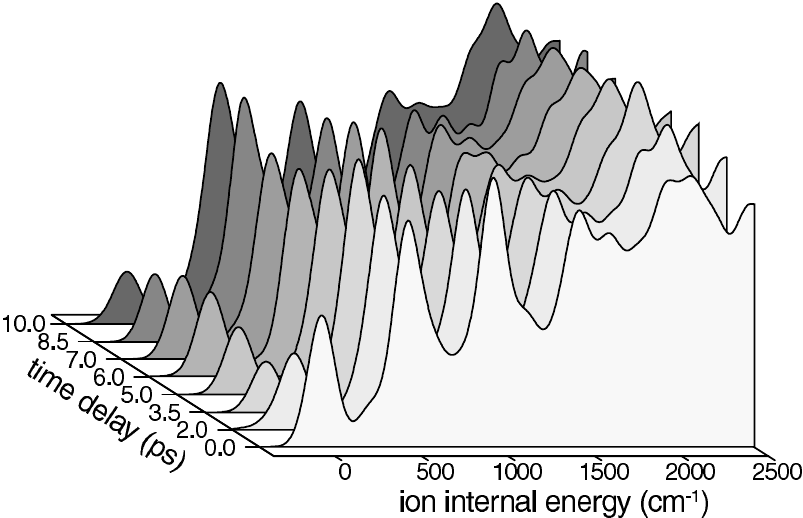}

  \includegraphics[width=6cm]{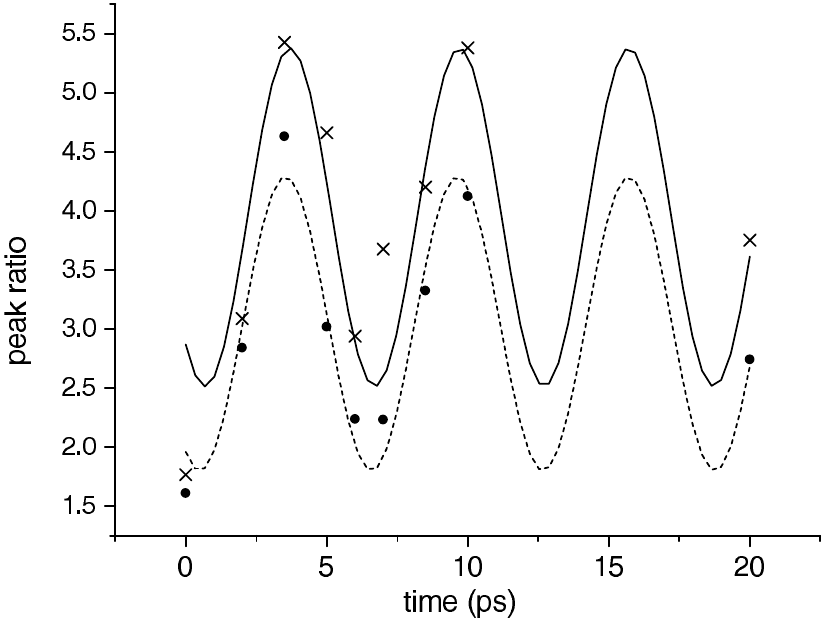}
  \caption{(Top) Photoelectron spectra following preparation of the
    $6a^1+10b^1 6b^1$ Fermi resonance at 457~cm$^{-1}$ in the S$_1$ state of
    toluene with a 1~ps pump pulse. PES are shown as a function of time delay
    between the pump pulse and a 1~ps probe pulse. (Bottom) Fits to the
    oscillation of the PES according to two different models of the data shown
    in the top panel showing clearly a period of oscillation of $\sim6$~ps,
    corresponding to the energy separation of the two vibrational states
    comprising the wavepacket. For more detail see
    ref.~\cite{Hammond2006}. Reproduced with permission from
    ref.~\cite{Hammond2006}.}
  \label{fig:fermi}
\end{figure}

In a second example of the utility of TRPES for the study of vibrational
dynamics, Reid and co-workers studied the dynamics associated with a Fermi
resonance between two near-degenerate vibrational modes in the S$_1$ state of
toluene~\cite{Hammond2006}. In this study, the pump pulse prepared a coherent
superposition of the $6a^1$ state, corresponding to one quanta of vibrational
excitation in the totally symmetric ring breathing mode, and the $10b^16b^1$
state, corresponding to one quanta in the CH$_3$ wagging mode ($10b$) and one
quanta in the C--H out-of-plane bending mode ($16b$). The anharmonic coupling
between these two states gives rise to an oscillation in the form of the PES,
as shown in \figref{fig:fermi}, the timescale of which corresponds to the
energy separation of the two vibrational modes.

\begin{figure}
  \includegraphics[height=3in]{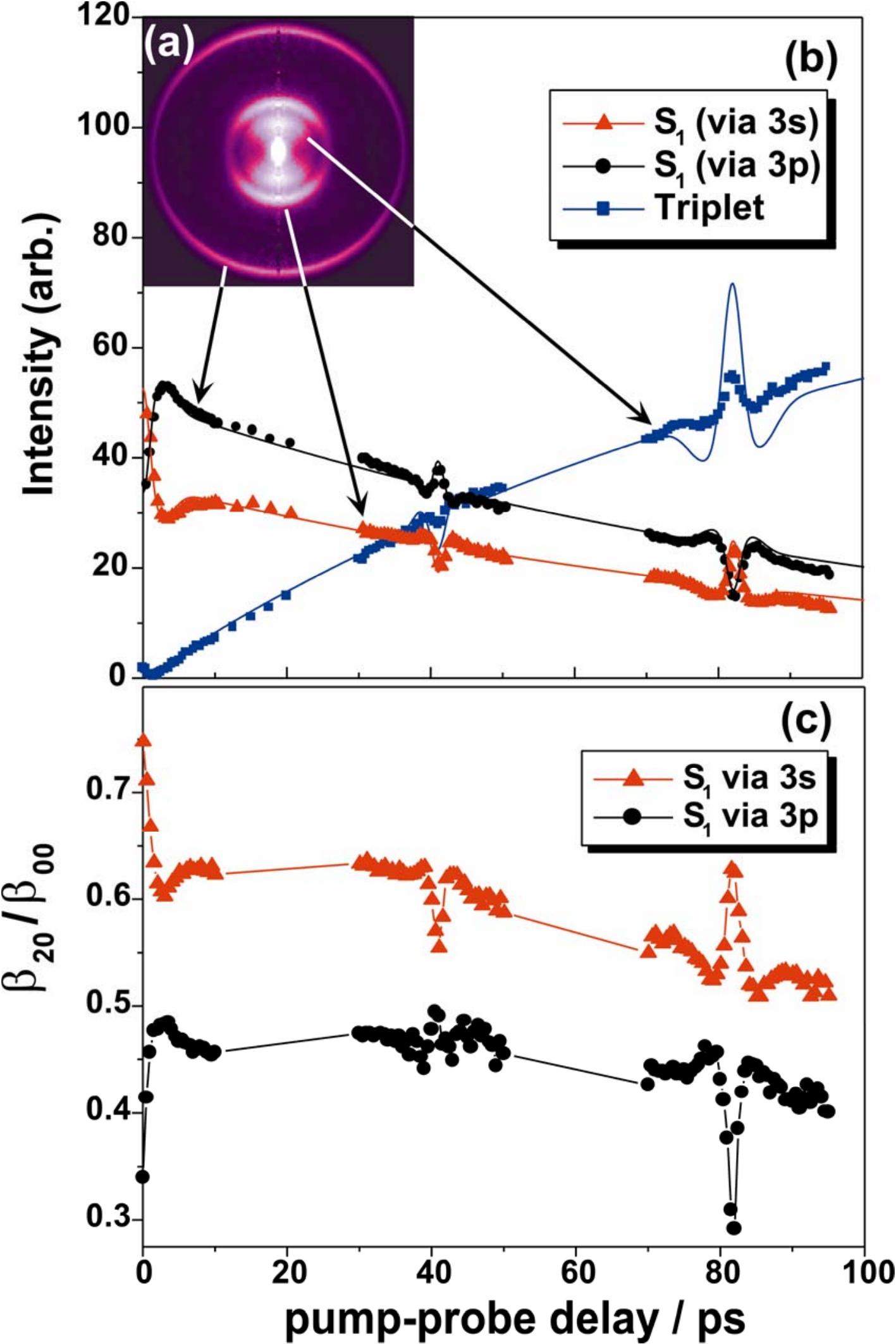}
  \caption{(a) Inverse Abel transformed photoelectron image showing the lab
    frame PAD for ionization of pyrazine with a pump pulse at 323~nm and a
    probe pulse at 401~nm. The laser pulses had parallel linear polarizations
    and a temporal separation of 30~ps.  The outer two rings correspond to two
    photon ionization of the S$_1$ electronic state via3s and 3p Rydberg
    states, and the inner ring corresponds to two-photon ionization of the
    triplet state manifold T$_1$ in the neutral formed by inter-system
    crossing from the S$_1$ state. (b) Time-dependence of the angle-integrated
    signals in (a). (c) Time-dependence of the PAD anisotropy for the 3
    signals in (b) as monitored by the ratio of the PAD parameters
    $\beta_{20}/\beta_{00}$ from a fit to an expansion in spherical harmonics,
    \eqref{eq:betalm}.}
  \label{fig:pads_suzuki}
\end{figure}

As discussed in \secref{sec:photoionzn} time-resolved PAD (TRPAD) measurements
are sensitive to molecular rotational motion by virtue of their geometric
dependence upon the molecular axis distribution in the LF. An elegant
experimental demonstration of this has been performed by Suzuki and co-workers
who measured the PAD temporal evolution from excited state
pyrazine~\cite{Tsubouchi2001}. In these experiments, the origin of the S$_1$
electronic state of pyrazine was excited by a pump pulse at 323~nm, and
subsequently probed by a time delayed probe pulse via a two photon ionization.
In this energy region, the excited state dynamics of pyrazine involve a well
studied intersystem crossing caused by strong spin-orbit coupling of the
S$_1$ $B_{3u}(n\pi^\ast)$ state with a manifold of triplet states, denoted
T$_1$, resulting in a complex energy spectrum~\cite{McDonald1981, Knee1985,
  Lorincz1985, Felker1986}. While earlier studies focussed on the monitoring
of fluorescence from the S$_1$ state, this work directly measured the S$_1$
decay and the T$_1$ formation via TRPES and TRPAD measurements. Two photon
ionization in these experiments proceeded via 3s and 3p$_z$ Rydberg states of
the neutral, producing well resolved bands attributable to the S$_1$ and T$_1$
states (see \figref{fig:pads_suzuki}). 

In these experiments, the pump and probe pulse were linearly polarized with
their electric fields vectors mutually parallel. The pump pulse created an
initially aligned $\cos^2\theta$ distribution of principle molecular axes,
with $\theta$ the polar angle between the principle molecular axis and the
laser field polarization. This initially prepared rotational wavepacket
subsequently evolved with time, and the ionization yield (shown in panel (b)
of \figref{fig:pads_suzuki}) from the singlet state directly reflected this
wavepacket evolution, exhibiting the expected rotational recurrence behaviour
of a near oblate symmetric top~\cite{Felker1995, Felker1994, Felker1992,
  Felker1992a, Baskin1987, Baskin1986}. The sensitivity of the photoelectron
yield to the molecular axis alignment arises due to the well defined molecular
frame direction of the transition dipole momement for excitation of the
intermediate Rydberg states in the probe step -- the 3p$_z\leftarrow$S$_1$ and
the 3p$_z\leftarrow$S$_1$ transitions being perpendicular and parallel to the
principle molecular axis respectively, resulting in the opposite (``out of
phase'') behaviours in the black and red lines showin in panel (b) of
\figref{fig:pads_suzuki}. Interestingly, the rotational coherence is also
directly observed in the signal representing the formation of the T$_1$
manifold (blue line in \figref{fig:pads_suzuki}), demonstrating that
rotational coherence is (perhaps partially) preserved upon internal
conversion~\cite{Tsubouchi2004, Suzuki2006}. Additionally, the PAD also
reflected the wavepacket evolution with the time dependence of the value of
$\beta_{20}/\beta_{00}$ mapping the rotational recurrence behaviour, as shown
in panel (c) of \figref{fig:pads_suzuki}. In this case the different
dependence upon molecular axis alignment of the LF PAD for ionization via the
3s vs. the 3p$_z$ Rydberg states reflects the different MF PADs for ionization
of these two Rydberg states. These measurements demonstrate the utility of
TRPAD measurements as a probe of rotational dynamics. Such measurements are
sensitive to vibration-rotation coupling~\cite{Reid1999, Althorpe2000,
  Seideman2000a, Althorpe1999}.

\section{Excited State Intramolecular Proton Transfer}
\begin{figure}
  \includegraphics[height=5in]{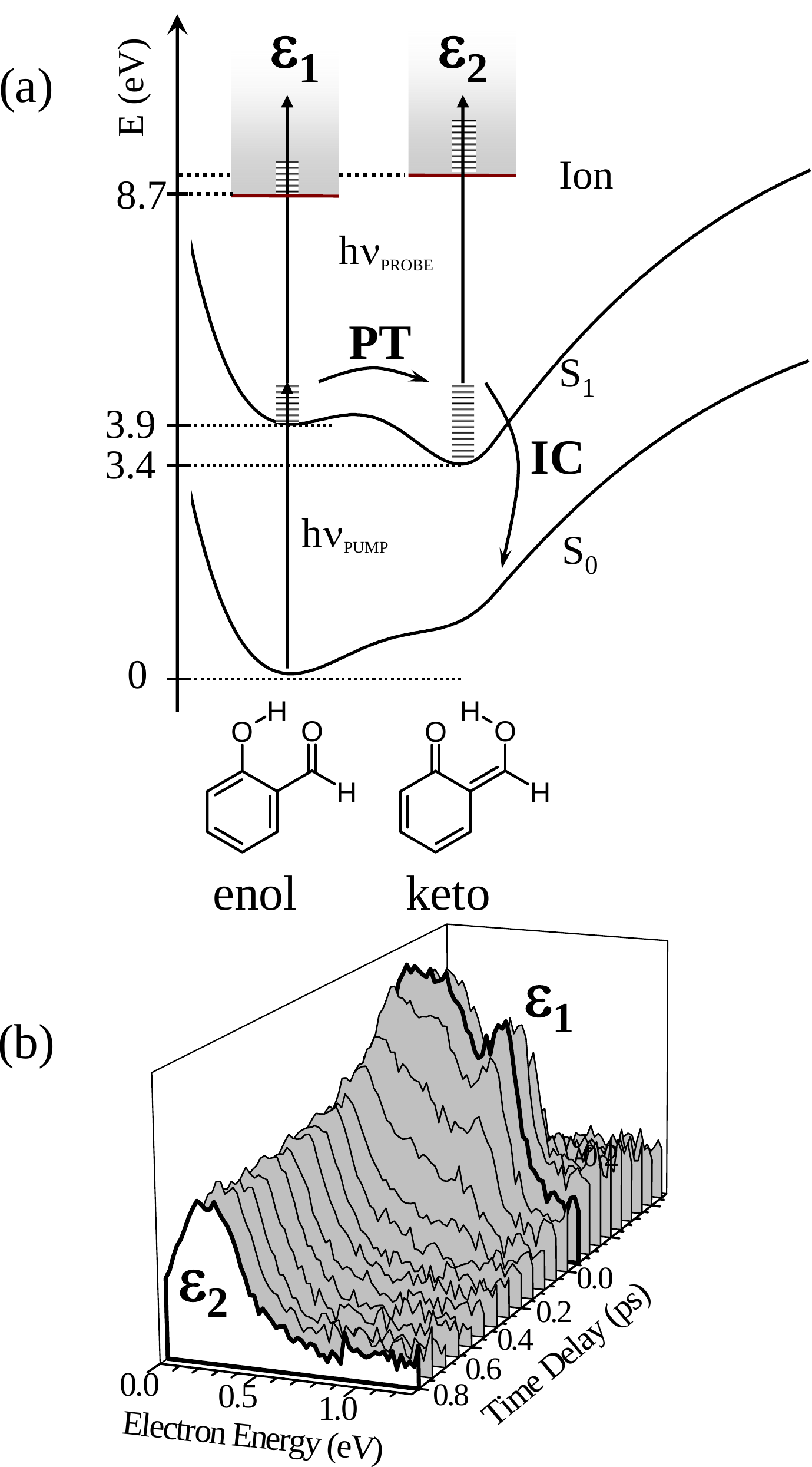}
  \caption{(Top) Energetics for excited state intramolecular proton transfer
    (ESIPT) in OHBA, showing the enol and keto forms. Excitation with a pump
    laser forms the enol tautomer in the S$_{1}(\pi\pi^\ast)$ state.  ESIPT
    leads to ultrafast population transfer from the S$_{1}$ enol to the
    S$_{1}$ keto tautomer.  On a longer time scale, the keto S$_{1}$
    population decays via internal conversion to the keto ground state. Both
    the enol and keto excited state populations are probed via TRPES,
    producing the two photoelectron bands $\varepsilon_{1}$ and
    $\varepsilon_{2}$.  (Bottom) TRPES spectra of OHBA at an excitation
    wavelength of 326~nm and a probe wavelength of 207~nm. Two photoelectron
    bands were observed: $\varepsilon_{1}$ due to ionization of the S$_{1}$
    enol, and $\varepsilon_{2}$ due to ionization of the S$_{1}$ keto.  Band
    $\varepsilon_{1}$ was observed only when the pump and probe laser beams
    overlapped in time, indicating a sub-50~fs timescale for the proton
    transfer. Band $\varepsilon_{2}$ displayed a pump wavelength dependent
    lifetime in the picosecond range corresponding to the energy dependent
    internal conversion rate of the dark S$_{1}$ keto state formed by the
    proton transfer. }
  \label{fig:OHBA_TRPES}
\end{figure}

Excited state intramolecular proton transfer (ESIPT) processes are important
for both practical and fundamental reasons.  o-Hydroxybenzaldehyde (OHBA) is
the simplest aromatic molecule displaying ESIPT and serves as a model system
for comparison with theory. TRPES was used to study ESIPT in OHBA,
monodeuterated ODBA and an analogous two-ring system hydroxyacetonaphtone
(HAN) as a function of pump laser wavelength, tuning over the entire enol
S$_{1}(\pi\pi^\ast)$ absorption band of these
molecules~\cite{Lochbrunner2000a, Lochbrunner2001}.  The experimental scheme
is depicted in \figref{fig:OHBA_TRPES}, showing energetics for the case of
OHBA. Excitation with a tuneable pump laser h$\nu_{\mathrm{pump}}$ forms the
enol tautomer in the S$_{1}(\pi\pi^\ast)$ state.  ESIPT leads to ultrafast
population transfer from the S$_{1}$ enol to the S$_{1}$ keto tautomer.  On a
longer time scale, the S$_{1}$ keto population decays via internal conversion
to the ground state. Both the enol and keto excited state populations are
probed by photoionization with a probe laser h$\nu_{\mathrm{probe}}$,
producing the two photoelectron bands $\varepsilon_{1}$ and $\varepsilon_{2}$.

In \figref{fig:OHBA_TRPES} are TRPES spectra of OHBA at an excitation
wavelength of 326~nm. Two photoelectron bands $\varepsilon_{1}$ and
$\varepsilon_{2}$ with distinct dynamics were observed.  Band
$\varepsilon_{1}$ is due to photoionization of the initially populated S$_{1}$
enol tautomer, and band $\varepsilon_{2}$ is due to the photoionization of the
S$_{1}$ keto tautomer. The decay of band $\varepsilon_{1}$ yields an estimated
upper limit of 50~fs for the lifetime of the S$_{1}$ enol tautomer. Proton
transfer reactions often proceed via tunneling of the proton through a
barrier.  Deuteration of the transferred proton should then significantly
prolong the lifetime of the S$_{1}$ enol tautomer. In experiments with ODBA,
an isotope effect was not observed -- i.e. the ESIPT reaction was again
complete within the laser cross-correlation.

\begin{figure}
  \includegraphics[width=6cm]{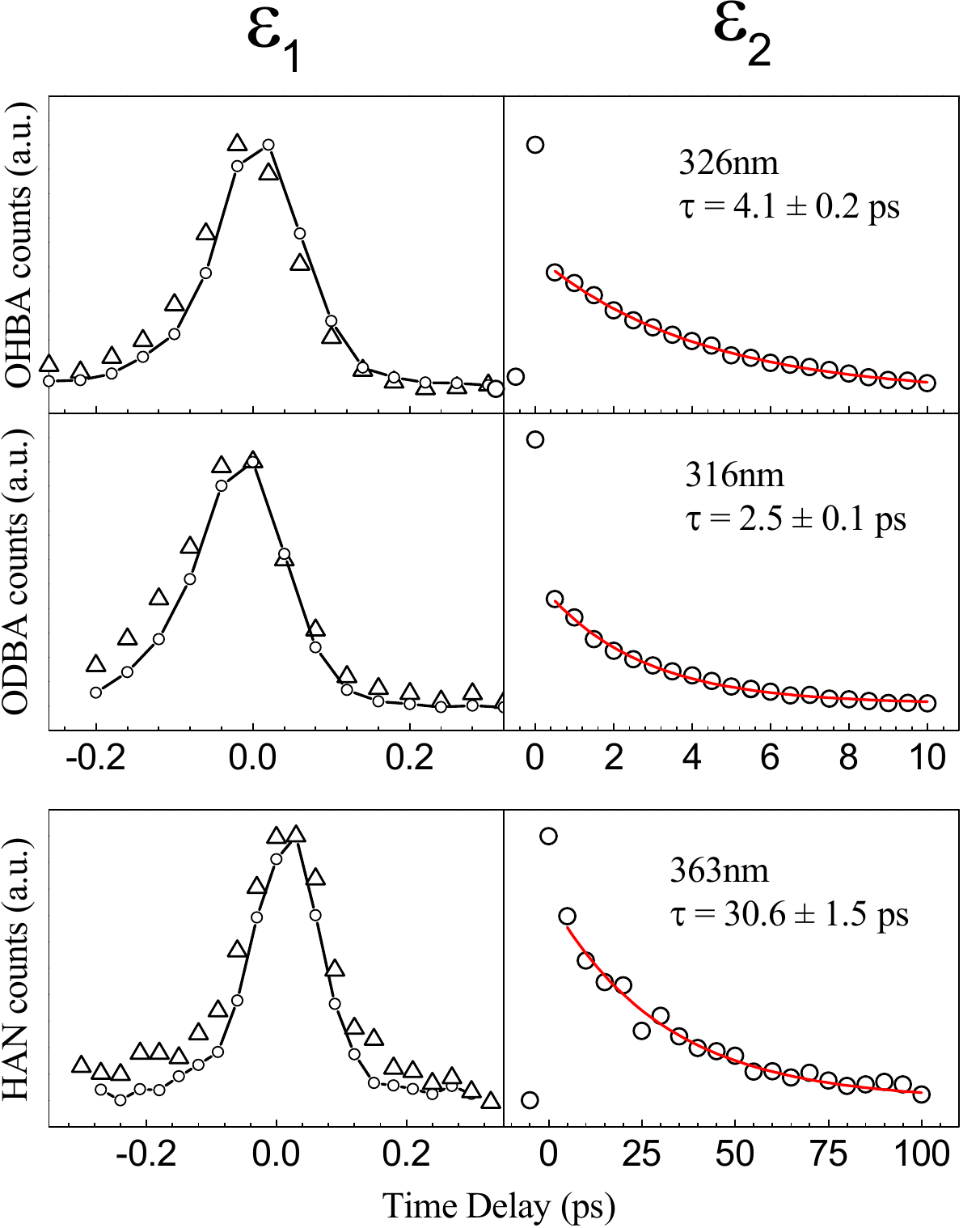}
  \caption{Integrated signals $\varepsilon_{1}$ and $\varepsilon_{2}$ for OHBA
    (top), ODBA (middle), and HAN (bottom) plotted as a function of the time
    delay at the indicated excitation wavelength. Note the change in ordinate
    time scales. Signal $\varepsilon_{1}$ always followed the laser
    cross-correlation, indicating a rapid proton transfer reaction. The decay
    of signal $\varepsilon_{2}$ was fitted via single exponential decay,
    yielding the time constant for internal conversion of the S$_{1}$ keto
    state in each molecule.}
  \label{fig:OHBA_decayrates}
\end{figure}

In \figref{fig:OHBA_decayrates} are examples of fits to OHBA at 326~nm, ODBA
at 316~nm, and HAN at 363~nm. The proton transfer rates for all three
molecules were sub-50~fs over their entire S$_{1}$ enol absorption bands. It
was concluded that the barrier in the OH stretch coordinate must be very small
or non-existent. This interpretation is consistent with ab initio calculations
which predict no barrier for the proton transfer~\cite{Sobolewski1994,
  Sobolewski1999}. An estimate of the corresponding reaction rate using an
instanton calculation, which takes into account the multi-mode character of
proton transfer, resulted in S$_{1}$ enol lifetimes of $\sim20$~fs for the
transfer of a proton and $<$50~fs for the transfer of a deuteron when the
barrier was lowered to 2.4~kcal/mol~\cite{Lochbrunner2000a, Lochbrunner2001}.
This value was considered to be an upper limit for the proton transfer
barrier.

\begin{figure}
  \includegraphics[width=7cm]{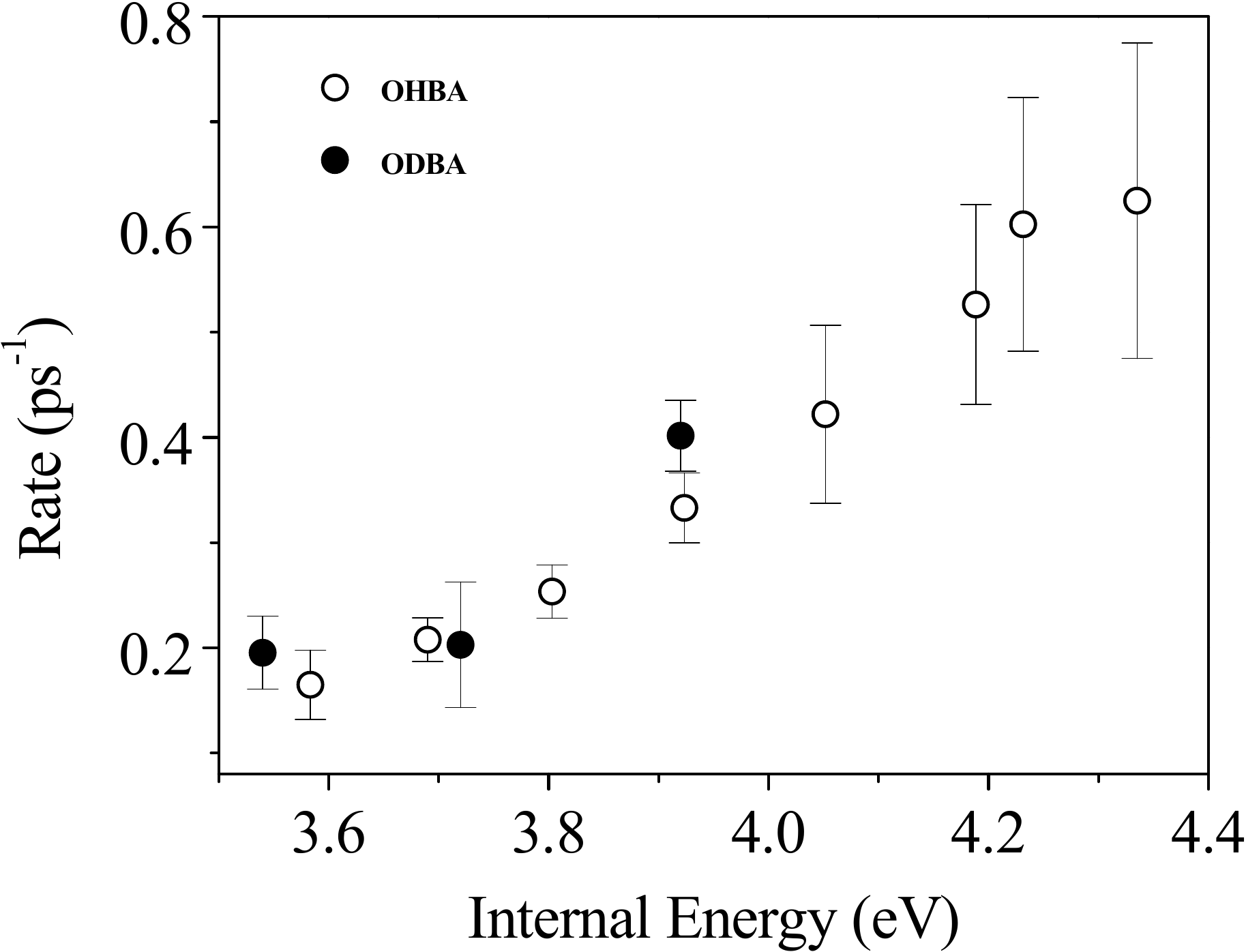}
  \caption{Internal conversion rates of the S$_{1}$ keto state of OHBA (open
    circles) and ODBA (filled circles), determined by single exponential fits
    to the $\varepsilon_{2}$ band decay. Both show a monotonic increase in
    rate as a function of the excitation energy, but without a significant
    isotope effect.}
  \label{fig:OHBA_ODBA}
\end{figure}

As is common in TRPES, these spectra also give insights into the dynamics on
the ``dark'' S$_{1}$ keto state. The picosecond decay of band
$\varepsilon_{2}$ corresponds to S$_{1}$ keto internal conversion to the
ground state.  The wavelength-dependent S$_{1}$ keto internal conversion rates
for OHBA and ODBA shown in \figref{fig:OHBA_ODBA} revealed no significant
isotope effect.  Interestingly, the measured internal conversion rates for
OHBA/ODBA are very fast (1.6--6~ps over the range 286--346~nm) considering the
large energy gap of 3.2~eV between the ground and excited state. One
possibility is that fast internal conversion in such systems is due to an
efficient conical intersection involving a $\pi\pi^\ast$ state with a
$\pi\sigma^\ast$ via large amplitude hydroxy H-atom
motion~\cite{Sobolewski1994, Sobolewski1999}. However, the observed absence of
an isotope effect on S$_{1}$ keto internal conversion rates in ODBA does not
support this mechanism. A clue is found in the comparison with internal
conversion rates of OHBA/ODBA with the larger HAN, shown in
\figref{fig:OHBA_decayrates}. HAN has both a smaller S$_{1}$--S$_{0}$ energy
gap and a higher density of states, leading to the expectation that its
internal conversion rate should be faster than that of OHBA. Surprisingly, it
is about ten times slower, indicating that some other effect must be
operative. A major difference between the two molecules is the position of a
n$\pi^\ast$ state, which is almost isoenergetic with the $\pi\pi^\ast$ state
in OHBA, but more than 0.5~eV higher in HAN. The coupling of the $\pi\pi^\ast$
and n$\pi^\ast$ states, mediated by out-of-plane vibrations, greatly increases
the internal conversion rate in OHBA. The local mode character of the OH
out-of-plane bending vibration makes this mode inefficient for the coupling of
the n$\pi^\ast$ and $\pi\pi^\ast$ states. As a result, the bending modes of
the aromatic ring dominate this interaction, which explains the absence of an
isotope effect~\cite{Lochbrunner2000a, Lochbrunner2001}. This example serves
to illustrate how TRPES can be used to study the dynamics of biologically
relevant processes such as ESIPT and that it reveals details of both the
proton transfer step and the subsequent dynamics in the ``dark'' state formed
after the proton transfer.

\section{Dynamics of Molecular Electronic Switches}
\label{sec:app_switches}
The burgeoning area of active molecular electronics involves the use of
molecules or molecular assemblies acting as switches, transistors or
modulators.  A central theme is that structural rearrangement processes such
as isomerization should lead to changes in either optical or electrical
properties, generating the desired effect. It is often proposed that these
structural rearrangements be induced via electronic excitation.  The rational
design of active molecular electronic devices must include a detailed
consideration of the dynamics of the ``switching'' process for several
reasons. Foremost is that activation of the device (e.g. by a photon) must
indeed lead to the desired change in optical or electrical properties and
therefore this basic mechanism must be present.  Two other issues, however,
are of great practical significance.  The efficiency of the molecular
electronic process is an critical element because excited organic molecules
often have a variety of complex decay paths that compete with the desired
process.  The efficiency of a device can be defined simply as the rate of the
desired process divided by the sum of the rates of all competing processes.
As certain of these competing processes can occur on ultrafast time scales
(e.g.  dissipation, dissociation), the rate of the desired process must be
very fast indeed, even if the required overall response is slow.  A directly
related issue is that of stability.  A molecular modulator that operates at
1~GHz and lasts for three years must ``switch'' $\sim10^{17}$ times without
malfunction. The quantum yields of any ``harmful'' processes must therefore be
exceedingly small.  Unfortunately, excited organic molecules have a number of
destructive decay pathways such as photodissociation and triplet formation
(often leading to reaction). The relative rates and quantum yields of these
processes as well as their dependence on substituent and environmental
effects, will be critical elements in the design of efficient, stable active
molecular devices.  Trans-azobenzene is often considered the canonical
molecular switch and its photoisomerization is the basis for numerous
functional materials~\cite{Rau1990}. Azobenzene provides an important example
for the study of the dynamics of Molecular Electronic switches via
TRPES~\cite{Schultz2003}.

\begin{figure}
  \includegraphics[width=8cm]{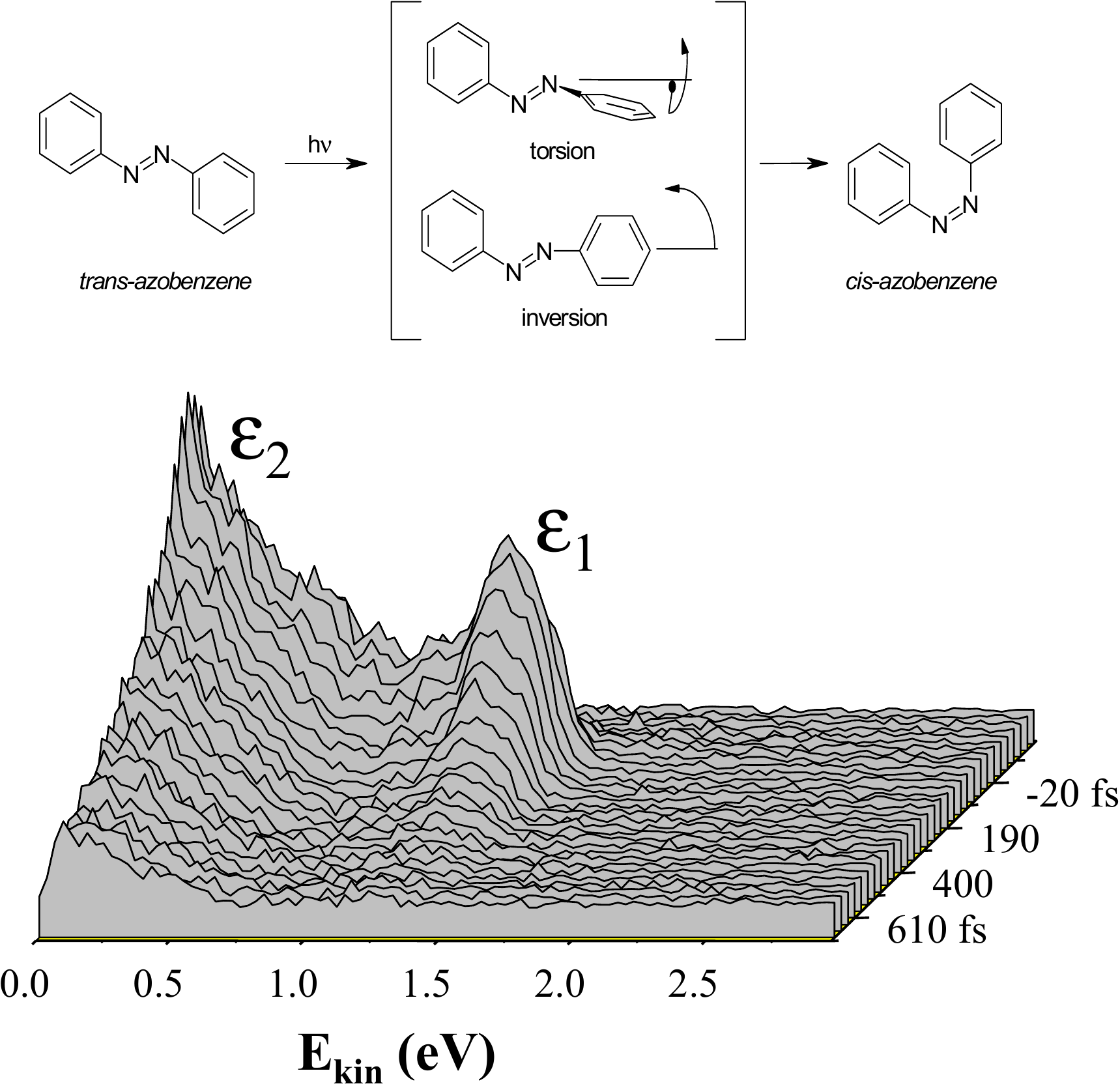}
  \caption{ (Top) Photoisomerization dynamics of trans- to cis-azobenzene,
    indicating torsional and inversion pathways. (Bottom).  TRPES spectra of
    trans-azobenzene excited at 330~nm and probed at 207~nm.  Two
    photoelectron bands $\varepsilon_{1}$ and $\varepsilon_{2}$ were observed,
    having identical laser-limited risetimes but differing decay rates
    ($\tau_1=130$~fs, $\tau_2=410$~fs) and, importantly differing Koopmans'
    ionization correlations. These results indicate that there is a previously
    unrecognized $\pi\pi^\ast$ state, S$_{3}$ (centered on the aromatic
    rings), involved in the dynamics.}
  \label{fig:azobenzene}
\end{figure}

Despite great interest in azobenzene photophysics, the basic
photoisomerization mechanism remains disputed~\cite{Tamai2000}: in contrast to
the expectations of Kasha's rule, the isomerization quantum yield decreases
rather than increases with increasing photon energy.  In
\figref{fig:azobenzene}, the two possible isomerization channels, proceeding
via either a planar pathway (inversion) or a non-planar, twisted pathway
(torsion) are shown.  Previous studies determined that isomerization in the
first excited state S$_{1}$ state proceeds along the inversion
coordinate~\cite{Rau1990}. The second excited state
S$_{2}(\pi\pi^{\ast}_{\mathrm{N=N}})$ is generally thought to be the N=N
analogue of the C=C $\pi\pi^\ast$-state in stilbene and that, somehow, motion
along the torsional coordinate in S$_{2}(\pi\pi^{\ast}_{\mathrm{N=N}})$ is
responsible for the observed reduction in isomerization yield. Time-resolved
studies suggested that photoisomerization proceeds via the inversion
coordinate in S$_{1}$~\cite{Rau1990}.  The role of the torsional isomerization
pathway remains controversial. Theoretical studies have supported both torsion
and inversion pathways but disagreed on the states involved in the excited
state relaxation. Any successful model considering $\pi\pi^\ast$ state
relaxation in AZ must address three puzzling features~\cite{Schultz2003}: (A)
The violation of Kasha's rule, i.e. $\Phi_{\mathrm{isom}}\sim25\%$ for
S$_{1}(\mathrm{n}\pi^\ast)$ but drops to $\Phi_{\mathrm{isom}}\sim12\%$ for
the higher lying $\pi\pi^\ast$ state(s); (B) Inhibition of the torsional
coordinate in sterically restrained AZ increases $\Phi_{\mathrm{isom}}$ of the
$\pi\pi^\ast$ states to a level identical to that observed for S$_{1}$
photoexcitation; (C) The observation of efficient relaxation of
S$_{2}(\pi\pi^\ast)$ to the S$_{1}$ state via planar geometries.

In \figref{fig:azobenzene} a time-resolved photoelectron spectrum for
excitation of AZ to the origin of its S$_{2}(\pi\pi^{\ast}_{\mathrm{N=N}})$
state is shown. Two photoelectron bands $\varepsilon_{1}$ and
$\varepsilon_{2}$ with differing lifetimes and differing Koopmans'
correlations were observed.  Due to these two differences, the
$\varepsilon_{1}$ and $\varepsilon_{2}$ bands must be understood as arising
from the ionization of two different electronic states. Furthermore, as both
bands rise within the laser cross-correlation, they are due to direct
photoexcitation from S$_{0}$ and not to secondary processes.  Therefore, in
order to account for different lifetimes, different Koopmans' correlations and
simultaneous excitation from S$_{0}$, the existence of an additional state,
labeled S$_{3}(\pi\pi^{\ast}_{\phi})$, which overlaps spectroscopically with
S$_{2}(\pi\pi^{\ast}_{\mathrm{N=N}})$ must be invoked. According to the
Koopmans' analysis (based upon assignment of the photoelectron bands) and to
high level, large active space CASSCF calculations, this new state
S$_{3}(\pi\pi^{\ast}_{\phi})$ corresponds to $\pi\pi^\ast$ excitation of the
phenyl rings~\cite{Schultz2003}, as opposed to the
S$_{2}(\pi\pi^{\ast}_{\mathrm{N=N}})$ state where excitation is localized on
the N=N bond. Therefore, $\pi\pi^\ast$-excitation in the phenyl rings does not
directly ``break'' the N=N bond and leads to reduced isomerization quantum
yields.

A new model for AZ photophysics was proposed as a result of these TRPES
studies. The S$_{2}(\pi\pi^{\ast}_{\mathrm{N=N}})$ state internally converts
to S$_{1}$ in a planar geometry, explaining puzzle (C) above. The subsequent
relaxation of S$_{1}$ does indeed follow Kasha's rule and yields
$\Phi_{\mathrm{isom}}\sim25\%$ for the population originating from
S$_{2}(\pi\pi^{\ast}_{\mathrm{N=N}})$. Different dynamics are observed in the
TRPES experiments for the S$_{3}(\pi\pi^{\ast}_{\phi})$ state, indicating a
different relaxation pathway. To explain puzzle (A), relaxation of
S$_{3}(\pi\pi^{\ast}_{\phi})$ with reduced isomerization must be assumed: the
ring-localized character of S$_{3}(\pi\pi^{\ast}_{\phi})$ suggests a
relaxation pathway involving phenyl-ring dynamics. This could involve torsion
and lead directly to the trans-AZ ground state -- explaining both puzzles (A)
and (B). Ab initio Molecular Dynamics (AIMD) simulations~\cite{Schultz2003}
starting from the Franck-Condon geometry in
S$_{2}(\pi\pi^{\ast}_{\mathrm{N=N}})$ agree with result (C) and predict that
the molecule quickly ($<50$~fs) samples geometries near conical intersections
while still in a planar geometry, with no evidence for torsion or
inversion. For S$_{1}$, the AIMD simulations predict that a conical
intersection involving inversion is approached within
50~fs~\cite{Schultz2003}. This mechanism differs greatly from that of all
earlier models in that those always assumed that only a single bright state,
S$_{2}(\pi\pi^{\ast}_{\mathrm{N=N}})$, exists in this wavelength region. This
example shows how TRPES can be used to study competing electronic relaxation
pathways in a model molecular switch, revealing hidden yet important
electronic states that can be very hard to discern via conventional means.

\section{Photodissociation Dynamics}
\label{sec:app_photodis}
From the point of view of chemical reaction dynamics, the most interesting
case is that of unbound excited states or excited states coupled to a
dissociative continuum -- i.e. photodissociation dynamics. The dissociative
electronically excited states of polyatomic molecules can exhibit very complex
dynamics, usually involving non-adiabatic processes. TRPES and TRCIS may be
used to study the complex dissociation dynamics of neutral polyatomic
molecules, and below we'll give two examples of dissociative molecular systems
that have been studied by these approaches, NO$_2$ and (NO)$_2$.

TRCIS was first applied to dissociative multiphoton ionization of NO$_{2}$ at
375.3~nm~\cite{Davies1999}. This was identified as a three-photon transition
to a repulsive surface correlating with NO(C~$^{2}\Pi$)+O($^{3}P$)
fragments. The NO(C) was subsequently ionizing by a single photon, yielding
NO$^{+}$(X$^{1}\Sigma^{+}$)+e$^{-}$.

\begin{figure}
  \includegraphics[width=7cm]{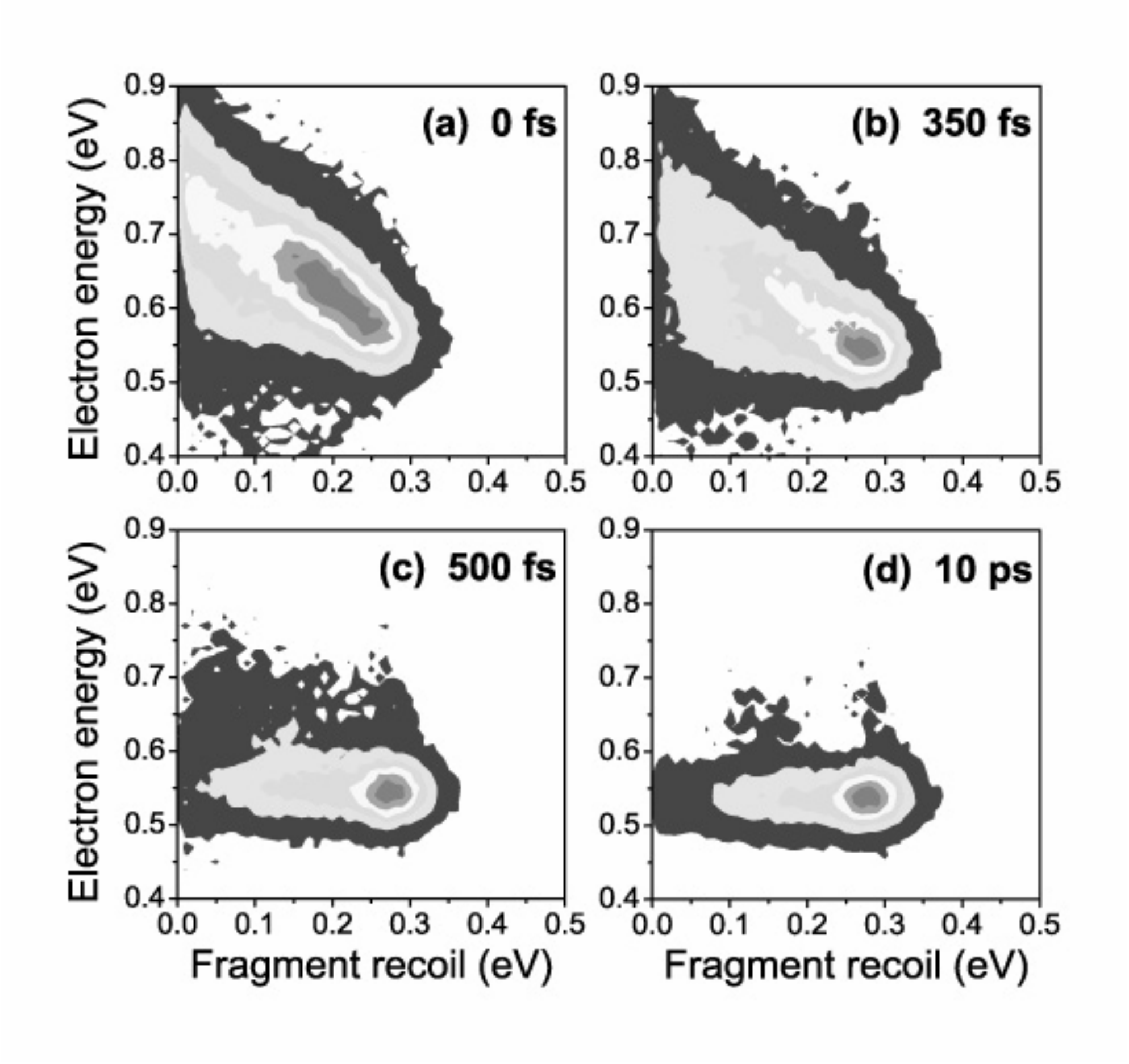}
  \caption{Time-Resolved Coincidence-Imaging Spectroscopy (TRCIS) of
    dissociative multiphoton ionization processes in NO$_{2}$ using ~100~fs
    laser pulses at 375.3~nm, using energy-energy correlations. The two
    dimensional maps show, at time delays of 0~fs, 350~fs, 500~fs and 10~ps,
    the correlation between the photoelectron kinetic energy (abscissa) and NO
    photofragment recoil energy (ordinate).  The intensity distributions
    change from a negative correlation at early times to uncorrelated at later
    times, yielding information about the molecule as it
    dissociates. Reproduced with permission from ref.~\cite{Davies1999}}
  \label{fig:TRCIS_NO2_KE}
\end{figure}

\begin{figure}
  \includegraphics[height=3in]{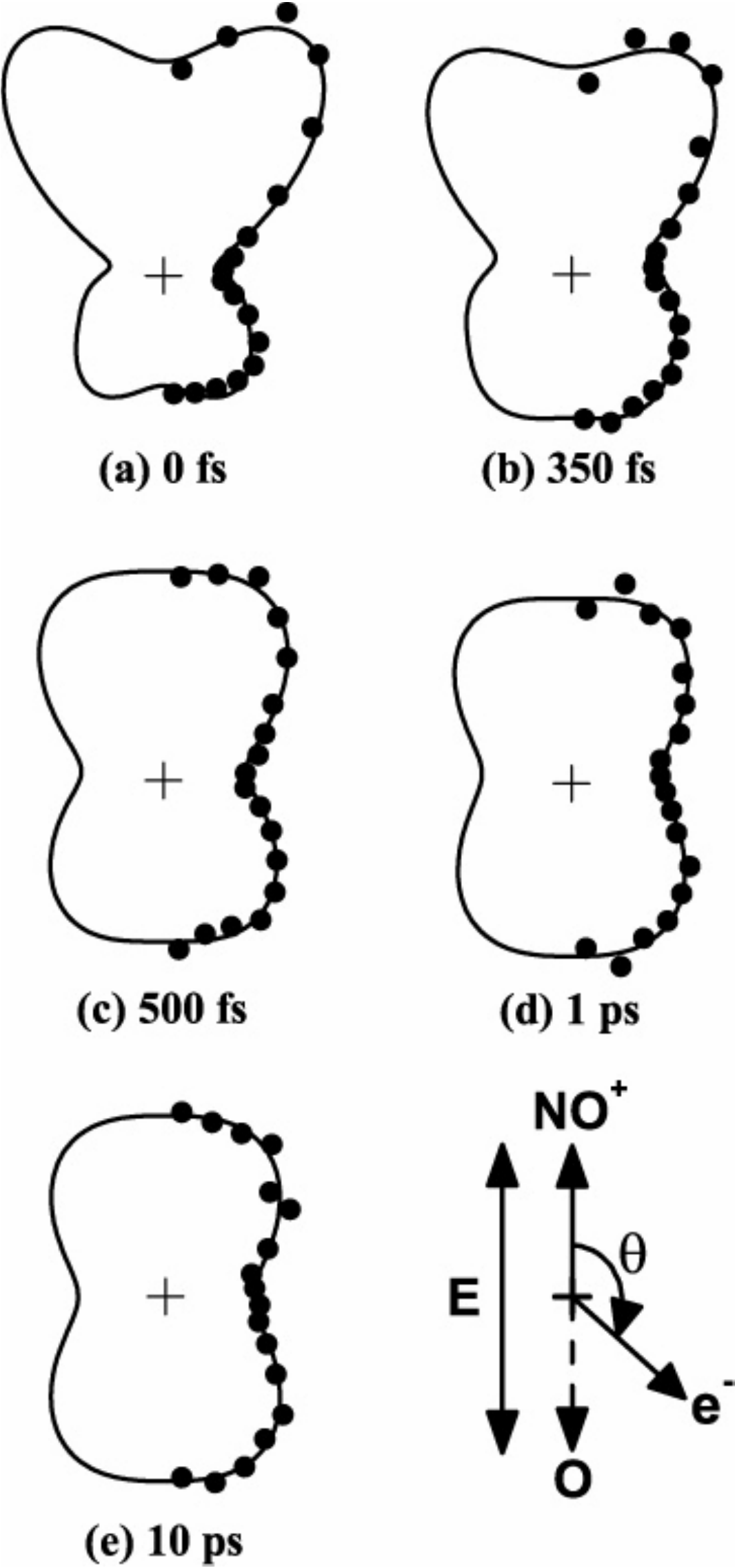}
  \caption{Coincidence-Imaging Spectroscopy of dissociative multiphoton
    ionization processes in NO$_{2}$ with $\sim100$~fs laser pulses at
    375.3~nm, using angle-angle correlations. The polar plots show, at time
    delays of 0~fs, 350~fs, 500~fs, 1~ps and 10~ps, the angular correlation
    between the ejected electron and NO photofragment when the latter is
    ejected parallel to the laser field polarization vector.  The intensity
    distributions change from a forward-backward asymmetric distribution at
    early times to a symmetric angular distribution at later times, yielding
    detailed information about the molecule as it dissociates. Reproduced with
    permission from ref.~\cite{Davies2000}}
  \label{fig:TRCIS_NO2_PADS}
\end{figure}

As an illustration of the multiply differential information obtained via
TRCIS, energy-energy correlations plotting photoelectron kinetic energy
vs. NO(C) photofragment kinetic energy, as a function of time, are shown in
\figref{fig:TRCIS_NO2_KE}.  At early times, 0~fs and 350~fs, there is a
negative correlation between electron and fragment recoil energy. This is the
form expected for a molecule in the process of dissociating where there is a
trade-off between ionization energy and fragment recoil energy. At longer time
delays, 500~fs and 10~ps, the NO(C) fragment is no longer recoiling from the O
atom -- it is a free particle -- and the photoelectron spectrum obtained is
simply that of free NO(C) and, hence, the negative correlation
vanishes~\cite{Davies1999}. By measuring the angle of recoil of both
photoelectron and photofragment in coincidence, the PAD may be transformed
into the RF at each time delay~\cite{Davies2000}. In
\figref{fig:TRCIS_NO2_PADS}, the time-resolved RF PADs are shown for the case
of photofragments ejected parallel to the laser polarization axis.  It can be
seen that at early times, 0 and 350~fs, the PAD is highly asymmetric.  The
breaking of forward-backward symmetry in the RF originates from NO(C)
polarization due to the presence of the O atom from which it is recoiling.  At
longer times, 1~ps and 10~ps, this forward-backward asymmetry vanishes, as the
NO(C) becomes a free particle.  This once again shows the power of TRCIS in
obtaining highly detailed information about molecules in the process of
dissociating.

\begin{figure}
  \includegraphics[height=4in]{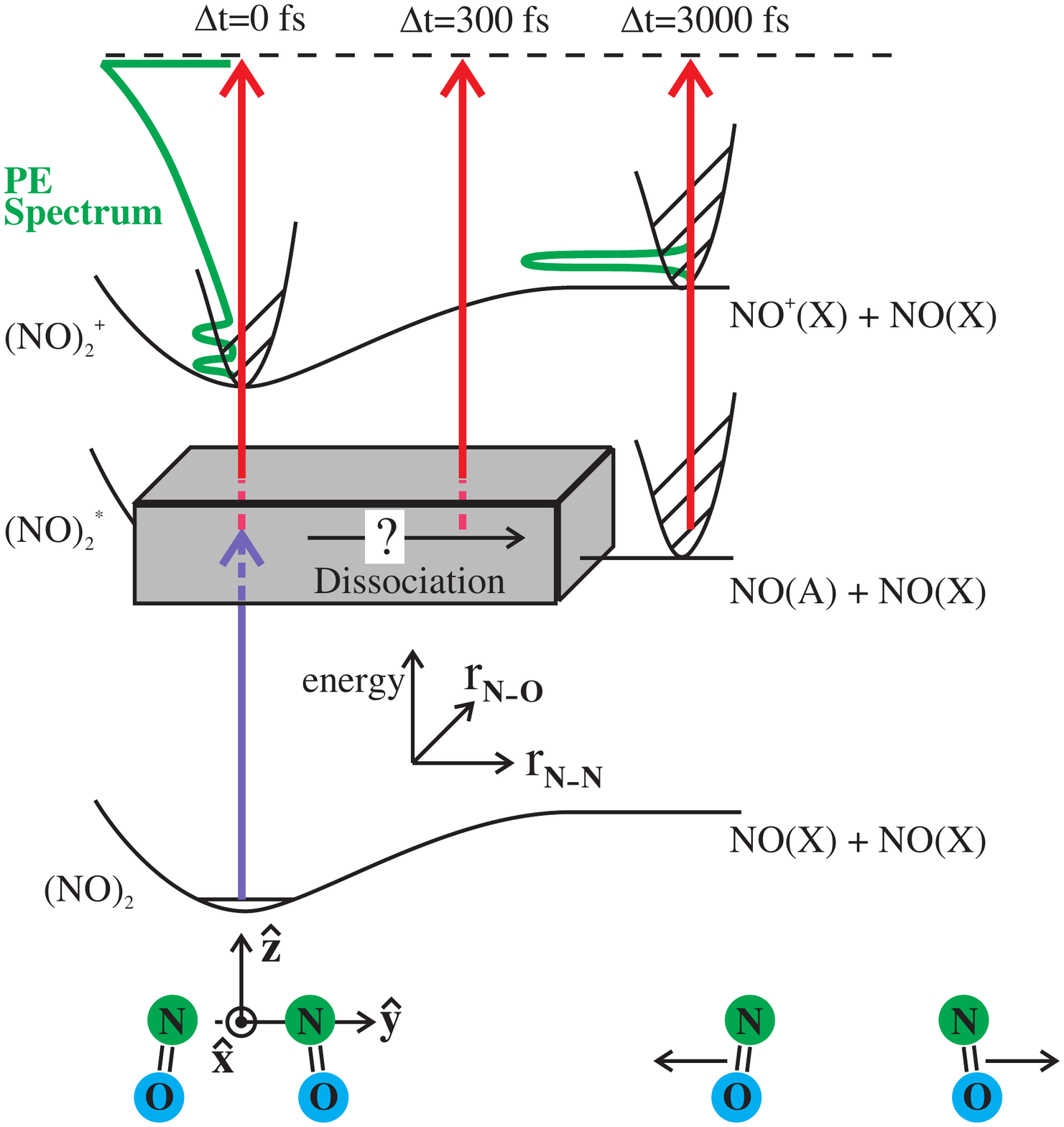}
  \caption{A femtosecond TRPES scheme for studying NO dimer photodissociation.
    A UV pump pulse creates the excited state (NO)$_{2}^{\ast}$. Its
    subsequent evolution is monitored all the way from initial excitation to
    final product emission via a UV probe pulse, projecting the wavepacket
    onto the ionization continuum.  The resulting photoelectron spectrum,
    reflecting vibrational and electronic changes during dissociation, is
    depicted in green.}
  \label{fig:NOdimer_scheme}
\end{figure}

A second illustrative example of the utility of TRPES and TRCIS for studying
complex molecular photodissociation dynamics which involve multiple electronic
state is the case of the weakly bound cis-planar $C_{2v}$ nitric oxide
dimer~\cite{Blanchet1998a}. The weak (D$_{0}$ = 710~cm$^{-1})$ $^{1}A_{1}$
ground state covalent bond is formed by the pairing of two singly occupied
$\pi^\ast$ orbitals, one from each NO(X$^{2}\Pi)$ monomer. The very intense UV
absorption spectrum of the NO dimer appears broad and featureless and spans a
190--240~nm range, with a maximum at $\sim205$~nm. This transition was
assigned as $^{1}B_{2}\leftarrow\null^{1}A_{1}$ and therefore has a transition
dipole along the N--N bond direction (with $B_2$ symmetry).  Recent ab initio
studies of the excited electronic states of the dimer revealed a complex set
of interactions between two very strongly absorbing states of mixed
valence/Rydberg character that play a central role in the photodissociation
dynamics~\cite{Levchenko2006}. As we shall see from the following
measurements, these ``diabatic'' states are roughly comprised of a diffuse
3p$_{y}$ Rydberg function (the $y$-axis is along the N--N bond) and a
localized valence function which has charge transfer character and therefore
carries most of the oscillator strength in the Franck-Condon region, as the
oscillator strengths are much too high for a pure Rydberg
state~\cite{Levchenko2006}. At 210~nm excitation one product channel is
dominant:
\begin{equation}
  (\mathrm{NO})_{2}^\ast \rightarrow 
  \mathrm{NO}(\mathrm{A}^{2}\Sigma^{+}, v, J)
  + \mathrm{NO}(\mathrm{X}^{2}\Pi, v', J').
\end{equation}
The fragment excited state $\mathrm{NO}(\mathrm{A}^{2}\Sigma^{+})$ is a
molecular 3s Rydberg state, and we shall refer to this as NO(A, 3s).  The
observed NO(A, 3s) product state distributions supported the notion of a
planar dissociation involving restricted intramolecular vibrational energy
redistribution (IVR)~\cite{Demyanenko2002}. A scheme for studying NO dimer
photodissociation dynamics via TRPES is depicted in
\figref{fig:NOdimer_scheme}. The NO(A, 3s)+NO(X) product elimination channel,
its scalar and vector properties and its evolution on the femtosecond time
scale have been discussed in a number of recent publications (see
ref.~\cite{Levchenko2006} and references therein).

The first TRPES study of NO dimer photodissociation at 210~nm excitation (and
287~nm probe) showed that the decaying (NO)$_{2}^{+}$ parent ion signal
disappeared more rapidly (when fit to a single exponential decay of 0.3~ps)
than the NO(A, 3s) state product signal appeared to rise (when fit to a single
exponential growth of 0.7~ps)~\cite{Blanchet1998a}. This result shows once
again that the time dependence of the parent ion signal alone can be
misleading. Due to its Rydberg character, the NO(A 3s, $v$, $J$) products
produced a single sharp peak in the photoelectron spectrum, due to the well
known NO(A$^{2}\Sigma^{+}$, $v$, $J$) $\rightarrow$ NO$^{+}$(X,
$^{1}\Sigma^{+}$, $v$) ionizing transition which has predominantly $\Delta
v=0$. The dissociation dynamics was interpreted in terms of a two step
sequential process involving an unknown intermediate configuration. Subsequent
femtosecond time-resolved ion and photoelectron imaging studies further
considered the dissociation dynamics of the NO dimer~\cite{Tsubouchi2003,
  Tsubouchi2003a, Tsubouchi2005}. These reported the observation that both the
decaying NO dimer cation signal and the rising NO(A) photoelectron signal
could be fit using single exponential functions. Furthermore, the emerging
NO(A, 3s) photoelectron peak changed shape and shifted in energy (by
15-20~meV) at early times.  This was taken as evidence for formation of a
dimer 3s Rydberg state which was expected to correlate directly to NO(A, 3s) +
NO(X) products. It was argued that when the shifting of this peak is taken
into consideration, the decay of the parent signal and the rise of the product
signal could be fit with the same single exponential time constant, suggesting
no need for an intermediate configuration.

\begin{figure}
  \includegraphics[height=5in]{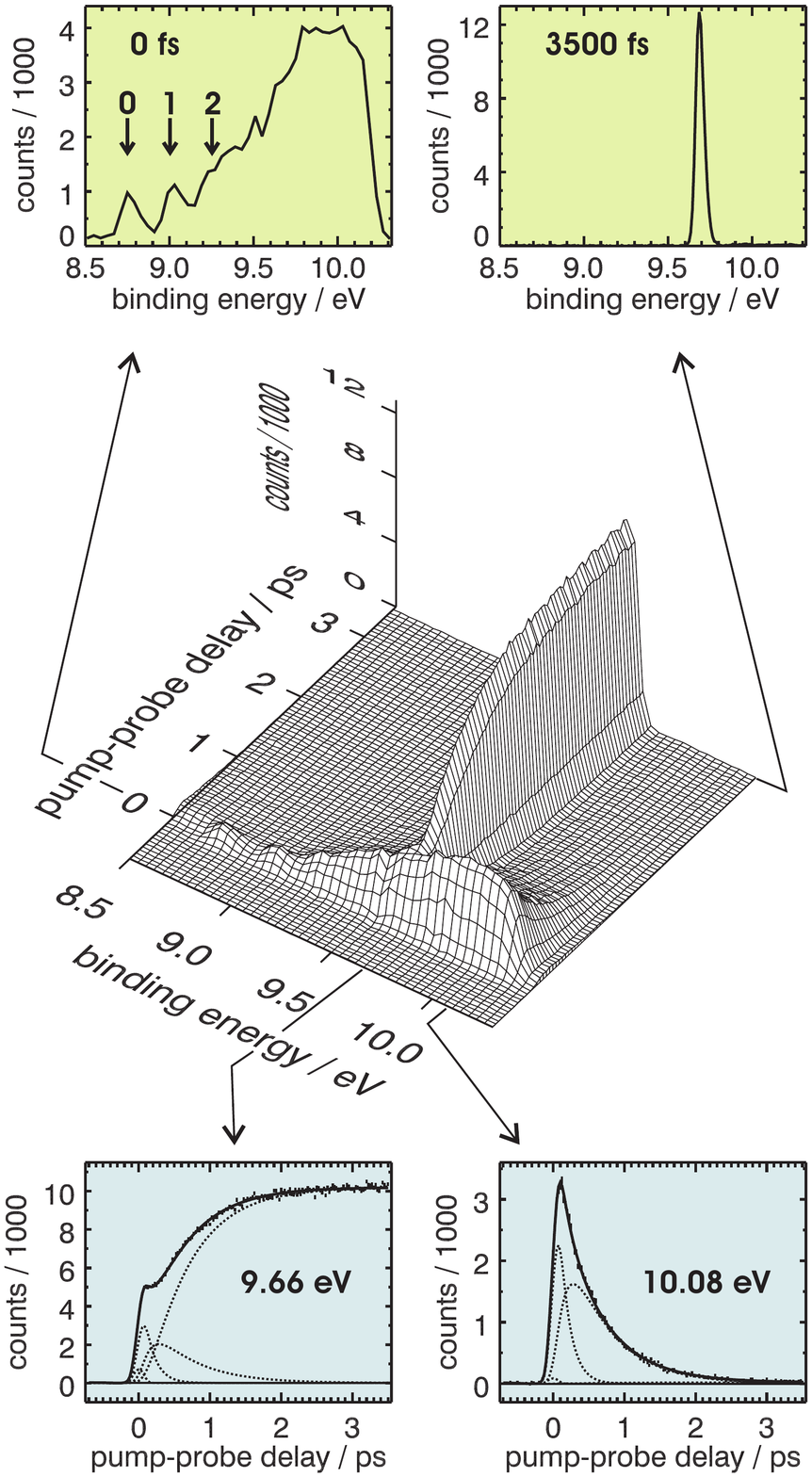}
  \caption{TRPES of NO dimer photodissociation at 210nm
    excitation~\cite{Gessner2006}. The broad, decaying spectrum apparent at
    early times is due to photoionization of the dissociating excited parent
    molecule. The sharp peak emerging with time is due to growth of the free
    NO(A, 3s) products. These 2D data are globally fit at all energies and
    time delays simultaneously.  The green inserts (top) are 1D cuts along the
    energy axis, showing photoelectron spectra at two selected time delays.
    The blue inserts (bottom) are 1D cuts along the time axis, showing the
    evolution of the photoelectron intensity at two selected binding
    energies. The solid lines in the blue graphs are from the 2D fits to the
    sequential two-step dissociation model discussed in the text.  The dashed
    lines are the respective initial, intermediate and final state signal
    components.}
  \label{fig:NO_dimer_TRPES}
\end{figure}

Recently, the photodissociation dynamics of the NO dimer was reinvestigated
using a high sensitivity magnetic bottle technique combined with Time-Resolved
Coincidence Imaging Spectroscopy (TRCIS, discussed
below)~\cite{Gessner2006}. In \figref{fig:NO_dimer_TRPES} we show a magnetic
bottle TRPES spectrum of (NO)$_{2}$ photodissociation. At $\Delta t=0$, a broad
spectrum due to photoionization of (NO)$_{2}^\ast$ shows two resolved
vibrational peaks assigned to 0 and 1 quanta of the cation N=O stretch mode
($\nu_{1}$). The $\nu_{1} = 2$ peaks merges with a broad, intense
Franck-Condon dissociative continuum.  At long times ($\Delta t=3500$~fs), a sharp
photoelectron spectrum of the free NO(A, 3s) product is seen. The 10.08~eV band
shows the decay of the (NO)$_{2}^\ast$ excited state.  The 9.66~eV band shows
both the decay of (NO)$_{2}^\ast$ and the growth of free NO(A, 3s) product. It
is not possible to fit these via single exponential kinetics. However, these
2D data are fit very accurately at all photoelectron energies and all time
delays simultaneously by a two-step sequential model, implying that an initial
bright state (NO)$_{2}^\ast$ evolves to an intermediate configuration
$\mathrm{(NO)}_{2}^{\ast\dag}$ which itself subsequently decays to yield free
NO(A, 3s) products~\cite{Gessner2006}
\begin{equation}
  \label{eq:NO dimer sequential}
  (\mathrm{NO})_{2}^{\ast}\rightarrow(\mathrm{NO})_{2}^{\ast\dag}
  \rightarrow ~\mathrm{NO}(\mathrm{A}, \mathrm{3s}) 
  + \mathrm{NO}(\mathrm{X})
\end{equation}

The requirement for a sequential model is seen in the 9.66~eV photoelectron
band, showing NO(A, 3s) product growth. The delayed rise of the free NO(A, 3s)
signal simply cannot be fit by a single exponential decay followed by single
exponential growth with the same time constant. The 10.08~eV dissociative
ionization band, dominant at early times, is revealing of (NO)$_{2}^\ast$
configurations preceding dissociation. Its time evolution, which also cannot
be fit by single exponential decay, provides another clear view of the
intermediate step. The decay time constant of the initial (NO)$_{2}^\ast$
state is 140$\pm$30~fs, which matches the rise time of the intermediate
(NO)$_{2}^{\ast\dag}$ configuration. This intermediate configuration has a
subsequent decay time of 590$\pm$20~fs. These two time constants result in a
maximum for (NO)$_{2}^{\ast\dag}$ at $\Delta t\sim330$~fs delay. The two
components can be seen as the dashed lines in the fits to the 10.08~eV data
(along with a small instrumental response signal). In the 9.66~eV band, the
dashed lines from the fits show that the rise of the NO(A, 3s) product channel
is first delayed by 140$\pm$30~fs but then grows with a 590$\pm$20~fs time
constant. Although only two cuts are shown, the data are fit at all time
delays and photoelectron energies simultaneously. These results show that the
decay of the parent molecule does not match the rise of the free products and,
therefore, an intermediate configuration which has differing ionization
dynamics is required to model the data. The nature of this
(NO)$_{2}^{\ast\dag}$ configuration cannot be discerned from TRPES data
alone. In order to uncover its character, this system was also studied using
the TRCIS technique~\cite{Gessner2006}.

\begin{figure}
  \includegraphics[angle=-90,width=2.0in]{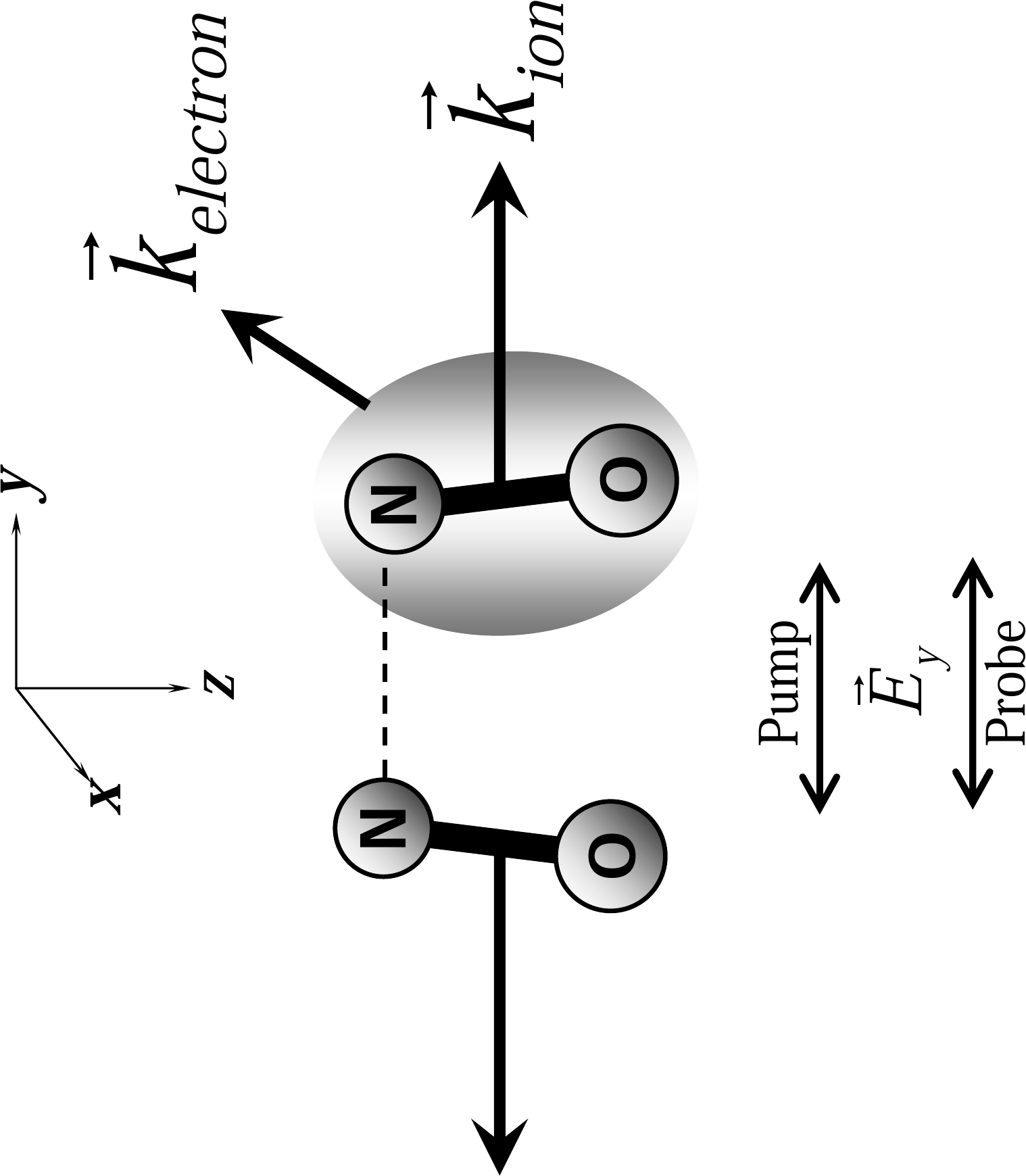}
  \caption{Molecular frame axis convention for the $C_{2v}$ NO dimer. The
    y-axis is along the N--N bond.  Both pump and probe laser polarizations are
    parallel to the y-axis. }
  \label{fig:NOdimer_vectors}
\end{figure}

The 6-dimensional fully correlated TRCIS data set may be cut, projected or
filtered to reveal both scalar and vector correlations as a function of
time. We restrict our discussion here to angular correlations. The molecular
frame axis convention for the NO dimer is shown in
\figref{fig:NOdimer_vectors}. Note that the pump and probe laser polarizations
were parallel to each other in these experiments.

The pump transition dipole is directed along the MF $y$-axis -- the N--N bond
axis. The pump transition therefore forms an anisotropic distribution of
excited (NO)$_{2}^\ast$ states in the LF with the N--N bond aligned along the
pump laser polarization axis. As we are concerned with intermediate
configurations in (NO)$_{2}^\ast$ evolution, we consider therefore the
photoionization probing of (NO)$_{2}^\ast$ , which leads predominately to
dissociative ionization as shown in \figref{fig:NO_dimer_TRPES}. The
dissociative ionization of (NO)$_{2}^\ast$ produces NO$^{+}$ fragments
strongly directed along the laser polarization axis. The NO$^{+}$ fragment
recoil direction therefore indicates the lab frame direction of the N--N bond
(MF $y$-axis) prior to ionization. Rotating the electron momentum vector into
the fragment recoil frame (RF) on an event-by-event basis allows for
reconstruction of the (NO)$_{2}^\ast$ photoelectron angular distribution in
the RF, rather than the LF. Here the RF coincides with the MF, differing only
by azimuthal averaging about the N--N bond direction. Out of all fragment
recoil events, only those directed (``up'' or ``down'') along the parallel
pump and probe laser polarization axis were selected.  Importantly, by
choosing events from this selected set, the data is restricted to excited
state ionization events arising from interactions with the MF $y$-component of
the ionization transition dipole only. As discussed below, this restriction
greatly limits the allowed partial waves for the emitted electron, especially
in the present case where only a single cation electronic continuum is
accessed~\cite{Gessner2006}.

\begin{figure}
  \includegraphics[height=5in]{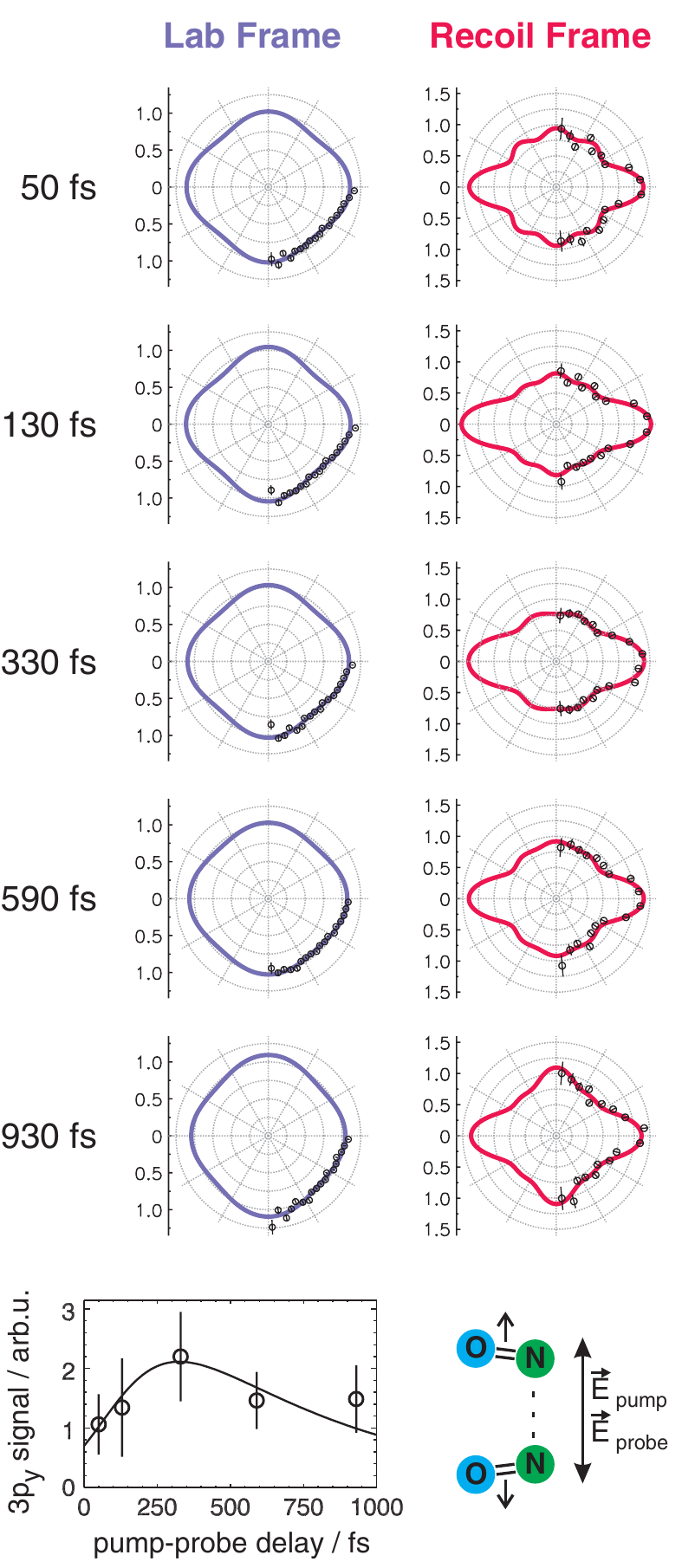}
  \caption{Coincidence-Imaging Spectroscopy of (NO)$_{2}$ photodissociation at
    210nm showing LF (left) and RF (right) photoelectron angular distributions
    (PADs) from the 9.9-10.3~eV dissociative ionization region of
    \figref{fig:NO_dimer_TRPES}. The laser polarizations and RF axes are along
    the y-direction, as shown (bottom right).  The LF PADs show featureless
    and almost invariant behaviour.  The RF PADs show strong anisotropies
    which vary with time.  The fit curves (solid lines) include even order
    Legendre polynomials $P_{L}(\cos\theta)$ up to $L = 4$ for the LF and up
    to $L = 8$ for the RF.  The average partial wave contribution expected
    from Rydberg 3p$_{y}$ ionization is plotted as a function of time (bottom
    left). The time dependence of the intermediate configuration extracted
    from the TRPES data of \figref{fig:NO_dimer_TRPES} and is plotted here as
    the solid line, agreeing well with the time dependence of the 3p$_{y}$
    ionization contribution.  This substantiates the intermediate
    configuration as being of Rydberg 3p$_{y}$ character.  For details see the
    text.}
  \label{fig:TRCIS_NO_dimer}
\end{figure}

In \figref{fig:TRCIS_NO_dimer} time-resolved lab and RF PADs arising from
photoionization of (NO)$_{2}^\ast$ in the 9.9--10.3~eV band of
\figref{fig:NO_dimer_TRPES} are presented. This dissociative ionization region
contains significant contributions from the intermediate (NO)$_{2}^{\ast\dag}$
configuration. In general, the time dependence of PADs relates to the
evolution of excited state electronic structure, as discussed in
\secref{sec:photoionzn}. Here, the LF PADs have a largely isotropic character
that show no discernible change with time, obscuring information about excited
state dynamics. By contrast, the RF PADs show a highly anisotropic character
and a variation with time delay. The solid lines in the polar plots of
\figref{fig:TRCIS_NO_dimer} are fits to an expansion in Legendre polynomials
$P_L(\cos\theta)$,
\begin{equation}
  \label{eq:legpol}
  I(\theta)=\sum_L \mathcal{B}_L P_L(\cos\theta).
\end{equation}
For the RF PADs only even $L$ terms were non-zero with $L\leq8$ in this
fit. Increasing the maximum value of $L$ did not improve the fit to the data,
and odd $L$ coefficients were found to converge to zero in the fits, in
agreement with the up-down symmetry of the RF PADs.

Interestingly, the RF PADs have dominant intensity perpendicular to the laser
polarization axis. An A$_{1}$ Rydberg 3s intermediate state would most likely
yield maximum intensity parallel to the laser polarization axis, contrary to
what is observed, since a 3s Rydberg state would ionize to primarily form
p-wave ($l=1$ electrons). As can be seen from visual inspection of the data,
the ratio of perpendicular to parallel photoelectron intensity varies with
time, going through a maximum at around 0.3~ps before decaying again to
smaller values. This ``model-free'' result rules out the A$_{1}$ Rydberg 3s
state as the intermediate configuration.  Corroborated by ab initio
calculations~\cite{Levchenko2006}, the RF PADs were modeled using states of
B$_{2}$ symmetry. It was also assumed that the molecule largely retains
$C_{2v}$ symmetry, supported by the retention of planarity during
dissociation~\cite{Gessner2006, Levchenko2006} as deduced from vector
correlation measurements.

To proceed further, detailed analysis of the RF PADs is required. The outgoing
free electron partial waves are decomposed into symmetry-adapted spherical
harmonics~\cite{Chandra1987, Underwood2000}, as given by
\eqref{eq:MFsymadapwfn}. For $C_{2v}$ , these harmonics are described by their
$C_{2v}$ symmetry and by $l_{|\lambda|}$ , where $l$, $|\lambda|$, are the
orbital angular momentum and projection quantum numbers, respectively. Values
of $l = 0, 1, 2\dots$ are labeled s, p, d$\dots$ whereas values of $|\lambda|
= 0,1,2\dots$ are labeled $\sigma,\pi, \delta\dots$. For the case of the NO
dimer, ionization of a $B_{2}$ electronic state to an $A_{1}$ cation state via
a y-polarized transition (also of $B_{2}$ symmetry) means that the free
electron must have $A_{1}$ symmetry in order to satisfy the requirement in
\eqref{eq:direct_product}. This significantly restricts the allowed free
electron states. Since the fit to Legendre polynomials required $L\leq8$,
partial waves with $l=0\dots4$ are required to model the data.  The $A_{1}$
symmetry partial waves with $l\leq4$ are: s$_{\sigma}$, p$_{\sigma}$,
d$_{\sigma}$, d$_{\delta}$, f$_{\sigma}$, f$_{\delta}$, g$_{\sigma}$,
g$_{\delta}$, and g$_{\gamma}$. In general the s, p, and d waves were
dominant. Modelling of the data would therefore require 9 partial wave
amplitudes and 8 relative phases, and so clearly a unique fit to the data was
not possible. However it was possible to determine the range of partial wave
amplitudes that could reproduce the shape of the RF PADs using the following
method to systematically vary the model parameters. From a starting set of
initial partial wave parameters (amplitudes and phases), the downhill simplex
method~\cite{Nelder1965} was employed to adjust the sum of differences between
the model and experimental $\mathcal{B}_L$ coefficients. This optimization
process adjusted the parameters such that the agreement between model and
experimental $B_L$ coefficients was better than the experimental
uncertainty. This optimization process was carried out in 3 stages: (i) only
s$_\sigma$, p$_\sigma$, d$_\sigma$ and d$_\delta$ amplitudes and phases were
optimized with all other parameters held constant (ii) s$_\sigma$, p$_\sigma$,
d$_\sigma$ and d$_\delta$ amplitudes and phases were held constant at the
optimized values found in the previous step (iii) all parameters were
optimized, starting with the values found in the two previous steps. This
process was carried out for 32 different sets of starting parameters using the
same set of initial parameters for the five time delays.

In order to calculate the RF PAD for a set of partial wave amplitudes and
phases we use \eqref{eq:MFbetalm1} to first calculate the MF PAD. The MF is
defined with the $z$-axis along the $C_{2v}$ symmetry axis, the $y$-axis along
the N--N bond, and the $x$-axis perpendicular to the molecular plane. The RF
plane is defined with the $z$-axis along the N--N bond direction. In order to
calculate the RF PAD from the MF PAD, a rotation is applied to bring the MF
$z$-axis to the RF $z$-axis. The resulting PAD is then azimuthally averaged
about the $z$-axis (the N--N direction),
\begin{equation}
  I(\theta)=\int\mathrm{d}\phi\sum_{LMM'}
  \beta^{\mathrm{M}}_{LM}
  \drot{L}{M}{M'}{\pi/2,\pi/2,0}Y_{LM'}(\theta,\phi).
\end{equation}
Performing the integration over $\phi$ analytically yields
\begin{equation}
  \mathcal{B}_L=2\pi\sum_M
  \beta^{\mathrm{M}}_{LM}
  Y^\ast_{LM}(\pi/2,0)
\end{equation}

In order to obviate the dependence of our conclusions upon any specific
partial wave amplitude, the amplitudes were contracted into two sets: those
expected from 3p$_{y}$ ionization and those not. Ionization of a dimer
3p$_{y}$ Rydberg state via a $y$-polarized transition would, in an ``atomic''
$\Delta l=\pm1$ picture of Rydberg orbital ionization, produce only electrons
with $s_{\sigma}$, $d_{\sigma}$, $d_{\delta}$ character. Therefore, the ratio
of [$s_{\sigma}+d_{\sigma}+d_{\delta}$] to the sum of all other contributions
$\Sigma_{pfg}$ is a measure of 3p$_{y}$ Rydberg character in the
(NO)$_{2}^\ast$ excited electronic states. In \figref{fig:TRCIS_NO_dimer} (bottom)
we plot the time dependence of this ratio, labelled "3p$_{y}$ signal", showing
that dimer 3p$_{y}$ Rydberg character rises from early times, peaks at ~330~fs
and then subsequently falls. The solid curve is the time dependence of the
intermediate configuration extracted from \figref{fig:NO_dimer_TRPES}, showing
that the 3p$_{y}$ character follows the time behaviour of the intermediate
(NO)$_{2}^{\ast\dag}$ configuration. The agreement substantiates
(NO)$_{2}^{\ast\dag}$ as being of 3p$_{y}$ character.

Ab initio studies fully support this picture~\cite{Levchenko2006}.  Briefly, a
very bright diabatic charge transfer (valence) state carries the transition
oscillator strength. At 210nm, a vibrationally excited (roughly estimated,
$\nu_{1}\sim4$) adiabatic (NO)$_{2}^\ast$ state of mixed
charge-transfer/Rydberg character is populated. This quickly evolves, via N=O
stretch dynamics, towards increasing 3p$_{y}$ Rydberg character, forming the
(NO)$_{2}^{\ast\dag}$ state. The 140~fs initial decay constant is the time
scale for the initial valence state to develop intermediate 3p$_{y}$ character
and explains the emergence of 3p$_{y}$ ionization dynamics seen in
\figref{fig:TRCIS_NO_dimer} at intermediate time scales
((NO)$_{2}^{\ast\dag})$. The 590~fs sequential time constant is the time scale
for evolution of the dimer 3p$_{y}$ configuration to free products via
intramolecular vibrational energy redistribution (IVR), coupling the N=O
stretch to the low frequency N--N stretch and other modes. Due to photofragment
indistinguishability, the dimer 3p$_{y}$ state correlates adiabatically to
free NO(A, 3s) + NO(X) products without any curve crossings. With respect to
the 3s Rydberg state, a dimer A$_{1}$ Rydberg 3s state was indeed found but at
lower energy than the bright valence state and does not cross the latter in
the FC region~\cite{Levchenko2006}. It is therefore likely that the dimer 3s
state does not participate in the dissociation dynamics except perhaps far out
in the exit valley where the dimer 3s and 3p states become degenerate and
strongly mix.

\section{Photostability of the DNA Bases}
The UV photostability of biomolecules is determined by the competition between
ultrafast excited state electronic relaxation processes. Some of these, such
as excited state reaction, photodissociation or triplet formation, can be
destructive to the molecule. In order to protect against these, nature
designed mechanisms which convert dangerous electronic energy to less
dangerous vibrational energy. However, in order to have non-zero efficiency,
any such protective mechanisms must operate on ultrafast time scales in order
to dominate over competing photochemical mechanisms that potentially lead to
destruction of the biomolecule.  In DNA, the nucleic bases are not only the
building blocks of genetic material but are also the UV chromophores of the
double helix. It has been suggested that DNA must have photoprotective
mechanisms which rapidly convert dangerous electronic energy into
heat~\cite{Kohler2004}.

The purine bases adenine and guanine and the pyrimidine bases cytosine,
thymine, and uracil are all heterocycles. They typically have strong
$\pi\pi^\ast$ UV absorption bands and, due to the lone electron pairs on the
heteroatoms, have additional low-lying n$\pi^\ast$ transitions. Furthermore,
for some bases $\pi\sigma^\ast$ states are also in a similarly low-energy
range. This can lead to rather complex photophysical properties. Of all the
bases, adenine has been most extensively studied~\cite{Kohler2004}. In the gas
phase, the 9H tautomer of isolated adenine is the lowest energy and most
abundant form. Two competing models were proposed to explain the photophysics
of isolated adenine involving these low-lying states. One predicted internal
conversion from the initially excited $\pi\pi^\ast$ state to the lower
n$\pi^\ast$ state along a coordinate involving six-membered ring
puckering~\cite{Broo1998}. This would be followed by further out-of plane
distortion, initiating relaxation back to the S$_{0}$ ground state. An
alternate model suggested that along the 9-N H-stretch coordinate a two-step
relaxation pathway evolves via conical intersections of the $\pi\pi^\ast$
state with a repulsive $\pi\sigma^\ast$ state, followed by decay back to the
S$_{0}$ ground state~\cite{Sobelewski2002}. Due to the repulsive character of
the $\pi\sigma^\ast$ state, this mechanism was suggested to be highly
efficient. More recently, various other possible relaxation pathways have been
suggested~\cite{Perun2005, Marian2005, Blancafort2006}.  The relative
importance of the electronic relaxation channels in adenine has been a matter
of some debate.

A time-resolved ion yield study of adenine excited state dynamics yielded an
excited state lifetime of $\sim1$~ps and seemed to support the model of
internal conversion via the n$\pi^\ast$ state along a coordinate involving
six-membered ring puckering~\cite{Kang2002}. In order to determine the global
importance of the $\pi\sigma^\ast$ channel, a comparison of the primary
photophysics of adenine with 9-methyl adenine will be useful, as the latter
lacks a $\pi\sigma^\ast$ channel at the excitation energies of concern here.
The first study of this type revealed no apparent changes in excited state
lifetime upon methylation at the N9 position~\cite{Kang2003}: a lifetime of
$\sim 1$~ps was observed for both adenine and 9-methyl adenine. This was
interpreted as evidence that the $\pi\sigma^\ast$ is not involved in adenine
electronic relaxation.

By contrast, the first TRPES studies compared adenine electronic relaxation
dynamics at two different wavelengths, 266~nm vs. 250~nm, and concluded that
the $\pi\sigma^\ast$ state may indeed be important~\cite{Ullrich2004,
  Ullrich2004a}. Additional evidence of $\pi\sigma^\ast$ state participation
obtained from H-atom loss experiments~\cite{Hunig2004, Zierhut2004}. Hydrogen
atom detection is highly sensitive and can reveal even minor H-atom loss
channels.  The observation of fast hydrogen atoms following UV excitation of
adenine is a compelling argument for the $\pi\sigma^\ast$ state: fast H atoms
result from an excited state potential which is repulsive in the N9H
coordinate. Although this shows that a $\pi\sigma^\ast$ channel exists, it
might play only a minor role since the H-atom quantum yield remains unknown. A
more detailed time-resolved ion yield study comparing adenine with 9-methyl
adenine photophysics revealed further insights~\cite{Canuel2005}. The excited
state decay dynamics of adenine at 266~nm excitation required a bi-exponential
fit using two time constants: a fast component decaying in 0.1~ps followed by
a slower component with a 1.1~ps lifetime. Interestingly, 9-methyl adenine
also exhibited that same two time constants of 0.1~ps and 1.1~ps. This again
led to the suggestion that the $\pi\sigma^\ast$ state was not strongly
involved in the dynamics, supporting the earlier ion yield experiments but
contradicting the TRPES results.

\begin{figure}
  \includegraphics[width=4in]{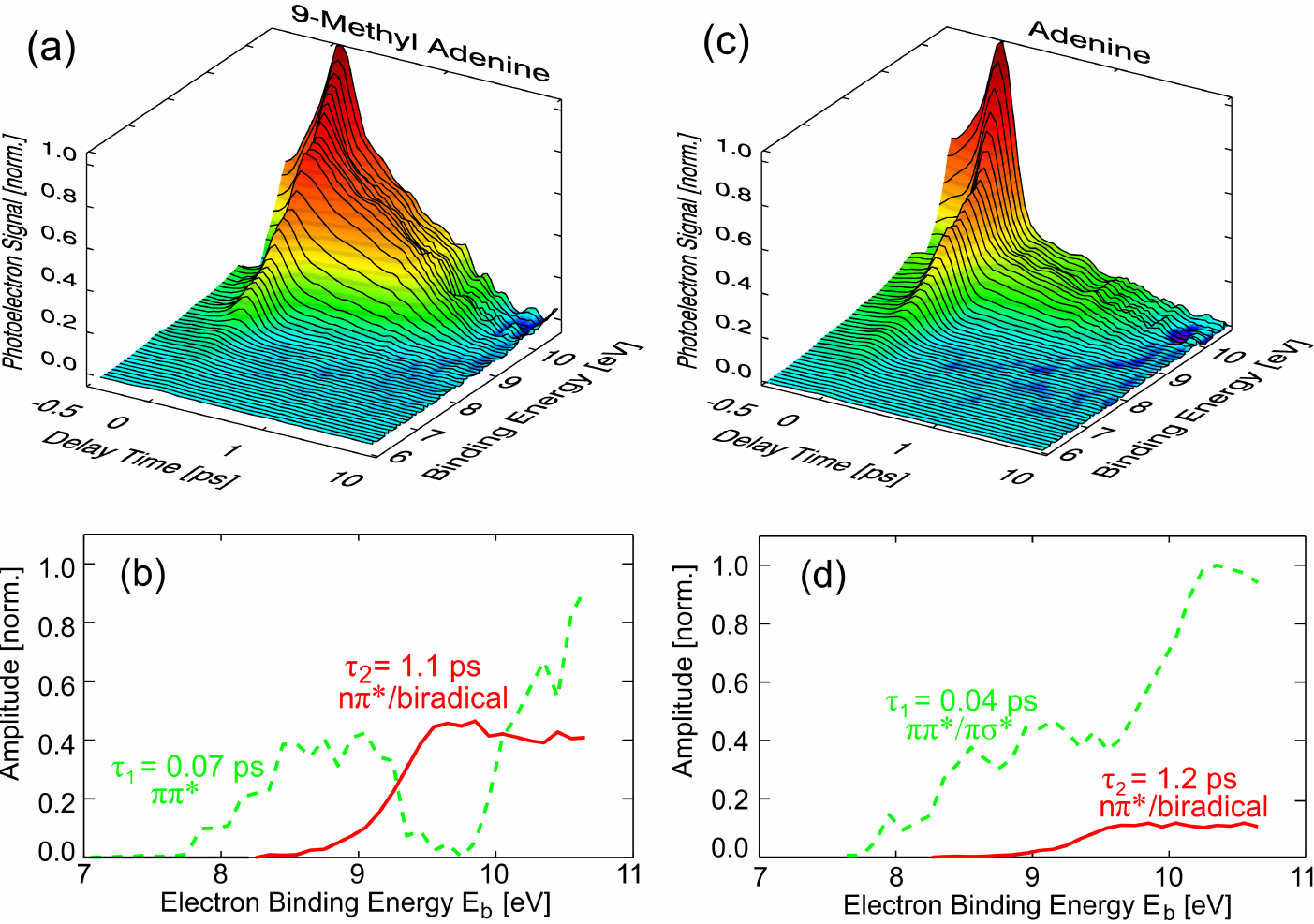}
  \caption{TRPES spectra for adenine (left) and 9-methyl adenine (right),
    pumped at $\lambda_{\mathrm{pump}}=267$~nm and and probed at
    $\lambda_{\mathrm{probe}}=200$~nm.  The time dependence is plotted using a
    linear/logarithmic scale with a linear scale in the region
    $-0.4$--$1.0$~ps and a logarithmic scale for delay times
    $1.0$--$10.0$~ps.}
  \label{fig:Adenine_TRPES}
\end{figure}

More recently, a new TRPES study compared adenine with 9- methyl
adenine~\cite{Satzger2006}, as shown in \figref{fig:Adenine_TRPES}. The
behaviour of the two molecules appears quite similar but there are important
differences, as discussed below. Both molecules exhibit a broad spectral
feature that covers the 7.5--10.8~eV electron binding energy (E$_b)$
range. This feature, peaking towards 10.8~eV, decays quickly and, beyond
500~fs, transforms into a second spectral feature spanning the 8.5--10.8~eV
(E$_b)$ range. This second spectrum grows smoothly between 8.5--9.6~eV and is
flat between 9.6--10.8~eV. This feature decays more slowly, in about
3~ps. Beyond $\sim6$~ps, no remaining photoelectron signal was
observed. Global 2D non-linear fitting algorithms determined that two
exponential time constants were needed to fit these data. For 9-methyl adenine
these were $\tau_1=70\pm25$~fs and $\tau_2=1.1\pm0.1$~ps. For adenine,
these were $\tau_1=40\pm20$~fs and $\tau_2=1.2\pm0.2$~ps. Note that the
two time constants for these molecules are the same within errors and agree
quantitatively with the two time constants previously reported in the ion
yield experiments~\cite{Canuel2005}.

\begin{figure}
  \includegraphics[width=6cm]{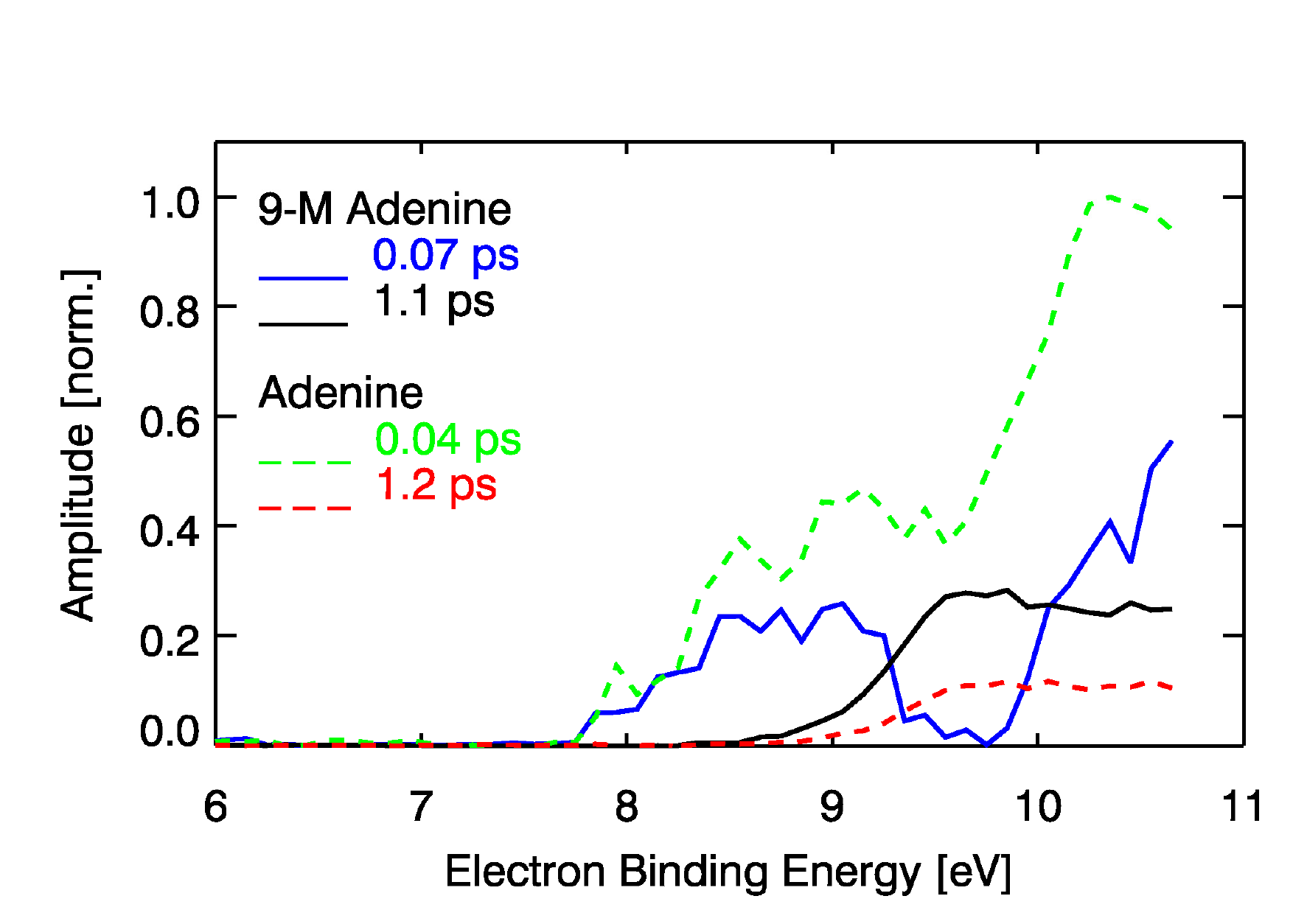}
  \caption{Decay associated spectra for adenine (dashed lines) and 9-methyl
    adenine (solid lines), extracted from the 2D TRPES spectra using global
    fitting procedures. Both molecules were fit by the same two time
    constants: $\tau_1\leq0.1$~ps and $\tau_2\sim1.1$~ps, agreeing
    quantitatively with previous results. The spectra, however, are very
    different for adenine as compared to 9-methyl adenine. For details, see
    the text.}
  \label{fig:Adenine_spectra}
\end{figure}

Although the time constants for adenine and 9-methyl adenine are very similar,
the associated photoelectron spectra reveal important differences that are
obscured in ion-yield measurements. The decay associated spectra obtained from
the fitting algorithm are shown in \figref{fig:Adenine_spectra}. The spectra
of the fast ($<0.1$~ps) components are shown for adenine (dashed green) and
9-methyl adenine (solid blue). Likewise, the spectra of the 1.1~ps components
for adenine (dashed red) and 9-methyl adenine (solid black) are given.  The
electronic states of the cations are D$_0(\pi^{-1})$, D$_1$(n$^{-1})$ and
D$_2$ ($\pi^{-1})$. The expected Koopmans' correlations would therefore be:
$\pi\pi^\ast\rightarrow$D$_0(\pi^{-1})$, D$_2(\pi^{-1})$ and
n$\pi^\ast\rightarrow$D$_1($n$^{-1})$. As detailed
elsewhere~\cite{Satzger2006}, the spectra of the 1.1~ps component correspond
to the n$\pi^*\rightarrow$D$_1$(n$^{-1})$+e$^-$ ionizing transitions. Although
the form of the n$\pi^\ast$ spectra are similar, the yield (amplitude) of
n$\pi^\ast$ state is considerably reduced in adenine as compare to 9-methyl
adenine. The most significant difference lies in the form of the spectra of
the short-lived 0.1~ps component: the spectrum of 9-methyl adenine (blue
solid) appears as two lobes with a gap in between whereas the spectrum of
adenine (dashed green) appears as a broad spectrum without a gap.

\begin{figure}
  \includegraphics[width=\columnwidth]{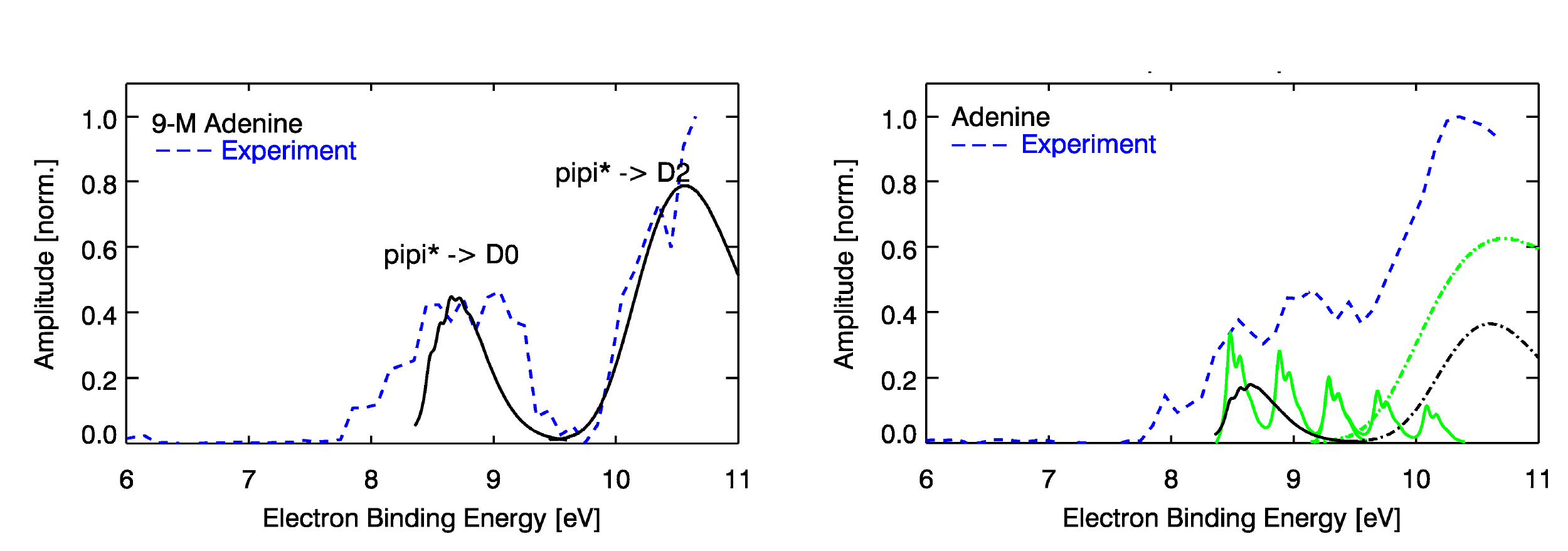}
  \caption{Decay associated spectra of the short-lived state compared with
    calculated FC spectra for 9-methyl adenine (right) and adenine (left).  In
    9-methyl adenine, the $\pi\pi^\ast\rightarrow$D$_0(\pi^{-1})$,
    D$_2(\pi^{-1})$ transitions leave a FC gap. In adenine, this gap is filled
    by the $\pi\sigma^*$ ionizing transitions.}
  \label{fig:Adenine_fits}
\end{figure}

In \figref{fig:Adenine_fits}, we compare the associated spectrum of the fast
component in 9-methyl adenine with calculated~\cite{Satzger2006} Franck-Condon
(FC) structures for the $\pi\pi^*\rightarrow$D$_0(\pi^{-1})$+e$^-$ (solid
line) and $\pi\pi^*\rightarrow$D$_2(\pi^{-1})$+e$^-$ (dash-dotted line)
ionizing transitions. The two separated peaks agree well with the FC
calculations, strongly suggesting that the short-lived state in 9-methyl
adenine is the $\pi\pi^\ast$ state. By contrast, adenine contains an
additional contribution which fills in the gap between the
$\pi\pi^*\rightarrow$D$_{0}(\pi^{-1})$+e$^-$ (black solid) and
$\pi\pi^*\rightarrow$D$_{2}(\pi^{-1})$+e$^-$ (black dotted) transitions. This
gap is filled from the left by transitions due to
$\pi\sigma^*\rightarrow$D$_{0}(\pi^{-1})$+e$^-$ (green solid) and from the
right by $\pi\sigma^*\rightarrow$D$_{2}(\pi^{-1})$+e$^-$ (green dashed)
ionizing transitions. These calculated FC structures provide strong evidence
that the $\pi\sigma^*$ is present in adenine but absent in 9-methyl
adenine~\cite{Satzger2006}. Adenine has two fast relaxation channels from the
$\pi\pi^*$ state, whereas 9-methyl adenine has only one. This also explains
why the yield (amplitude) of n$\pi^\ast$ state is reduced in adenine as
compare to 9-methyl adenine. The fact that the two fast relaxation channels in
adenine have very similar time constants is the reason why the ion yield
experiments showed no apparent difference in lifetimes between adenine and
9-methyl adenine. Once again, the importance of measuring (dispersed)
photoelectron spectra as opposed to (integrated) ion yield spectra is
apparent.

%%% Local Variables: 
%%% mode: latex
%%% TeX-master: "trpes"
%%% End: 

\chapter{Conclusion}
Our goals were to elucidate important physical concepts in energy-angle
resolved TRPES and to illustrate the range of its applicability to problems in
molecular dynamics. We discussed general aspects of femtosecond
pump-probe experiments from both the wavepacket and the frequency domain point
of view.  Experimentalists are, in principle, free to choose a final state
through which to observe the wavepacket dynamics of interest. We emphasized
the critical role of the choice of final state in determining both the
experimental technique (e.g., collection of photons or particles) and the
information content of an experiment (averaged or state-resolved). The
molecular ionization continuum has rich structure which can act as a template
onto which multi-dimensional wavepacket dynamics may be projected. The set of
electronic states of the cation are sensitive to both the electronic
population dynamics and the vibrational dynamics in the excited state, whereas
the free electron continua are sensitive to the electronic population dynamics
and the molecular frame alignment dynamics. In sum, TRPES and its variants are
well suited to the study of excited state polyatomic dynamics because of their
sensitivity to both electronic configurations and vibrational dynamics, the
universal nature of photoionization as a probe, and the dispersed (energy- and
angle-resolved) nature of the measurement.

A powerful variant, TRCIS, measures energy-resolved and 3D angle-resolved
photoions and photoelectrons in coincidence, yielding unprecedented details
about complex molecular photodissociations.  However, TRCIS has potential
beyond the ability to observe time-resolved molecular frame excited state
dynamics. For example, in even more complex dissociation problems, it may be
very difficult to ``follow'' the excited state dynamics all the way from initial
excitation to final product emission. In such cases, one is tempted to resort
to statistical models of the dynamics such as phase space theory. TRCIS
provides a new opportunity to follow the time evolution of the product states
distributions. For example, product attributes such as photofragment kinetic
energy and angular distributions, photofragment angular momentum polarization
and $\mu-v-J$ correlations may all now be measured as a function of time. We
expect that the time evolution of these will be related to the divergence of
phase space flux during dissociation and may well provide new insights into
the timescales for the onset of and the extent of statisticality in energized
molecules.

Future applications of TRPES and its variants will undoubtedly benefit from
ongoing developments in detector technologies, femtosecond and attosecond
laser sources, nonlinear optical frequency conversion schemes, and
developments in free electron lasers and forth generation synchrotron light
sources. TRPES research will include molecular-frame measurements,
photofragment-photoelectron scalar and vector correlations, extreme time
scales, and innershell dynamics. The use of shaped, intense nonresonant laser
fields to create field-free alignment in polyatomic systems~\cite{Lee2006,
  Friedrich1995, Seideman1995, Peronne2003, Lee2004, Bisgaard2004,
  Berakdar2003, Dooley2003, Larsen2000, Machholm2001, Stapelfeldt2003,
  Sugny2004, Underwood2003, Underwood2005, Jauslin2005, Tanji2005, Sakai2003}
will combine with TRPES and TRCIS to help probe molecular-frame
dynamics. Further development of the multiply differential
photoelectron-photofragment coincidence and coincidence-imaging methods will
permit highly detailed investigation of statistical and nonstatistical
photoinduced charge and energy flow, an area of fundamental dynamical interest
and of interest in applications to the gas-phase photophysics of
biomolecules. The development of high average power femtosecond VUV/XUV
sources and the dawn of attosecond science present the possibility of probing
highly excited states, core dynamics, and electron correlation in real
time. Equally important are ongoing theoretical developments in ab initio
molecular dynamics methods for studying non-adiabatic processes in polyatomic
molecules (see, for example ref.~\cite{Martinez2006} and references
therein). New methods for calculating photoionization differential cross
sections (see, for example ref.~\cite{Suzuki2004} and references therein) will
play an increasingly important role in the future of TRPES. These experimental
and theoretical challenges will, we expect, be met by many researchers, surely
leading to exciting new developments in the dynamics of polyatomic molecules.

%%% Local Variables:
%%% mode: latex
%%% TeX-master: "trpes"
%%% End:

\chapter*{Acknowledgements}
\addcontentsline{toc}{chapter}{Acknowledgements} 

We thank our co-workers and collaborators who have contributed both materially
and intellectually to the work describered here: C.~Bisgaard, V.~Blanchet,
A.~Boguslavskiy, A.~L.~L.~East, N.~Gador, O.~Gessner, C.~C.~Hayden, A.~Krylov,
A.~M.~D.~Lee, S.~Lochbrunner, T.~J.~Martinez, K.~L.~Reid, H.~Reisler,
H.~Satzger, M.~Schmitt, T.~Schultz, T.~Seideman, J.~P.~Shaffer, D.~Townsend,
S.~Ullrich and M.~Z.~Zgierski. We thank T.~Suzuki for permission to use Figure
18, K.~L.~Reid for permission to use figures 15, 16, 17, and C.~C.~Hayden for
permission to use Figure 11.

%%% Local Variables: 
%%% mode: latex
%%% TeX-master: "trpes"
%%% End: 

\appendix

\numberwithin{equation}{chapter}
\chapter{Derivation of \eqref{eq:betalm1}}
\label{app:betalm1}
Expanding \eqref{eq:dcs1} and substituting in Eqs~(\ref{eq:mtxel}),
(\ref{eq:density_matrix}), (\ref{eq:multipole_expansion}) and
(\ref{eq:rotintegral}) yields
\begin{equation}
  \label{eq:start}
  \begin{split}
    \sigma(\epsilon, \kL;t)&\propto
    \frac{1}{64\pi^2}
    \sum_{n_{\alpha}n_{\alpha'}}
    \sum_{n_{\alpha_+}}\sum_{\Kpl\Mpl}
    \sum_{lm}\sum_{l'm'}\sum_{\lambda\lambda'}
    \sum_{\Ka\Ma}\sum_{\Kap\Map}
    \sum_{KQ}
    \sum_{pp'}
    \\&\times
    \sum_{qq'}
    \sum_{j_tj'_t}\sum_{k_tk'_t}\sum_{m_tm'_t}
    \phase{\Ja+q+q'+\Ma-2\Ka}
    [j_t,j'_t,\Jpl]
    [K,\Ja,\Jap]^{1/2}
    \\&\times
    \threej{\Jpl}{\Ja}{j_t}{-\Mpl}{\Ma}{m_t}
    \threej{\Jpl}{\Jap}{j'_t}{-\Mpl}{\Map}{m'_t}
    \threej{l}{1}{j_t}{m}{-p}{m_t}
    \\&\times
    \threej{l'}{1}{j'_t}{m'}{-p'}{m'_t}
    \threej{\Jpl}{\Ja}{j_t}{-\Kpl}{\Ka}{k_t}
    \threej{\Jpl}{\Jap}{j'_t}{-\Kpl}{\Kap}{k'_t}
    \\&\times
    \threej{l}{1}{j_t}{\lambda}{-q}{k_t}
    \threej{l'}{1}{j'_t}{\lambda'}{-q'}{k'_t}
    \threej{J_{\alpha}}{J'_{\alpha'}}{K}{M_{\alpha}}{-M'_{\alpha'}}{-Q}
    \\&\times
    Y_{lm}(\uvect{k})
    Y^\ast_{l'm'}(\uvect{k})
    \multmom{n_{\alpha}}{n'_{\alpha'};t}{K}{Q}
    e_{-p}e_{-p'}^\ast
    \\&\times
    a_{K_\alpha}^{J_\alpha\tau_\alpha}
    a_{K'_{\alpha'}}^{J'_{\alpha'}\tau'_{\alpha'}}
    \left|
      a_{K_\alp}^{J_\alp\tau_\alp}
    \right|^2
    \mathcal{E}(n_\alp, n_\alpha,\epsilon)
    \mathcal{E}^\ast(n_\alp, n'_{\alpha'},\epsilon).
    \\&\times
    \sum_{\Gamma\mu h}
    \sum_{\Gamma'\mu'h'}
    b^{\Gamma\mu}_{hl\lambda}
    b^{\Gamma'\mu'\ast}_{h'l'\lambda'}
    \mathrm{(-i)}^{l-l'}\exp{\mathrm{i}(\sigma_l(\epsilon)-\sigma_{l'}(\epsilon))}
    \\&\times
    D_{\Gamma\mu h l}^{{\alpha}v_{\alpha}{\alp}v_\alp}(q)
    D_{\Gamma'\mu' h' l'}^{{\alpha'}v'_{\alpha'}{\alp}v_{\alp}\ast}(q')
  \end{split}
\end{equation}
The various angular momentum algebraic manipulations outlined below draw on
the text by Zare~\cite{Zare1988}.  The two spherical harmonics in
\eqref{eq:start} may be combined using the Clebsch-Gordan series,
\begin{multline}
  \label{eq:shmcombo}
  Y_{lm}(\kL)
  Y^\ast_{l'm'}(\kL)=
  \sqrt{\frac{[l,l']}{4\pi}}
  \phase{m}
  \sum_{L}[L]^{1/2}
  \\\times
  \threej{l}{l'}{L}{-m}{m'}{M}
  \threej{l}{l'}{L}{0}{0}{0}
  Y_{LM}(\kL).
\end{multline}
Eq. (4.16) of Zare~\cite{Zare1988} is used to perform the following
manipulations,
\begin{multline}
  \threej{\Jpl}{\Ja}{j_t}{-\Mpl}{\Ma}{m_t}
  \threej{\Jpl}{\Jap}{j'_t}{-\Mpl}{\Map}{m'_t}
  =
  \\
  \sum_X[X]\phase{J+j_t-\Jpl+j'_t+\Jap+X-\Ma+m'_t}
  \sixj{\Ja}{j_t}{\Jpl}{j'_t}{\Jap}{X}
  \\\times
  \threej{\Jap}{\Ja}{X}{-\Map}{\Ma}{x}
  \threej{j_t}{j'_t}{X}{m_t}{-m'_t}{-x},
\end{multline}
\begin{multline}
  \threej{l}{1}{j_t}{m}{-p}{m_t}
  \threej{j_t}{j'_t}{X}{m_t}{-m'_t}{-x}
  =
  \sum_Y[Y]\phase{l+1+2j'_t+2X+Y-m-m'_t}
  \\\times
  \sixj{l}{1}{j_t}{j'_t}{X}{Y}
  \threej{X}{l}{Y}{x}{m}{y}
  \threej{1}{j'_t}{Y}{-p}{m'_t}{-y},
\end{multline}
\begin{multline}
  \threej{l'}{1}{j'_t}{m'}{-p'}{m'_t}
  \threej{1}{j'_t}{Y}{-p}{m'_t}{-y}
  =
  \sum_P[P]\phase{l'-j'_t+Y+P-m'-p}
  \\\times
  \sixj{l'}{1}{j'_t}{1}{Y}{P}
  \threej{Y}{l'}{P}{y}{m'}{p-p'}
  \threej{1}{1}{P}{-p'}{p}{p'-p},
\end{multline}
\begin{multline}
  \threej{Y}{l'}{P}{y}{m'}{p-p'}
  \threej{X}{l}{Y}{x}{m}{y}
  =
  \sum_G[G]\phase{X+l+G+m'-x}
  \\\times
  \sixj{l'}{P}{Y}{X}{l}{G}
  \threej{l}{l'}{G}{m}{-m'}{g}
  \threej{P}{X}{G}{p'-p}{x}{-g}.
\end{multline}
The orthogonality of the Wigner $3j$ symbols is then used to perform the
summations over $m$, $m'$, $\Ma$ and $\Map$,
\begin{equation}
  \sum_{mm'}
  \threej{l}{l'}{G}{m}{-m'}{g}
  \threej{l}{l'}{L}{-m}{m'}{M}
  =\phase{l+l'+L}[L]^{-1}\delta_{LG}\delta_{-Mg},
\end{equation}
\begin{multline}
  \sum_{\Ma\Map}
  \threej{\Jap}{\Ja}{X}{-\Map}{\Ma}{x}
  \threej{J_{\alpha}}{J'_{\alpha'}}{K}{M_{\alpha}}{-M'_{\alpha'}}{-Q}=
  \\\phase{\Ja+\Jap+K}[K]^{-1}\delta_{KX}\delta_{-Qx}.
\end{multline}
The sum over $Y$ is carried out analytically using the following identity
relating the Wigner $9j$ symbol to Wigner $6j$ symbols:
\begin{equation}
  \sum_Y\phase{2Y}[Y]
  \sixj{l'}{1}{j'_t}{1}{Y}{P}
  \sixj{l'}{P}{Y}{K}{l}{L}
  \sixj{l}{1}{j_t}{j'_t}{K}{Y}
  =
  \ninej{1}{1}{P}{j_t}{j_t'}{K}{l}{l'}{L}.
\end{equation}

\chapter{Derivation of \eqref{eq:betalm2}}
\label{app:betalm2}
Eq. (4.16) of Zare~\cite{Zare1988} is used to perform the following
manipulations, 
\begin{multline}
  \threej{\Jpl}{\Ja}{j_t}{-\Kpl}{\Ka}{k_t}
  \threej{\Jpl}{\Jap}{j'_t}{-\Kpl}{\Kap}{k'_t}
  =\\
  \sum_R[R]\phase{\Ja+j_t-\Jap+j'_t+\Jap+R-\Ka+k'_t}
  \sixj{\Ja}{j_t}{\Jpl}{j'_t}{\Jap}{R}
  \\\times
  \threej{\Jap}{\Ja}{R}{-\Kap}{\Ka}{r}
  \threej{j_t}{j'_t}{R}{k_t}{-k'_t}{-r},
\end{multline}
\begin{multline}
  \threej{l}{1}{j_t}{\lambda}{-q}{k_t}
  \threej{j_t}{j'_t}{R}{k_t}{-k'_t}{-r}
  =\sum_S[S]\phase{l+1-j_t+R+j'_t+S-\lambda-r}
  \\\times
  \sixj{l}{1}{j_t}{R}{j'_t}{S}
  \threej{j'_t}{l}{S}{k'_t}{\lambda}{s}
  \threej{1}{R}{S}{-q}{r}{-s},
\end{multline}
\begin{multline}
  \threej{l'}{1}{j'_t}{\lambda'}{-q'}{k'_t}
  \threej{j'_t}{l}{S}{k'_t}{\lambda}{s}
  =\sum_T[T]\phase{l'+1-j'_t+S+l+T-\lambda'+s}
  \\\times
  \sixj{l'}{1}{j'_t}{S}{l}{T}
  \threej{l}{l'}{T}{-\lambda}{\lambda'}{t}
  \threej{1}{S}{T}{-q'}{-s}{-t},
\end{multline}
\begin{multline}
  \threej{1}{R}{S}{-q}{r}{-s}
  \threej{1}{S}{T}{-q'}{-s}{-t}
  =\sum_U[U]\phase{R+2T+1+U+q-t}
  \\\times
  \sixj{1}{R}{S}{T}{1}{U}
  \threej{1}{1}{U}{q'}{-q}{q-q'}
  \threej{R}{T}{U}{r}{t}{q'-q}.
\end{multline}
The sum over $S$ can be carried out analytically by relating the Wigner $6j$
symbols to the Wigner $9j$ symbol,
\begin{equation}
  \sum_S
  \phase{2S}[S]
  \sixj{l}{1}{j_t}{R}{j'_t}{S}
  \sixj{l'}{1}{j'_t}{S}{l}{T}
  \sixj{1}{R}{S}{T}{1}{U}
  =\ninej{1}{1}{U}{l}{l'}{T}{j_t}{j'_t}{R}.
\end{equation}
The summation over $\Jpl$ is then completed using the orthogonality of the
Wigner $6j$ symbols,
\begin{equation}
  \sum_{\Jpl}[\Jpl,R]
  \sixj{\Ja}{j_t}{\Jpl}{j'_t}{\Jap}{R}
  \sixj{\Ja}{j_t}{\Jpl}{j'_t}{\Jap}{K}
  =\delta_{RK}.
\end{equation}
together with the fact that
$\sum_{\Jpl}\left|a_{K_\alp}^{J_\alp\tau_\alp}\right|^2=1$.  Rearranging the
Wigner $9j$ symbol in \eqref{eq:betalm1},
\begin{equation}
  \ninej{1}{1}{P}{j_t}{j_t'}{K}{l}{l'}{L}
  =\phase{l+l'+L+P+j_t+j'_t+K}
  \ninej{1}{1}{P}{l}{l'}{L}{j_t}{j_t'}{K},
\end{equation}
allows the use of the orthogonality of the Wigner $9j$ symbols to remove the
summation over $j_t$ and $j'_t$,
\begin{equation}
  \sum_{j_tj'_t}[j_t,j'_t,L,P]
  \ninej{1}{1}{P}{l}{l'}{L}{j_t}{j_t'}{K}
  \ninej{1}{1}{U}{l}{l'}{T}{j_t}{j'_t}{K}
  =\delta_{PU}\delta_{LT}.
\end{equation}

\chapter{Derivation of \eqref{eq:MFbetalm}}
\label{app:mfpad}
Substitution of \eqref{eq:MFsymadapwfn}, \eqref{eq:erotMF} and
\eqref{eq:MFd_dot_e} into \eqref{eq:dcsMF} yields
\begin{equation}
  \begin{split}
    \sigma(\epsilon, \kM;t)
    &\propto
    \sum_{pp'}
    \sum_{qq'}
    \sum_{ll'}
    \sum_{\lambda\lambda'}
    \phase{q+q'}
    Y_{l\lambda}(\uvect{k})
    Y^\ast_{l'\lambda'}(\uvect{k})
    \Drot{1}{-p}{-q}
    \\&\times
    \Drot{1\ast}{-p'}{-q'}
    e_{-p}e^\ast_{-p'}
    \mathrm{(-i)}^{l-l'}
    \exp{\mathrm{i}(\sigma_l(\epsilon)-\sigma_{l'}(\epsilon))}
    \\&\times
    \sum_{{\alpha}v_{\alpha}}
    \sum_{{\alpha'}v'_{\alpha'}}
    C_{{\alpha}v_{\alpha}}(t)
    C_{{\alpha'}v'_{\alpha'}}^\ast(t)
    \sum_{\Gamma\mu h}
    \sum_{\Gamma'\mu'h'}
    b^{\Gamma\mu}_{hl\lambda}
    b^{\Gamma'\mu'\ast}_{h'l'\lambda'}
    \\&\times
    \sum_{\alp v_\alp}
    D_{\Gamma\mu h l}^{{\alpha}v_{\alpha}{\alp}v_\alp}(q)
    D_{\Gamma'\mu' h' l'}^{{\alpha'}v'_{\alpha'}{\alp}v_{\alp}\ast}(q')
    \\&\times
    \mathcal{E}(\alpha,v_\alpha,\alp,v_\alp, \epsilon)
    \mathcal{E}^\ast(\alpha',v'_{\alpha'},\alp,v_\alp, \epsilon).
  \end{split}
\end{equation}
This equation may be simplified using the Clebsch-Gordan series,
\begin{multline}
  Y_{l\lambda}(\kM)
  Y^\ast_{l'\lambda'}(\kM)=
  \sqrt{\frac{[l,l']}{4\pi}}
  \phase{\lambda}
  \sum_{L}[L]^{1/2}
  \\\times
  \threej{l}{l'}{L}{-\lambda}{\lambda'}{M}
  \threej{l}{l'}{L}{0}{0}{0}
  Y_{LM}(\kM),
\end{multline}
\begin{multline}
  \Drot{1}{-p}{-q}
  \Drot{1\ast}{-p'}{-q'}=\\
  \phase{p+q}\sum_P[P]
  \threej{1}{1}{P}{p}{-p'}{p'-p}
  \threej{1}{1}{P}{q}{-q'}{q'-q}
  \\\times
  \Drot{P}{p'-p}{q'-q}.
\end{multline}

%%% Local Variables: 
%%% mode: latex
%%% TeX-master: "trpes"
%%% End: 

\bibliography{refs}

\end{document}